\documentclass[cernpreprint,atlasdraft=true,texlive=2016,british,texmf]{atlasdoc}
 
\usepackage[subcaption]{atlaspackage}
\usepackage{atlasbiblatex}
 
\usepackage{atlasphysics}
 
\graphicspath{{logos/}{figures/}}
 
\usepackage{ANA-JETM-2018-05-PAPER-defs}

\AtlasTitle{Jet energy scale and resolution measured in proton--proton collisions at $\sqrt{s}=13$~TeV with the ATLAS detector}
 
\AtlasAbstract{
Jet energy scale and resolution measurements with their associated uncertainties are reported for jets using 36--81~fb$^{-1}$ of proton--proton collision data with a centre-of-mass energy of $\sqrt{s}=13$~\TeV{} collected by the ATLAS detector at the LHC. Jets are reconstructed using two different input types: \topos\ formed from energy deposits in calorimeter cells, as well as an algorithmic combination of charged-particle tracks with those \topos, referred to as the ATLAS particle-flow reconstruction method. The anti-$k_t$ jet algorithm with radius parameter $R=0.4$ is the primary jet definition used for both jet types. This result presents new jet energy scale and resolution measurements in the high \pileup conditions of late LHC \RunTwo as well as a full calibration of particle-flow jets in ATLAS.
Jets are initially calibrated using a sequence of simulation-based corrections.
Next, several \insitu techniques are employed to correct for differences between data and simulation and to measure the resolution of jets.
The systematic uncertainties in the jet energy scale for central jets ($|\eta|<1.2$) vary from 1\% for a wide range of high-\pt jets ($250<\pT<2000~\GeV$), to 5\% at very low \pt ($20~\GeV$) and 3.5\% at very high \pt ($>2.5~\TeV$). The relative jet energy resolution is measured and ranges from ($24 \pm 1.5$)\% at 20~\GeV{} to ($6 \pm 0.5$)\% at 300~\GeV.
}
 
\author{The ATLAS Collaboration}
 
\AtlasRefCode{JETM-2018-05}

\PreprintIdNumber{CERN-EP-2020-083}

\AtlasJournalRef{Eur. Phys. J. C 81 (2021) 689}
\AtlasDOI{10.1140/epjc/s10052-021-09402-3}

\hypersetup{pdftitle={ATLAS document},pdfauthor={The ATLAS Collaboration}}
 
\begin{document}
 
\maketitle
 
\tableofcontents

\newpage

\section{Introduction}
\label{sec:intro}
The energetic proton--proton (\pp) collisions produced by the Large Hadron Collider (LHC) yield final states that are predominantly characterized by jets, or collimated sprays of charged and neutral hadrons. Jets constitute an essential piece of the physics programme carried out using the ATLAS detector due to their presence in the signal processes being measured and searched for, the various background processes that hide those signals, and the additional activity due to simultaneous \pp collisions. Measurements of the energy scale and resolution of these complex objects, as well as their associated systematic uncertainties, are therefore essential both for precision measurements of the Standard Model (SM) and for sensitive searches for new physics beyond it. This paper presents the strategy used for the determination of the jet energy scale (JES) and resolution (JER) by the ATLAS experiment and its implementation as it pertains to the analysis of data from \RunTwo of the LHC. Results for the JES and JER are presented using data collected during 2015--2017, corresponding to integrated luminosities in the range 36--81~\ifb, depending on the analysis method and its goals. This publication focuses on calibrating jets reconstructed with the \antikt~\cite{Cacciari:2008gp} algorithm with radius parameter $R=0.4$.
 
The ATLAS Collaboration has published previous calibrations and uncertainties of the energy scale and resolution for this jet definition with data taken in 2010~\cite{PERF-2011-03,PERF-2011-05,PERF-2011-04}, 2011~\cite{PERF-2012-01}, 2012~\cite{PERF-2014-02}, and 2015~\cite{PERF-2016-04}.  Additionally, some ATLAS publications have targeted different jet definitions. In particular, the \RunOne papers include dedicated calibrations\footnote{Comparisons in \RunOne between $R=0.4$ and $R=0.6$ jets confirm the need for dedicated calibrations for different jet radii~\cite{ATLAS-CONF-2015-037}.} of jets reconstructed with the \antikt algorithm with $R=0.6$ and $R=1.0$, and a dedicated \insitu calibration of large-radius jets has also been completed in \RunTwo data~\cite{JETM-2018-02}. This publication extends and improves on previous calibrations of \antikt $R=0.4$ jets, taking full advantage of the larger dataset recorded over the period of 2015--2017.
The significant increase in the number of proton collisions per bunch crossing in 2016 and 2017 data-taking leads to a correspondingly more difficult environment for jet reconstruction, and this result presents new jet energy scale and resolution measurements in these unique high pile-up conditions.
 
\Sect{\ref{sec:detector}} describes the ATLAS detector, and \Sect{\ref{sec:dataMC}} describes the recorded data and the Monte Carlo (MC) simulation samples used in this paper. \Sect{\ref{sec:reconstruction}} presents the inputs and algorithms used to reconstruct the jets. \Sect{\ref{sec:jes}} and \Sect{\ref{sec:jer}} present the methods used and the result of both the calibration and the resulting systematic uncertainties of the JES and the JER, respectively.
 
\section{The ATLAS detector}
\label{sec:detector}
 
\newcommand{\AtlasCoordFootnote}{ATLAS uses a right-handed coordinate system with its origin at the nominal
interaction point in the centre of the detector. The positive $x$-axis is defined by the direction from the interaction point to the centre of the LHC ring, with the positive $y$-axis pointing upwards, while the beam direction defines the $z$-axis. Cylindrical coordinates $(r,\phi)$ are used in the transverse plane, $\phi$ being the azimuthal angle around the $z$-axis. The pseudorapidity $\eta$ is defined in terms of the polar angle $\theta$ by $\eta=-\ln\tan(\theta/2)$. Rapidity is defined as $y$ = $0.5 \ln[(E + p_z)/(E-p_z)]$, where $E$ denotes the energy and $p_z$ is the component of the momentum along the beam direction. The angular distance $\DeltaR$ is defined as $\sqrt{(\Delta y)^2 + (\Delta \phi)^2}$.}
 
The ATLAS detector~\cite{PERF-2007-01} at the LHC covers nearly the entire solid angle around the collision point.\footnote{\AtlasCoordFootnote}
It consists of an inner tracking detector surrounded by a thin superconducting solenoid, electromagnetic and hadronic calorimeters, and a muon spectrometer incorporating three large superconducting toroidal magnets.
The inner-detector system (ID) is immersed in a \SI{2}{\tesla} axial magnetic field and provides charged-particle tracking in the range \(|\eta| < 2.5\).
 
The silicon pixel detector covers the vertex region and typically provides four measurements per track,
with the innermost space-point provided by the insertable B-layer that was installed before Run~2~\cite{ATLAS-TDR-2010-19,PIX-2018-001}.
The pixel detector is followed by the silicon microstrip tracker, which usually yields eight measurements per track.
The silicon-based detectors are complemented by the transition radiation tracker (TRT),
which enables radially extended track reconstruction up to \(|\eta| = 2.0\).
The TRT also provides electron identification information
based on the fraction of hits (typically 30 in total) above a higher energy-deposit threshold corresponding to transition radiation.
 
The calorimeter system covers the pseudorapidity range \(|\eta| < 4.9\).
Within the region \(|\eta|< 3.2\), high-granularity lead/liquid-argon (LAr) calorimeters with both barrel and endcap sections provide electromagnetic calorimetry.
An additional thin LAr presampler covers \(|\eta| < 1.8\),
and is used to correct for energy loss in materials traversed by particles prior to reaching the calorimeters.
Hadronic calorimetry is provided by the steel/scintillator-tile calorimeter,
segmented into three barrel structures within \(|\eta| < 1.7\), and two copper/LAr hadronic endcap calorimeters cover the range $1.5 < |\eta|< 3.2$.
The solid angle coverage between $3.2 < |\eta|< 4.9$ is completed with forward copper/LAr and tungsten/LAr calorimeter modules
optimized for electromagnetic and hadronic measurements respectively.
Interfaces that exist between each of these components, in particular between the barrel and endcap regions, provide for space to route various services and infrastructure, such as electrical and fiber-optic cabling, cooling, and  support structures.
However, these so-called \textit{transition regions} also create discontinuities in the response of the calorimeter to both charged and neutral particles due to energy absorption in the inactive materials and changes in the geometry of the active materials of the calorimeters.
The calibrated response and resolution of the calorimeter must therefore either correct for these features, or account for them when establishing systematic uncertainties.
\Fig{\ref{fig:calorimeters}} shows the many components of the calorimeter system, with reference pseudorapidities and various relevant transition regions marked as well~\cite{,Group_2008,PERF-2007-01,ATLAS-TDR-1}.
 
\begin{figure}[hbp!]
\begin{center}
\includegraphics[width=\linewidth]{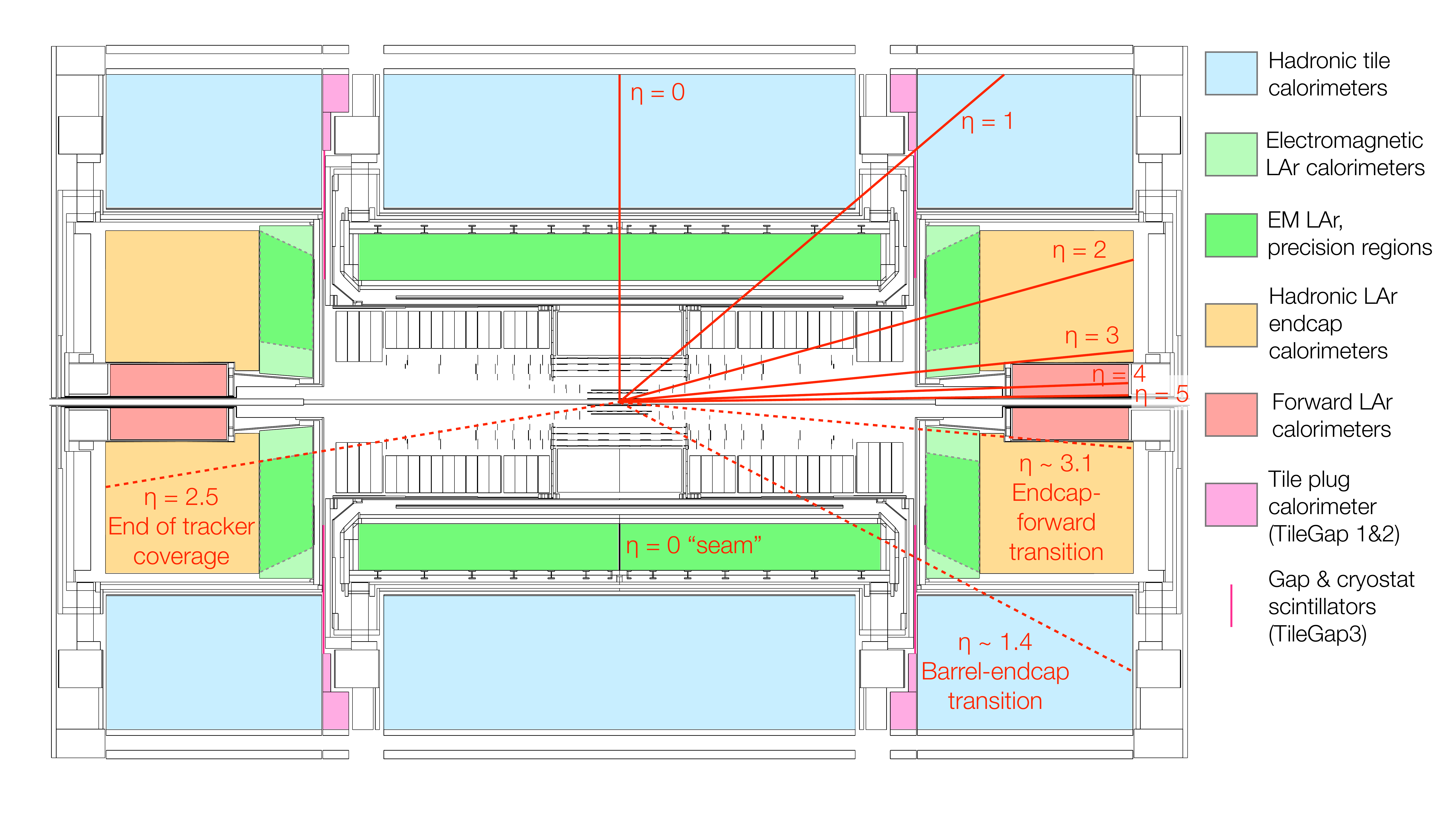}\\
\caption{Layout of the ATLAS calorimeters with pseudorapitidy ($\eta$) values marked for reference. The inner detector systems can be seen in black-and-white in the center of the diagram; tracking is provided up to $\eta = 2.5$. The electromagnetic (EM) barrel and endcap calorimeters are shown in green. The EM barrel has consistent performance throughout, but has a seam in the construction at $\eta=0$ which can impact jet energy resolution. The EM endcap has a precision region marked in darker green and an extended region in light green, and the transition from one to the other around $\eta\sim2.5$ involves a dramatic change in the material layers. The hadronic Tile calorimeter is shown in light blue while the hadronic endcap calorimeters based on liquid argon are illustrated in light orange. The forward calorimeters are shown in dark orange.  Pink filled regions represent the tile plug calorimeter, often referred to as TileGap1 and TileGap2. The thin hot pink line marks the location of the very narrow gap and cryostat scintillators (TileGap3). The regions corresponding to the transition from barrel to endcap ($\eta\sim1.4$) and from endcap to forward calorimeter ($\eta\sim3.1$) are given for reference.}
\label{fig:calorimeters}
\end{center}
\end{figure}
 
The muon spectrometer (MS) comprises separate trigger and
high-precision tracking chambers measuring the deflection of muons in a magnetic field generated by superconducting air-core toroids.
The field integral of the toroids ranges between \num{2.0} and \SI{6.0}{\tesla\metre}
across most of the detector.
A set of precision chambers covers the region \(|\eta| < 2.7\) with three layers of monitored drift tubes,
complemented by cathode-strip chambers in the forward region, where the background is highest.
The muon trigger system covers the range \(|\eta| < 2.4\) with resistive-plate chambers in the barrel, and thin-gap chambers in the endcap regions.
 
Interesting events are selected to be recorded by the first-level trigger system implemented in custom hardware,
followed by selections made by algorithms implemented in software in the high-level trigger~\cite{TRIG-2016-01}.
The first-level trigger accepts events from the \SI{40}{\MHz} bunch crossings at a rate below \SI{100}{\kHz},
which the high-level trigger reduces in order to record events to disk at about \SI{1}{\kHz}.
 
 
\section{Data and Monte Carlo simulated samples}
\label{sec:dataMC}
The data used for the measurements presented here were collected in \pp collisions at the LHC with a centre-of-mass energy of 13~\TeV\ and a 25~ns proton bunch crossing interval during 2015--2017. The integrated luminosities of the datasets used are in the range 36--81~\ifb~after requiring that all detector subsystems were operational during data recording.
 
Additional \pp collisions in the same and nearby bunch crossings are referred to as \textit{\pileup}.
The number of reconstructed primary vertices (\Npv) and the mean number of interactions per bunch crossing ($\mu$) are optimal observables to quantify the level of \pileup activity.
The average value of $\mu$ is 13.7, 24.9, and 37.8 in the 2015, 2016, and 2017 datasets, respectively~\cite{ATLAS-CONF-2019-021}. As described below, these conditions are accounted for in the production and reconstruction of simulated data.
 
\begin{table}[t!]
\centering
\caption{List of generators used for various processes. Information is given regarding the underlying-event tunes, the PDF parameter sets, and the perturbative QCD highest-order accuracy used in the matrix element. Abbreviations in the PDF names and matrix element orders are LO (leading order), NLO (next-to-leading order), and NNLO (next-to-next-to-leading order).}
{\small
\begin{tabular}{cccccc}
\toprule
Process &    Generator                             & Tune & PDF set  & Matrix element \\
& + fragmentation/hadronization  &          &                & order \\
\midrule
Dijet & \PYTHIA 8.186~\cite{Sjostrand:2007gs}
& A14~\cite{ATL-PHYS-PUB-2014-021}
& {\textsc NNPDF2.3LO}~\cite{Ball:2012cx}
& LO \\
\& & \POWPYTHIA 8.186~\cite{Alioli:2010xd,Frixione:2007vw,Sjostrand:2007gs}
& A14 & {\textsc NNPDF2.3LO}~\cite{Ball:2012cx}
& NLO \\
multijet & \SHERPA 2.1.1~\cite{Gleisberg:2008ta}
& \SHERPA-default
& CT10~\cite{Lai:2010vv}
& LO (2$\rightarrow$2+2$\rightarrow$3) \\
& \HERWIG 7.0.4~\cite{Bellm:2015jjp}
& H7UE~\cite{Bellm:2015jjp}
& \textsc{NNPDF3.0NLO}~\cite{Ball:2014uwa}
& LO \\
\midrule
\zjet & \POWPYTHIA 8.186~\cite{Alioli:2010xd,Frixione:2007vw,Sjostrand:2007gs}
& AZNLO~\cite{STDM-2012-23}
& CT10~\cite{Lai:2010vv}
& Z+0j@NLO \\
& \SHERPA 2.2.1~\cite{Gleisberg:2008ta}
& \SHERPA-default
& \textsc{NNPDF3.0NNLO}~\cite{Ball:2014uwa}
& Z+0,1,2j@NLO \\
\midrule
\gammajet &  \PYTHIA 8.186~\cite{Sjostrand:2007gs}
& A14~\cite{ATL-PHYS-PUB-2014-021}
& \textsc{NNPDF2.3LO}~\cite{Ball:2012cx}
& LO \\
& \SHERPA 2.1.1~\cite{Gleisberg:2008ta}
& \SHERPA-default
& CT10~\cite{Lai:2010vv}
& LO \\
\bottomrule
\end{tabular}
}
\label{tab:mcsamples}
\end{table}
 
Simulated dijet, multijet, \zjet, and \gammajet samples are used in determining the jet energy scale and its uncertainties. \Tab{\ref{tab:mcsamples}} summarizes the MC generators, adjustable sets of parameters (tunes), and parton distribution function (PDF) sets used for all nominal and alternative samples of the various simulated processes.  The nominal samples for the majority of analyses were generated with \PYTHIAV{8.186}~\cite{Sjostrand:2007gs} (from now on referred to as \PYTHIA{}8) or \POWPYTHIA{}8.186~\cite{Alioli:2010xd,Frixione:2007vw,Sjostrand:2007gs}. The multijet balance analysis uses \SHERPA 2.1.1~\cite{Gleisberg:2008ta} as the nominal generator since it incorporates up to three jets in the matrix element and is thus more suitable for multijet processes that have more than two jets in the final state. The dijet, multijet, and \gammajet nominal samples use the \textsc{NNPDF2.3LO} PDF set~\cite{Ball:2012cx} and the A14 set of tuned parameters~\cite{ATL-PHYS-PUB-2014-021}.  For the \zjet analysis, the dedicated AZNLO tune~\cite{STDM-2012-23} is used instead. Alternative samples for defining systematic variations use various generators and tunes.
 
Stable particles, defined as those with $c\tau > 10$~mm, output by the generators were passed through the $\GEANT4$-based simulation of the ATLAS detector~\cite{SOFT-2010-01,Agostinelli:2002hh}. This step simulates the interactions of the particles with matter in the detector and generates outputs which can be reconstructed in the same way as data. Hadronic showers were simulated using the FTFP BERT model as described in Ref.~\cite{PERF-2015-05}. A set of simulated dijet events using the less detailed Atlfast-II (AFII) are also studied to determine the difference in performance between full and fast simulation and provide appropriate calibrations for AFII samples in analyses~\cite{SOFT-2010-01}.
 
\Pileup is incorporated in the MC samples by overlaying simulated inelastic interactions on the generated hard-scatter interaction. The inelastic interactions were simulated in \PYTHIAV{8.210} using the A3 tune and the \textsc{NNPDF2.3LO} PDF set~\cite{ATL-PHYS-PUB-2016-017,Ball:2012cx}.
To determine the number of simulated \pp collisions to overlay onto a particular hard-scattering process, a random value is drawn from a Poisson distribution of the number of \pp collisions per bunch crossing with a mean given by the desired average number of collisions per crossing for a particular data period. Events simulated with a particular \pileup profile are then compared with data from the corresponding data period.
One set of MC samples was created using the \pileup profile of 2015+2016 data (average number of collisions 23.7) while a second independent set of samples used the profile of 2017 data. When data and simulation are compared in this paper, both sets of MC samples are used unless otherwise specified and are normalized to the luminosity of 2015+2016 data and 2017 data separately.
 
\section{Jet reconstruction}
\label{sec:reconstruction}
The primary jet definition used in the majority of physics analyses by the ATLAS Collaboration and in the studies presented here is the \antikt~\cite{Cacciari:2008gp} algorithm with a radius parameter $R=0.4$ as implemented in the \Fastjet 3.2.2~\cite{Fastjet,Cacciari:2005hq} software package. Four-vector objects are used as inputs to the algorithm, and may be stable particles defined by MC generators, charged-particle tracks, calorimeter energy deposits,
or algorithmic combinations of the latter two, as in the case of the particle-flow reconstruction technique~\cite{PERF-2015-09}.

For use in jet reconstruction, calorimeter cells are first clustered into three-dimensional, massless, topological clusters (\topos) using a nearest-neighbour algorithm~\cite{PERF-2014-07}. Cells are added to a \topo\ according to the ratio of the cell energy to the expected noise in each cell using thresholds that control the growth of each \topo. The resulting energy of the \topo\ is defined at the electromagnetic (EM) scale, which is the baseline calorimeter scale that correctly measures energy depositions from electromagnetic showers. Only positive-energy \topos\ are used as inputs to the jet reconstruction.  A jet produced in the hard-scatter process is expected to originate from the primary vertex, defined as the reconstructed vertex with at least two associated tracks and the largest sum of squared track momentum.  Therefore, an event-by-event correction to account for the position of the primary vertex in each event -- referred to as an \textit{origin correction} -- is applied to every \topo, based on its depth within the calorimeter and pseudorapidity. This method is to be contrasted with earlier approaches~\cite{PERF-2016-04} that applied this correction only to the jet four-momentum rather than to its constituents.

Jets reconstructed using only calorimeter-based energy information use the origin-corrected EM scale \topos\ and are referred to as \textit{\EMTopo jets}. This was the primary jet definition used in ATLAS physics results prior to the end of \RunTwo. \EMTopo jets exhibit robust energy scale and resolution characteristics across a wide kinematic range, and are independent of other reconstruction algorithms such as tracking at the jet-building stage.

Hadronic final-state measurements can be improved by making more complete use of the information from both the tracking and calorimeter systems. The particle flow (\textit{\pflow}) algorithm is based on Ref.~\cite{PERF-2015-09} and updated as described below. Particle flow directly combines measurements from both the tracker and the calorimeter to form the input signals for jet reconstruction, which are intended to approximate individual particles. Specifically, energy deposited in the calorimeter by charged particles is subtracted from the observed \topos\ and replaced by the momenta of tracks that are matched to those \topos. These resulting \textit{\pflow jets} exhibit improved energy and angular resolution, reconstruction efficiency, and \pileup stability compared to calorimeter jets~\cite{PERF-2015-09}. \EMTopo and \pflow jets are retained for the analyses discussed in this paper only if they have an uncalibrated $\pt > 7$~\GeV\ and $|\eta| < 4.5$.
 
The updates to the \pflow algorithm since its description in Ref.~\cite{PERF-2015-09} are as follows. The expected mean value of the energy deposited by pions, $\langle E_{\textrm{dep}} \rangle$, and its expected standard deviation, $\sigma(E_{\textrm{dep}})$, were recomputed using the updated simulation, geometry, and topo-cluster noise thresholds for \RunTwo~\cite{PERF-2016-04}. The shower profiles were similarly updated. The only algorithmic change was an improvement in the transition between using track energy and cluster energy in high-\pT jets.
Since energetic particles are often in the core of jets and thus poorly isolated from nearby activity, accurate removal of the calorimeter energy associated with the track can be difficult. Therefore, the \pflow algorithm prevents energy subtraction in these cases.
Formerly this was managed by applying a simple $\pT^{\textrm{trk}}<40~\GeV$~cut in the track selection.
In the updated algorithm, a more sophisticated procedure is used to prevent the subtraction in cases where the advantages of the tracker are smaller and where the particle shower falls in a region with significant energy depositions from other particles. For all tracks up to $\pT^{\textrm{trk}}=100~\GeV$, if the energy $E^{\textrm{clus}}$ in a cone of size $\Delta R=0.15$~around the extrapolated particle satisfies
\begin{equation}
\frac{E^{\textrm{clus}}-\langle E_{\textrm{dep}} \rangle}{\sigma(E_{\textrm{dep}})}>33.2\times\log_{10}(40~\GeV/\pT^{\textrm{trk}})\,,
\label{eq:subtraction}
\end{equation}
then the subtraction is not performed.
With this parameterization, the subtraction is performed at lower track momenta unless the calorimeter activity measured by $E^{\textrm{clus}}$ is very high, such as in very dense environments where the accuracy of the subtraction is degraded. Since the calorimeter provides a good energy measurement at high $\pT^{\textrm{trk}}$, this parameterization effectively slowly truncates the algorithm, yet allows the subtraction to continue to be performed for a small range above this cut-off even when the calorimeter energy deposition is low or near the expected value, $\langle E_{\textrm{dep}} \rangle$. The momentum range up to which the subtraction is still allowed to be performed is driven by the coefficient of 33.2 in \Eqn{(\ref{eq:subtraction})} and is typically about 20--50\% above the 40~\GeV\ cut-off previously used. Above $\pT^{\textrm{trk}}=100~\GeV$ no track information is used and the \pflow algorithm becomes equivalent to \EMTopo, benefitting from excellent calorimeter performance at high energies. The result of the improved subtraction method detailed here is that the energy resolution of \pflow jets becomes compatible with that of \EMTopo jets at high energies while remaining superior at low energies.
 
After the subtraction, two scalings are applied. These account for the difference in response, here defined as the ratio of measured to true particle energy, between \topos\ at the EM scale and tracks for which the energy scale is closer to the true particle energy.
The first scale factor applies only when no subtraction has been performed for a selected track. In this case the \pflow object includes both the full \topo\ energy and the track momentum. To avoid double-counting the energy while maintaining the contribution from the calorimeter measurement, the track momentum is scaled by a factor $(1 -\langle E_{\textrm{dep}} \rangle/p_{\text{trk}})$.
The resulting \pflow object uses the desired information and has a final energy of approximately $p_{\textrm{trk}}$, matching the response for the subtracted case.
The second scale factor is applied in both the subtracted and non-subtracted cases for all \pflow objects created from selected tracks below 100~\GeV. It smooths the transition between the lower-energy \pflow objects which are at the scale of the tracks and the higher-energy objects at the electromagnetic scale of the clusters.
The energy of these \pflow objects is scaled by unity for $\pT^{\textrm{trk}}$ below $30~\GeV$, by $(1- \langle E_{\textrm{dep}} \rangle/p_{\text{trk}})$~ for objects with $60~\GeV < \pt^{\textrm{trk}} < 100~\GeV$, and by a linearly descending scale factor in between.
This ensures that all objects are at the electromagnetic scale by 60~\GeV.

Tracks used in \pflow objects and in deriving calibrations for both \EMTopo and \pflow jets are reconstructed within the full acceptance of the inner detector ($|\eta| < 2.5$), required to have a $\pt > 500$~\MeV, and satisfy quality criteria based on the number of hits in the ID subdetectors. To suppress the effects of \pileup, tracks must satisfy $|z_0 \sin\theta|<2$~mm, where $z_0$ is the distance of closest approach of the track to the hard-scatter primary vertex along the $z$-axis and $\theta$ is the polar angle.
Tracks are matched to jets using ghost association~\cite{Cacciari:2007fd}, a procedure that treats them as four-vectors of infinitesimal magnitude during the jet reconstruction and assigns them to the jet with which they are clustered.

MC simulation is used to determine the energy scale and resolution of jets by comparing \pflow and \EMTopo jets with particle-level \textit{truth jets}. Truth jets are reconstructed using stable final-state particles and exclude muons, neutrinos, and particles from \pileup interactions. Truth jets are selected with the same $\pt > 7$~\GeV\ and $|\eta| < 4.5$ thresholds as \EMTopo and \pflow jets, and are geometrically matched to those jets using the angular distance \DeltaR with the requirement $\DeltaR < 0.3$.
 
\FloatBarrier
 
\section{Jet energy scale calibration}
\label{sec:jes}
The jet energy scale calibration restores the jet energy to that of jets reconstructed at the particle level.
The full chain of corrections is illustrated in \Fig{\ref{fig:jeschain}}.
All stages correct the four-momentum, scaling the jet \pT, energy, and mass.
 
\begin{figure}[hbp!]
\begin{center}
\includegraphics[width=\linewidth]{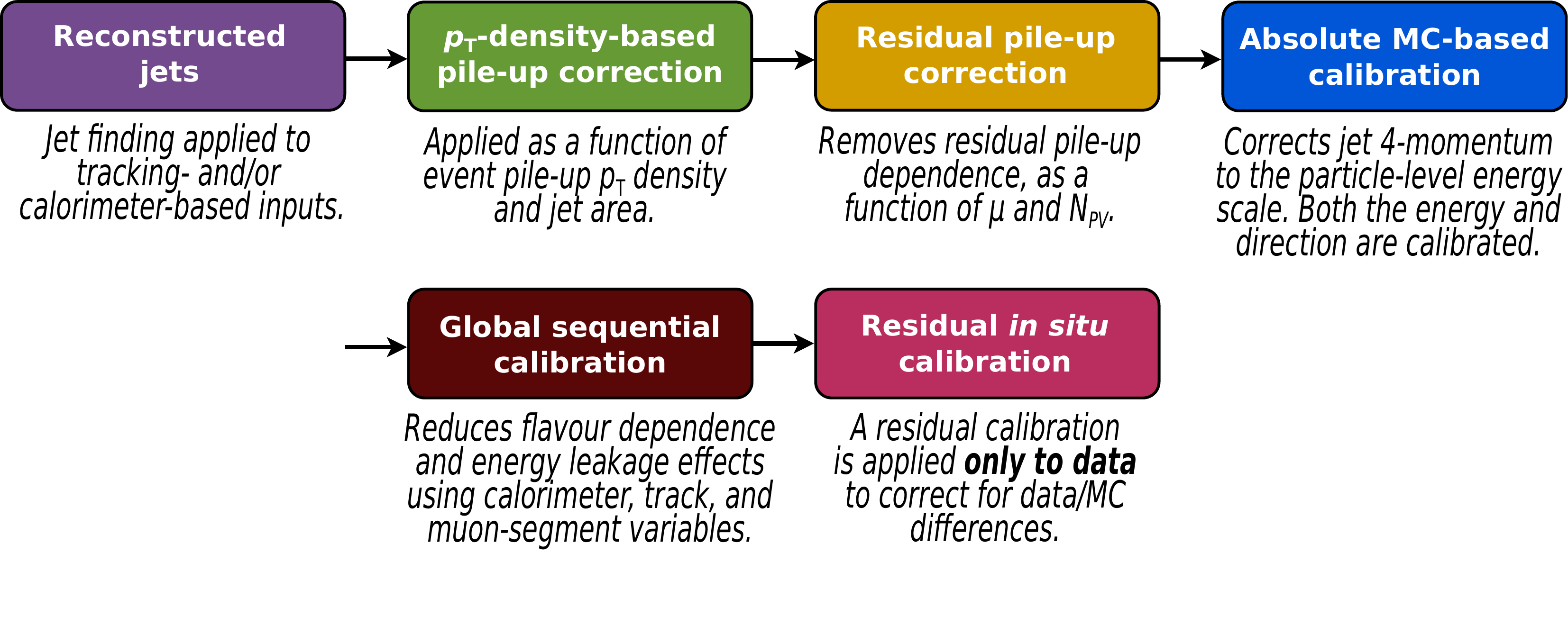}\\
\caption{Stages of jet energy scale calibrations. Each one is applied to the four-momentum of the jet.}
\label{fig:jeschain}
\end{center}
\end{figure}
 
At the beginning of the chain, the \pileup corrections remove the excess energy due to additional proton--proton interactions within the same (in-time) or nearby (out-of-time) bunch crossings.
These corrections consist of two components: a correction based on the jet area and transverse momentum density of the event, and a residual correction derived from MC simulation and parameterized as a function of the mean number of interactions per bunch crossing ($\mu$) and the number of reconstructed primary vertices in the event (\Npv).
These corrections are discussed in \Sect{\ref{sec:jes-pileupcorr}}.
The \textit{absolute} JES calibration corrects the jet so that it agrees in energy and direction with truth jets from dijet MC events, and is detailed in \Sect{\ref{sec:jes-mcjes}}.
Furthermore, the \textit{global sequential calibration} (derived from dijet MC events) improves the jet \pt resolution and associated uncertainties by removing the dependence of the reconstructed jet response on observables constructed using information from the tracking, calorimeter, and muon chamber detector systems, as introduced in \Sect{\ref{sec:jes-gsc}}. All these calibrations are applied to both data and MC simulation.
Finally, a residual \insitu calibration is applied to data only to correct for remaining differences between data and MC simulation.
It is derived using well-measured reference objects, including photons, \Zboson bosons, and calibrated jets, and for the first time benefits from a low-\pT measurement using the missing-$E_{\text{T}}$ projection fraction method for better \pileup robustness.
It is described in \Sect{\ref{sec:jes-insitu}}.
The full treatment and reduction of the systematic uncertainties is discussed in \Sect{\ref{sec:jes-systematics}}.

\subsection{Simulation-based jet calibrations}
\label{sec:jes-mccalib}
The derivation of the calibrations derived exclusively from MC simulation samples is described below.
 
\subsubsection{Pile-up corrections}
\label{sec:jes-pileupcorr}
As a result of the increase of the \topo ing \pT thresholds (to suppress electronic and \pileup noise) and in the instantaneous luminosity, the contribution from \pileup to the JES in the 2015--2017 data-taking period differs from the one observed in 2015.
The \pileup corrections are therefore evaluated using updated MC simulations of the software reconstruction and \pileup conditions.
These corrections are derived using the same methods employed in 2015~\cite{PERF-2016-04} and are summarized in the following paragraphs.
 
First, a jet \pt-density-based subtraction of the per-event \pileup contribution to the jet \pT is performed.
The jet area $A$ is a measure of the susceptibility of the jet to \pileup and is calculated by determining the relative number of \emph{ghost} particles associated with a jet after clustering.
Next, the \pileup contribution is estimated from the median \pT density, $\rho$, of jets in the $y$--$\phi$ plane, $\left<\pT/A\right>$.
The calculation of $\rho$ uses jets reconstructed using the $k_t$ algorithm~\cite{ktalgorithm} with radius parameter $R=0.4$ from positive-energy \topos{} with $|\eta|<2$.
The computation of $\rho$ in the central region of the detector gives a more meaningful measure of the \pileup activity than the median over the entire $\eta$ range, and this is because $\rho$ drops to nearly zero beyond $|\eta|\sim 2$.
This drop is due to the lower occupancy in the forward region relative to the central region, which is a result of a coarser segmentation in the forward region.
The $\kt$ algorithm is chosen due to its tendency to naturally reconstruct jets including an uniform soft background~\cite{Cacciari:2007fd}, while $\rho$ is used to reduce the bias from hard-scatter jets which populate the high-\pT tails of the distribution.
The distribution of $\rho$ in MC simulation for representative \Npv values is shown in \Fig{\ref{fig:rho}}.
The ratio of the $\rho$-subtracted jet \pT to the uncorrected jet \pT is applied as a scale factor to the jet four-momentum and does hence not affect its direction.
 
\begin{figure}[htp!]
\begin{center}
\includegraphics[width=0.5\linewidth]{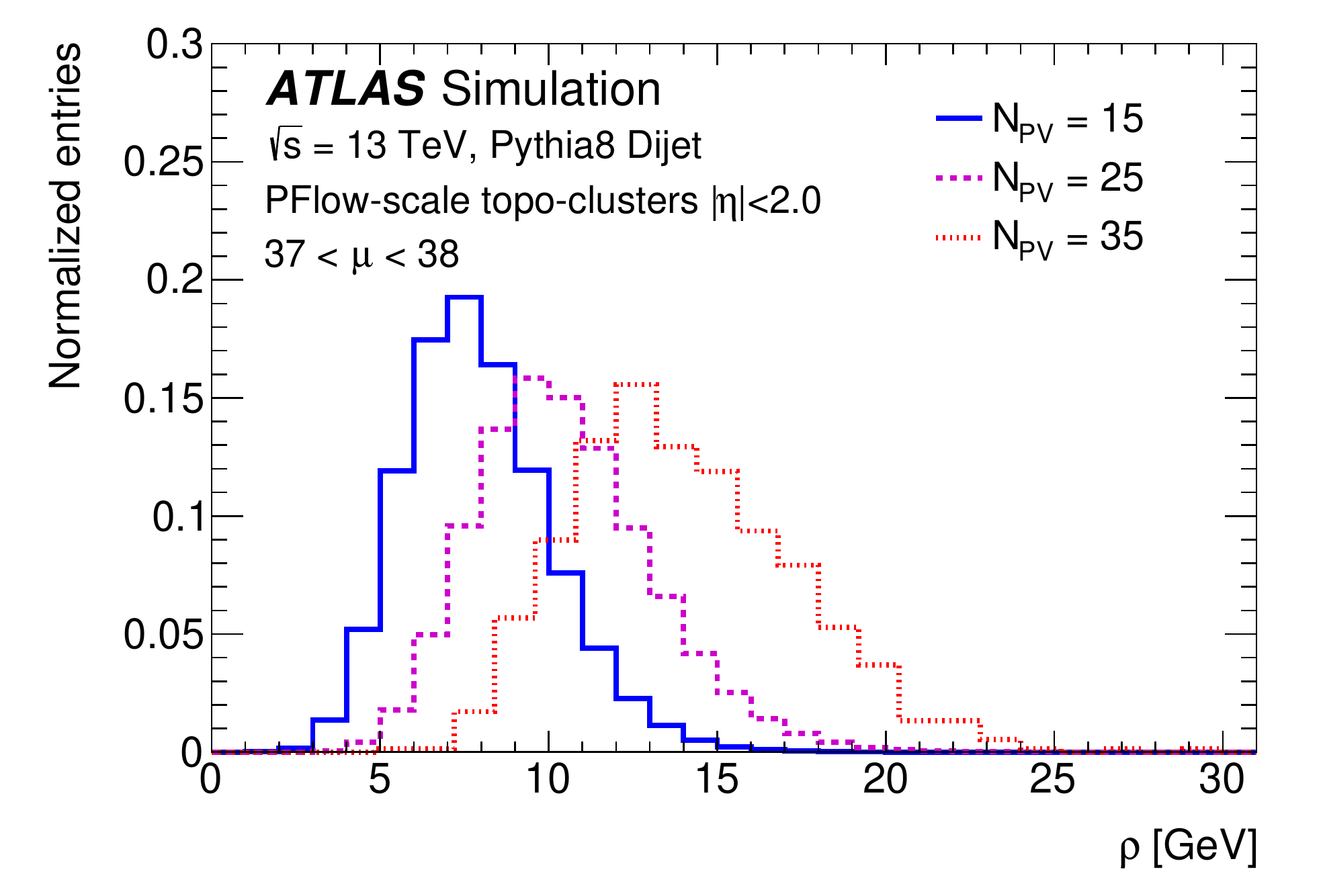}\\
\caption{Per-event median \pT density, $\rho$, at $\Npv=15$ (solid), $\Npv=25$ (long dashed), and $\Npv=35$ (short dashed) for $37<\mu<38$ as found in MC simulation.}
\label{fig:rho}
\end{center}
\end{figure}
 
The $\rho$ calculation is derived from the central, lower-occupancy regions of the calorimeter and does not fully describe the \pileup sensitivity in the forward calorimeter region or in the higher-occupancy core of high-\pT jets.
It is therefore observed that after this correction some dependence of the \antikt jet \pT on the \pileup activity remains, and consequently, a residual correction is derived.
This residual dependence is defined as the difference between the reconstructed jet \pT and truth jet \pT and it is observed as a function of both \Npv and $\mu$, which are sensitive to in-time and out-of-time \pileup respectively.
 
The jet \pT after all \pileup (\pt-density-based and residual) corrections is given by
\begin{equation}\nonumber
\pTcorr = \pTreco - \rho \times A - \alpha \times (\Npv-1) - \beta \times \mu\,,
\end{equation}
where \pTreco refers to the \pT of the reconstructed jet before any \pileup correction is applied.
Reconstructed jets with $\pT>7~\GeV$ are geometrically matched to truth jets within $\DeltaR = 0.3$.
The residual \pT dependences on \Npv ($\alpha$) and $\mu$ ($\beta$) are observed to be fairly linear and independent of one another.
Independent linear fits are used to derive $\alpha$ and $\beta$ coefficients in bins of \pTtruth and $|\etadet|$, where \pTtruth is the \pT of the truth jet that matches the reconstructed jet.
The jet $\eta$ pointing from the geometric centre of the detector, \etadet, is used to remove any ambiguity as to which region of the detector is measuring the jet.
Both the $\alpha$ and $\beta$ coefficients are seen to have a logarithmic dependence on \pTtruth, and logarithmic fits are performed in the range $20\,\GeV<\pTtruth<200$\,\GeV{} for each bin of $|\etadet|$.
In each $|\etadet|$ bin, the fitted values of the $\alpha$ and $\beta$ coefficients at $\pTtruth=25$\,\GeV{} are taken as their nominal values, reflecting their behaviour in the \pT region where \pileup effects are most relevant.
The differences between the logarithmic fits over the full \pTtruth range and the nominal fits are used for a \pT-dependent systematic uncertainty in the residual \pileup dependence.
Finally, linear fits are performed to the binned coefficients as a function of $|\etadet|$.
This reduces the effects of statistical fluctuations and allows the $\alpha$ and $\beta$ coefficients to be smoothly sampled in $|\etadet|$, particularly in regions of varying dependence.
 
The dependences of the \pt-density-based and residual corrections on \Npv and $\mu$ as a function of $|\etadet|$ for \pflow jets are shown in \Fig{\ref{fig:pileup_dependence}}.
The negative dependence on $\mu$ for out-of-time \pileup is a result of the liquid-argon calorimeter's pulse shape, which is negative during the period shortly after registering a signal~\cite{Abreu_2010}.
These corrections are similar to those derived for \EMTopo jets, although the \Npv-dependent corrections for \pflow jets in the $|\etadet|<2.5$ region are reduced by about 60\% relative to \EMTopo due to the usage of tracks in the \pflow algorithm.
For \EMTopo jets, the shape of the residual corrections is comparable to that found in 2015 MC simulation, except in the forward region ($|\etadet|>2.5$), where it is found to be smaller by 0.1\,\GeV.
This difference is primarily caused by higher \topo{} noise thresholds used in the full \RunTwo data.
 
\begin{figure}[htp!]
\begin{center}
\begin{subfigure}[b]{0.49\textwidth}
\centering
\includegraphics[width=\textwidth]{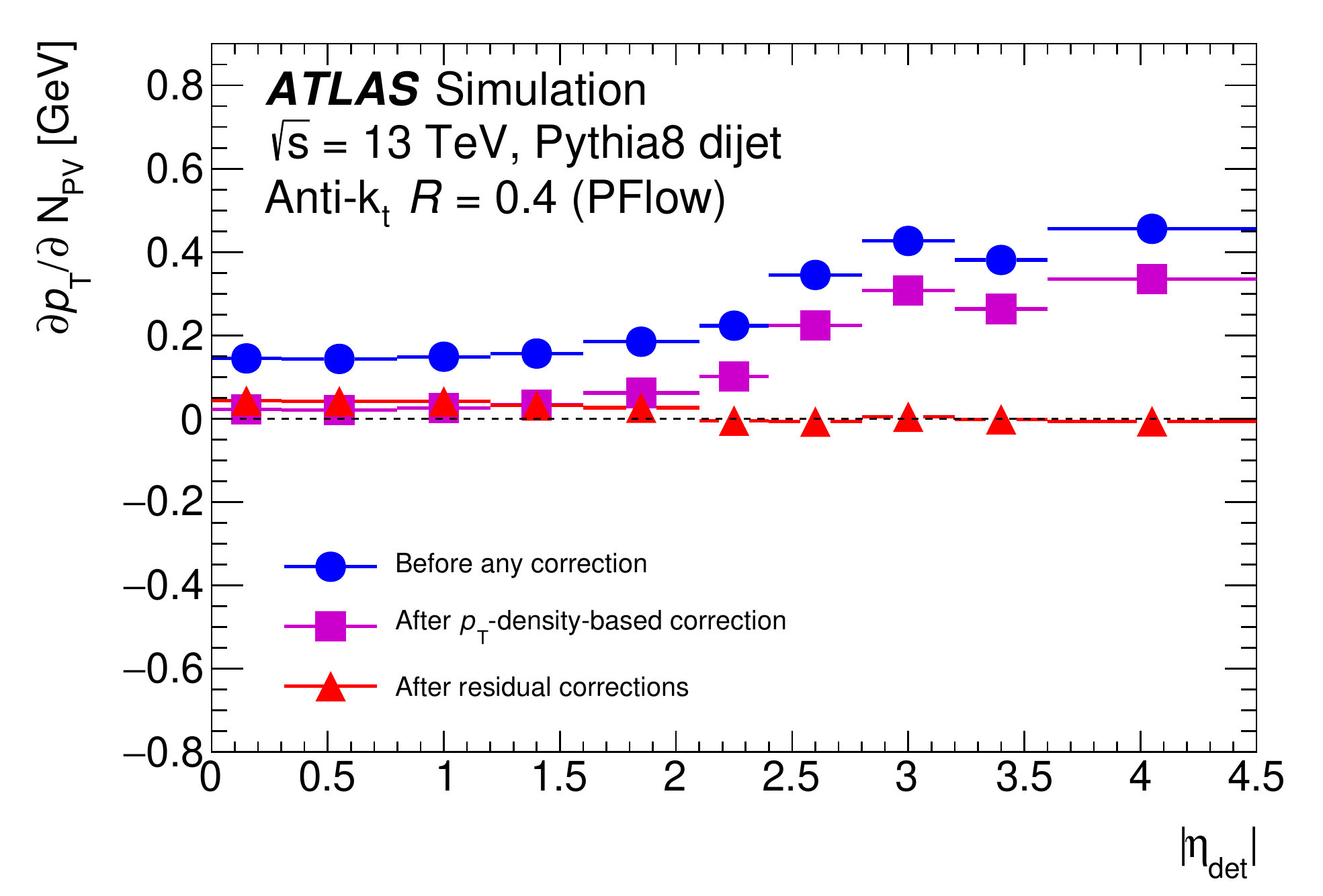}
\caption{In-time \pileup dependence}
\label{subfig:pileup_intime}
\end{subfigure}
\begin{subfigure}[b]{0.49\textwidth}
\centering
\includegraphics[width=\textwidth]{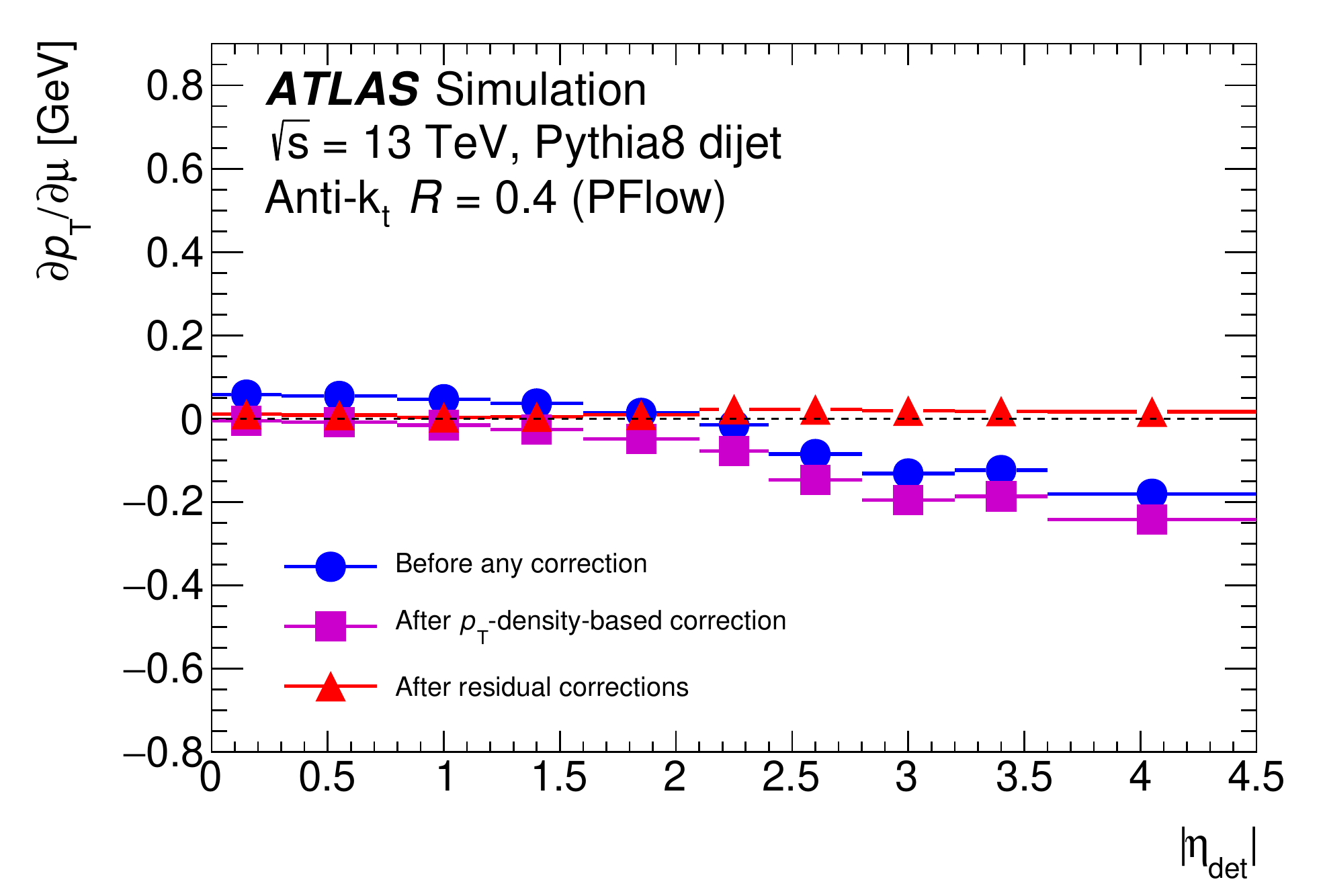}
\caption{Out-of-time \pileup dependence}
\label{subfig:pileup_outoftime}
\end{subfigure}
\caption{Dependence of \pflow jet \pt on \subref{subfig:pileup_intime} in-time \pileup (\Npv averaged over $\mu$) and \subref{subfig:pileup_outoftime} out-of-time \pileup ($\mu$ averaged over \Npv) as a function of $|\etadet|$ for $\pttruth=25$\,\GeV. Errors are taken from the fit results and are too small to be visible on the scale of the plot.}
\label{fig:pileup_dependence}
\end{center}
\end{figure}
 
Four systematic uncertainties are introduced to account for MC mis-modelling of \Npv, $\mu$, the $\rho$ topology, and the \pT dependence of the residual \pileup corrections.
The last of these is derived from the full logarithmic fits to $\alpha$ and $\beta$, as discussed previously.
Two \insitu methods are used to estimate uncertainties in the modelling of \Npv and $\mu$.
The first method uses jets reconstructed from tracks to provide a measure of the jet \pT{} independent of  \pileup.
This is only used for $|\eta|<2.1$.
The second method exploits the \pT balance between a reconstructed jet and a $Z$ boson and is used for $2.1<|\eta|<4.5$.
These systematic uncertainties are described in more detail in Ref.~\cite{PERF-2014-03}.
Finally, the $\rho$ topology uncertainty accounts for the uncertainty in the underlying event's contribution to $\rho$, and is discussed in detail in Section~\ref{subsubsec:pileup-insitu}.
 
\FloatBarrier
 
\subsubsection{Jet energy scale and $\eta$ calibration}
\label{sec:jes-mcjes}
The absolute jet energy scale and $\eta$ calibrations correct the reconstructed jet four-momentum to the particle-level energy scale accounting for non-compensating calorimeter response, energy losses in passive material, out-of-cone effects and biases in the jet $\eta$ reconstruction.
Such biases are primarily caused by the transition between different calorimeter technologies and sudden changes in calorimeter granularity.
The calibration is derived for $R=0.4$ \antikt jets from a \pythia{}8 MC simulation of dijet events after the application of the \pileup corrections.
Reconstructed jets are geometrically matched to truth jets within $\DeltaR=0.3$. 
In addition, reconstructed (truth) jets are required to have no other reconstructed (truth) jet of $\pT>7~\GeV$ within $\DeltaR=0.6$ ($\DeltaR=1.0$).
 
The average jet energy response $\mathcal{R}$, defined as the mean of a Gaussian fit to the core of the $\Ereco/\Etruth$ distribution, is measured in \Etruth and \etadet bins.
The decision to calculate the response as a function of \Etruth instead of \Ereco is motivated by the fact that for fixed \Etruth (\Ereco) bins the response distribution is (not) Gaussian.
The average response is parameterized as a function of \Ereco using a numerical inversion procedure, as detailed in Ref.~\cite{PERF-2011-03}, and the jet calibration factor is taken as the inverse of the average energy response.
The response is higher for \pflow jets than for \EMTopo jets at low energies since tracking information is used.
The response for \pflow jets as a function of \Ereco ($\etadet$) for representative \etadet (\Ereco) bins is shown in \Fig{\ref{fig:mcjes_energyCorr}}.

\begin{figure}[htp!]
\begin{center}
\begin{subfigure}[b]{0.49\textwidth}
\centering
\includegraphics[width=\textwidth]{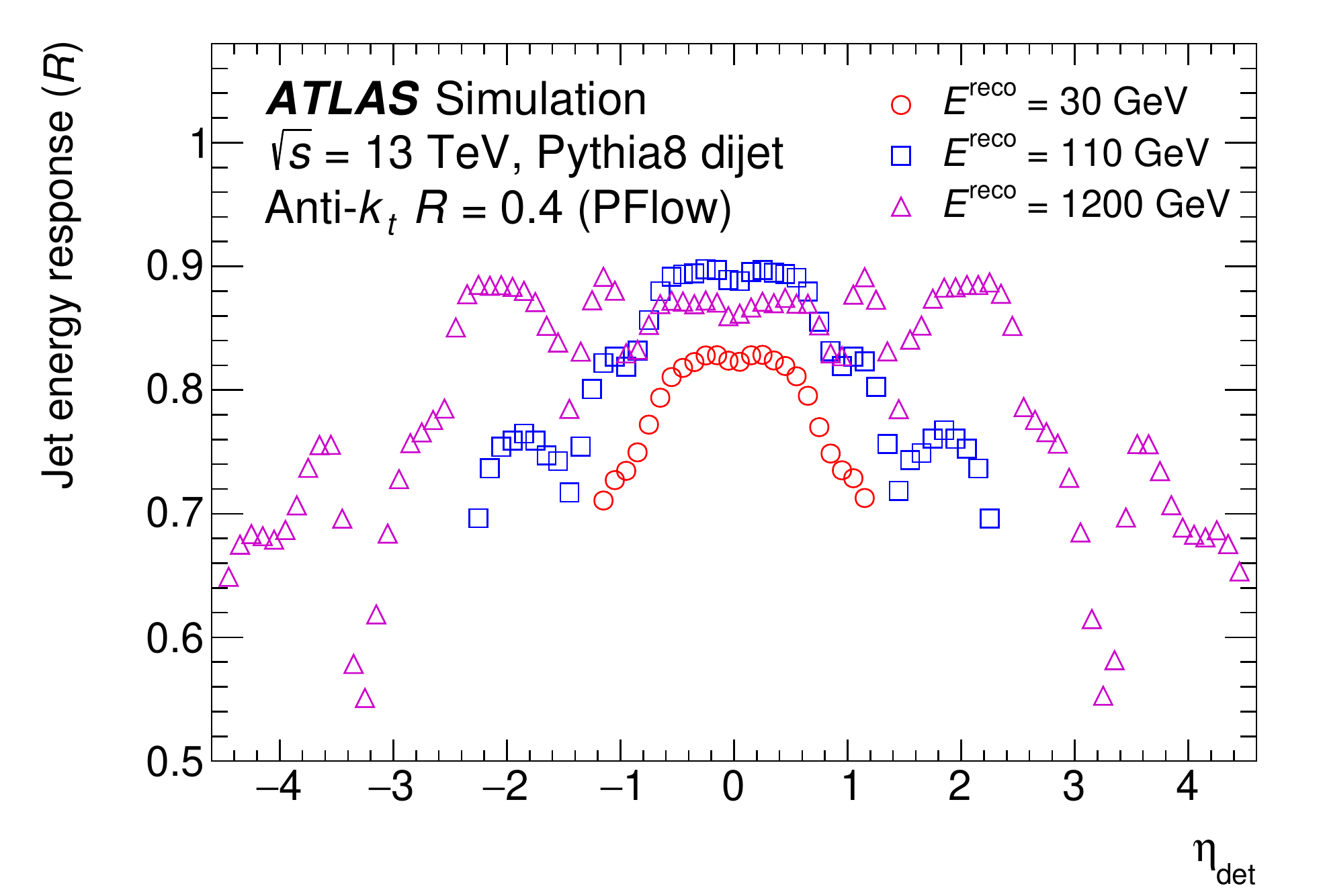}
\caption{}
\label{subfig:mcjes_response_vs_eta}
\end{subfigure}
\begin{subfigure}[b]{0.49\textwidth}
\centering
\includegraphics[width=\textwidth]{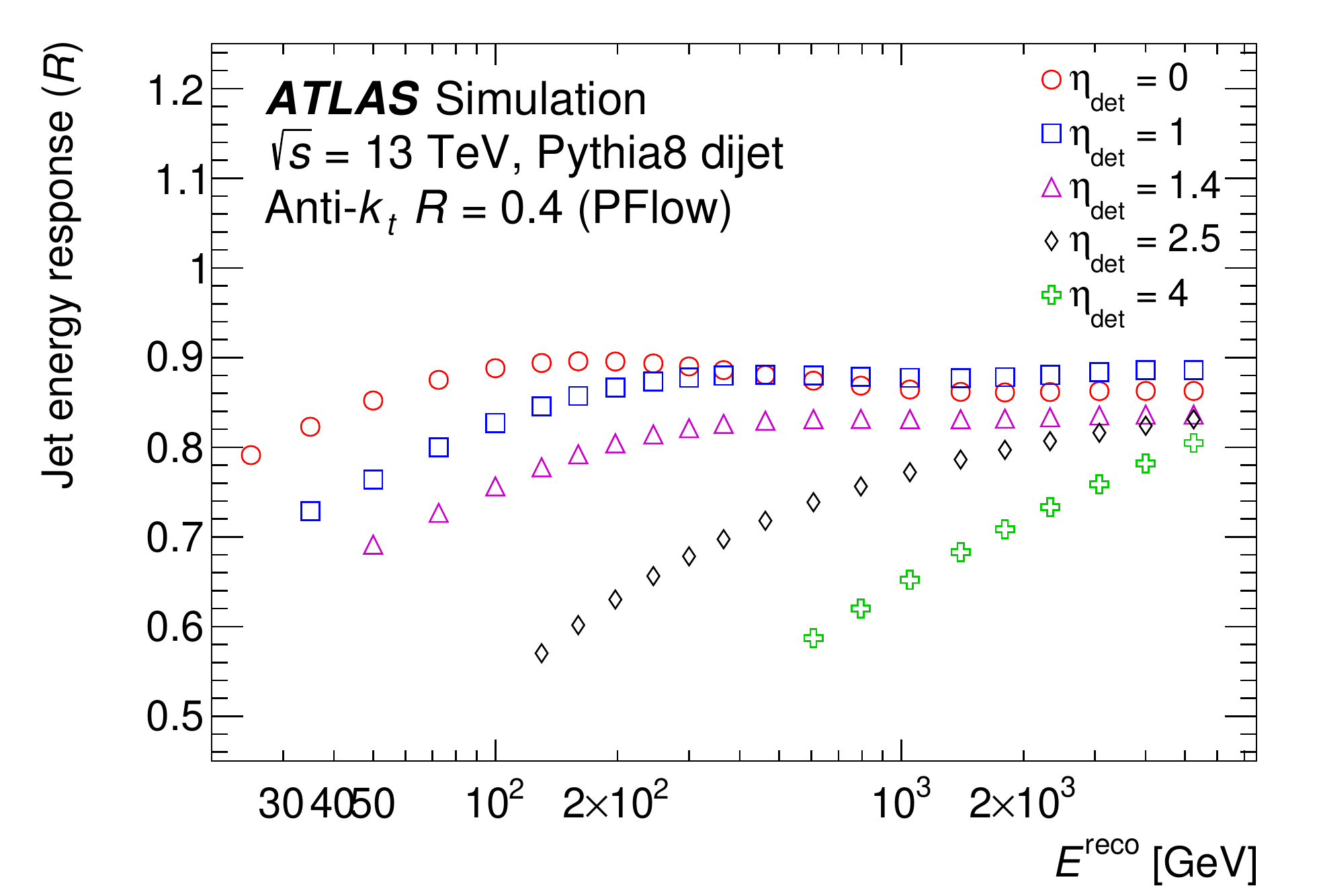}
\caption{}
\label{subfig:mcjes_response_vs_e}
\end{subfigure}
\caption{The average energy response as a function of reconstructed jet \subref{subfig:mcjes_response_vs_eta} $\etadet$ and \subref{subfig:mcjes_response_vs_e} energy \Ereco. Each value is obtained from the corresponding parameterized function derived with the \pythia{}8 MC sample and only jets satisfying $\pt>20$~\GeV\ are shown.}
\label{fig:mcjes_energyCorr}
\end{center}
\end{figure}
 
After the JES calibration based on the results in~\Fig{\ref{fig:mcjes_energyCorr}} is applied, the response diverges from 1 by a maximum of about 5\% (3\%, 1\%) at $\pttrue = 20~(30, 50)~\GeV$. This level of non-closure is observed across entire $\etadet$ range.
These small non-closures are seen for low-\pT jets due to a slightly non-Gaussian energy response and jet reconstruction threshold effects, both of which impact the response fits.
The closure in this result is an improvement with respect to the 2015 calibration and is thanks to advances in the fitting method and parameters.
 
\begin{figure}[!htb]
\begin{center}
\includegraphics[width=0.49\textwidth]{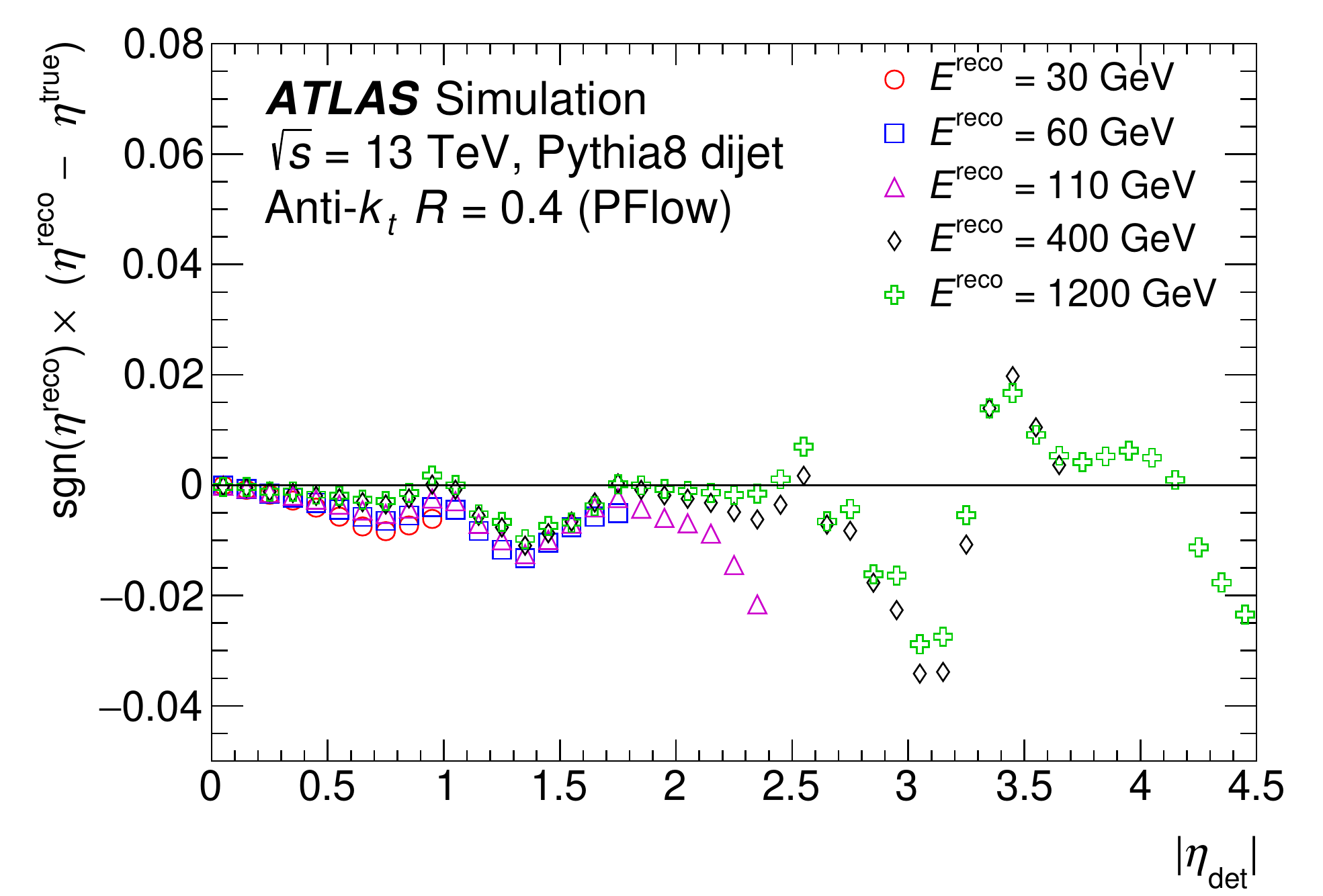}
\end{center}
\caption{The signed difference between the reconstructed and truth jet $\eta$, denoted by \etareco and \etatrue respectively. Each value is obtained from the corresponding parameterized function derived with the \pythia{}8 MC sample and only jets satisfying $\pt>20$~\GeV\ are shown.}
\label{fig:mcjes_etaCorr}
\end{figure}
 
A bias in the reconstructed jet $\eta$, defined as a significant deviation from zero in the signed difference between the reconstructed and truth jet $\eta$, denoted by \etareco and \etatrue respectively, is observed and shown in \Fig{\ref{fig:mcjes_etaCorr}} as a function of $|\etadet|$ for \pflow jets.
The bias for \EMTopo jets is similar, showing the same features.
It is largest in jets that encompass two calorimeter regions with different energy responses caused by changes in calorimeter geometry or technology.
This artificially increases the energy of one side of the jet relative to the other, altering the reconstructed four-momentum.
The barrel--endcap ($|\etadet|\sim1.4$) and endcap--forward ($|\etadet|\sim3.1$) transition regions can be clearly seen in \Fig{\ref{subfig:mcjes_response_vs_eta}} as susceptible to this effect.
A second correction is therefore derived as the difference between the reconstructed and truth $\eta$ (\etareco and \etatrue respectively) parameterized as a function of \Etruth and \etadet to remove such bias.
A numerical inversion procedure is again used to derive corrections in \Ereco from \Etruth.
This calibration only alters the jet \pT and $\eta$, not the full four-momentum.
\EMTopo and \pflow jets calibrated with the full jet energy scale and $\eta$ calibration are considered to be at the \EMJES scale and \EMPFlowJES scale, respectively.
 
The absolute JES and $\eta$ calibrations are also derived for a \pythia{}8 MC sample using AFII.
An additional systematic uncertainty is considered for these samples to account for a small non-closure in the calibration beyond $|\etadet|\sim3.2$, due to the approximate treatment of hadronic showers in the forward calorimeters.
This uncertainty is below 0.5\% for all central jets and is about 3\% for a forward jet of \pT$ = 20~\GeV$, falling rapidly with increasing \pT.
 
\FloatBarrier
 
\subsubsection{Global sequential calibration}
\label{sec:jes-gsc}
Even after the application of the previous jet calibrations (from now on referred to as MCJES), for a given (\pTtruth, $\etadet$) bin, the response can vary from jet to jet depending on the flavour and energy distribution of the constituent particles, their transverse distribution, and the fluctuations of the jet development in the calorimeter.
Furthermore, the average particle composition and shower shape of a jet varies between initiating particles, most notably between quark- and gluon-initiated jets.
A quark-initiated jet will often include hadrons with a higher fraction of the jet \pT{} that penetrate further into the calorimeter, while a gluon-initiated jet will typically contain more particles of softer \pT, leading to a lower calorimeter response and a wider transverse profile.
The global sequential calibration (GSC), a procedure used in the 2012~\cite{PERF-2014-02} and 2015~\cite{PERF-2016-04} calibrations, is a series of multiplicative corrections to reduce the effects from these fluctuations and improve the jet resolution without changing the average jet energy response.
The jet resolution \sigmaR is given by the standard deviation of a Gaussian fit to the jet \pt response distribution, where the \pt\ response is defined similarly to jet energy response as the ratio of \ptreco to \pTtruth.
 
The GSC is based on global jet observables such as the longitudinal structure of the energy depositions within the calorimeters, tracking information associated with the jet, and information related to the activity in the muon chambers behind a jet.
For these studies, reconstructed jets are geometrically matched to truth jets and a numerical inversion procedure is used, as explained in \Sect{\ref{sec:jes-mcjes}}.
Six observables are identified that improve the resolution of the JES through the GSC.
For each observable, an independent jet four-momentum correction is derived as a function of \pTtruth and $|\etadet|$ by inverting the reconstructed jet response in \pythia{}8 MC simulation events.
Corrections for each observable are applied independently and sequentially to the jet four-momentum for jets with $|\etadet|<3.5$ (unless stated otherwise).
No improvement in resolution was found from altering the sequence of the corrections.
 
The six stages of the GSC account for the dependence of the jet response on (in the order in which they are applied):
\begin{itemize}
\item \fch, the fraction of the jet \pT{} measured from ghost-associated tracks with $\pT>500$\,\MeV{} ($|\etadet|<2.5$);
\item \ftile, the fraction of jet energy measured in the first layer of the hadronic Tile calorimeter ($|\etadet|<1.7$);
\item \fem, the fraction of jet energy measured in the third layer of the electromagnetic LAr calorimeter ($|\etadet|<3.5$);
\item \nTrk, the number of tracks with $\pT>1$\,\GeV{} ghost-associated with the jet ($|\etadet|<2.5$);
\item \wTrk, also known as track width, the average \pT-weighted transverse distance in the $\eta$--$\phi$ plane between the jet axis and all tracks of $\pT>1$\,\GeV{} ghost-associated with the jet ($|\etadet|<2.5$);
\item \nseg, the number of muon track segments ghost-associated with the jet ($|\etadet|<2.7$).
\end{itemize}
The first correction is only applied to \pflow jets.
The \nseg correction, also known as the punch-through correction, reduces the tails of the response distribution caused by high-\pT jets that are not fully contained in the calorimeter.
All corrections are derived as a function of jet \pT, except for the punch-through correction, which is derived as a function of jet energy since this effect is more correlated with the energy escaping the calorimeters.

The underlying distributions of these observables are shown for \pflow jets in MC simulation and bins of equal statistics in Figure~\ref{fig:gsc_pflow_responses}.
Each observable has been studied in data and simulation and is found to be well modelled~\cite{PERF-2015-09,PERF-2014-02,PERF-2016-04}.
The spike at zero in the \ftile distribution at low \pTtruth, shown in Figure~\ref{subfig:gsc_pflow_responses_tile0}, corresponds to jets that are fully contained in the electromagnetic calorimeter and do not deposit energy in the Tile calorimeter.
The tail towards negative values in the \ftile and \fem distributions at low \pTtruth, shown in Figures~\ref{subfig:gsc_pflow_responses_tile0} and \ref{subfig:gsc_pflow_responses_em3}, respectively, reflects calorimeter noise fluctuations.
Slight differences with respect to data have a negligible impact on the GSC since the dependence of the average jet response on the observables is well modelled in MC simulation, as observed by an \insitu dijet tag-and-probe method described in Ref.~\cite{PERF-2011-03}.
In this method, the average \pT asymmetry between back-to-back jets is measured as a function of each observable.
 
The average jet \pT response for \pflow jets in MC simulation as a function of each of the GSC observables is shown in Figure~\ref{fig:gsc_pflow_responses} for representative \pTtruth ranges.
The dependence of the jet response on each observable is reduced to less than 2\% after the full GSC is applied, with small deviations from unity reflecting the correlations between observables that are unaccounted for in the corrections.
 
\begin{figure}[!htb]
\begin{center}
\begin{subfigure}[b]{0.49\linewidth}
\centering
\includegraphics[width=0.8\linewidth]{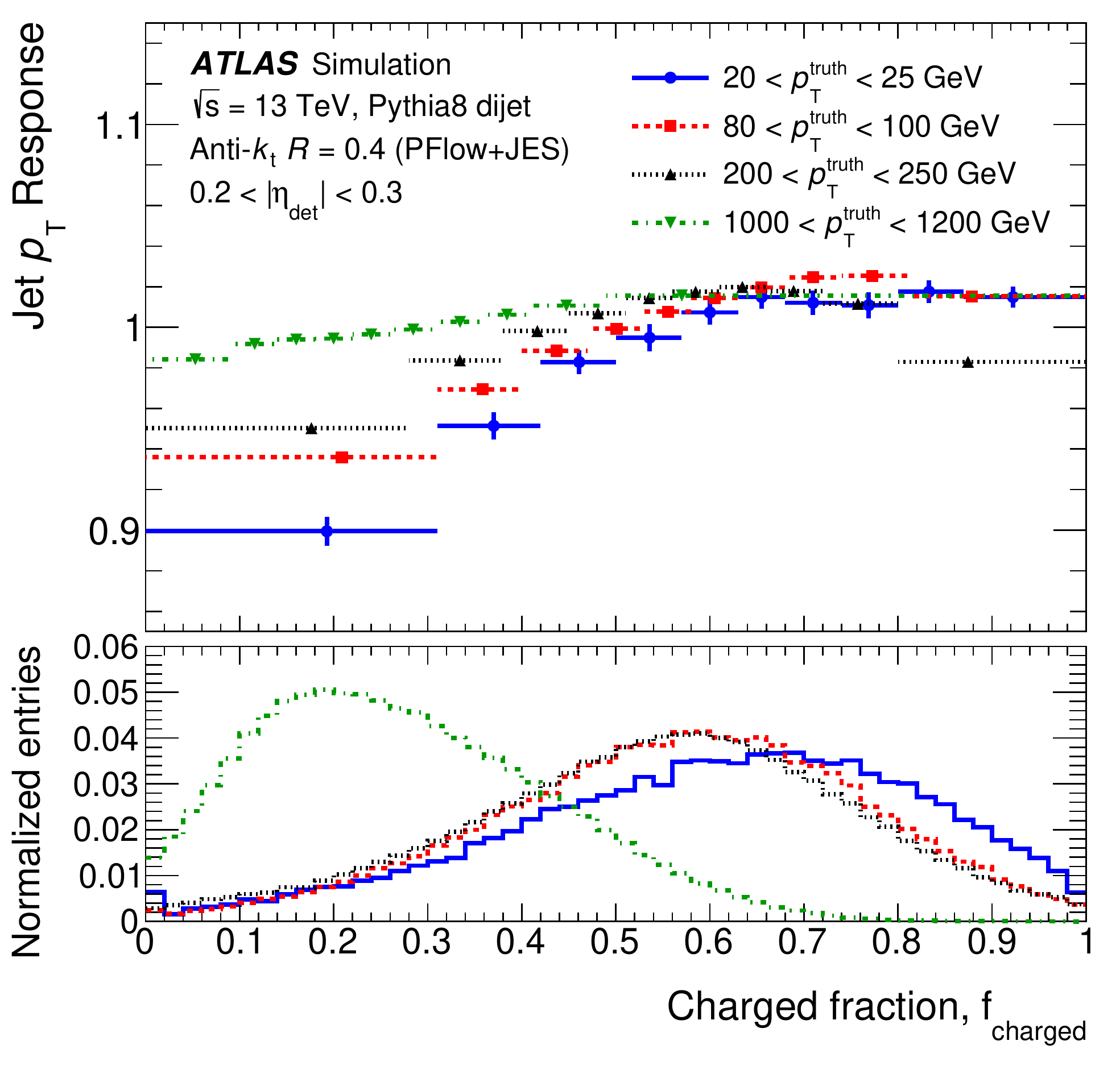}
\caption{\fch}
\label{subfig:gsc_pflow_responses_cf}
\end{subfigure}
\begin{subfigure}[b]{0.49\linewidth}
\centering
\includegraphics[width=0.8\linewidth]{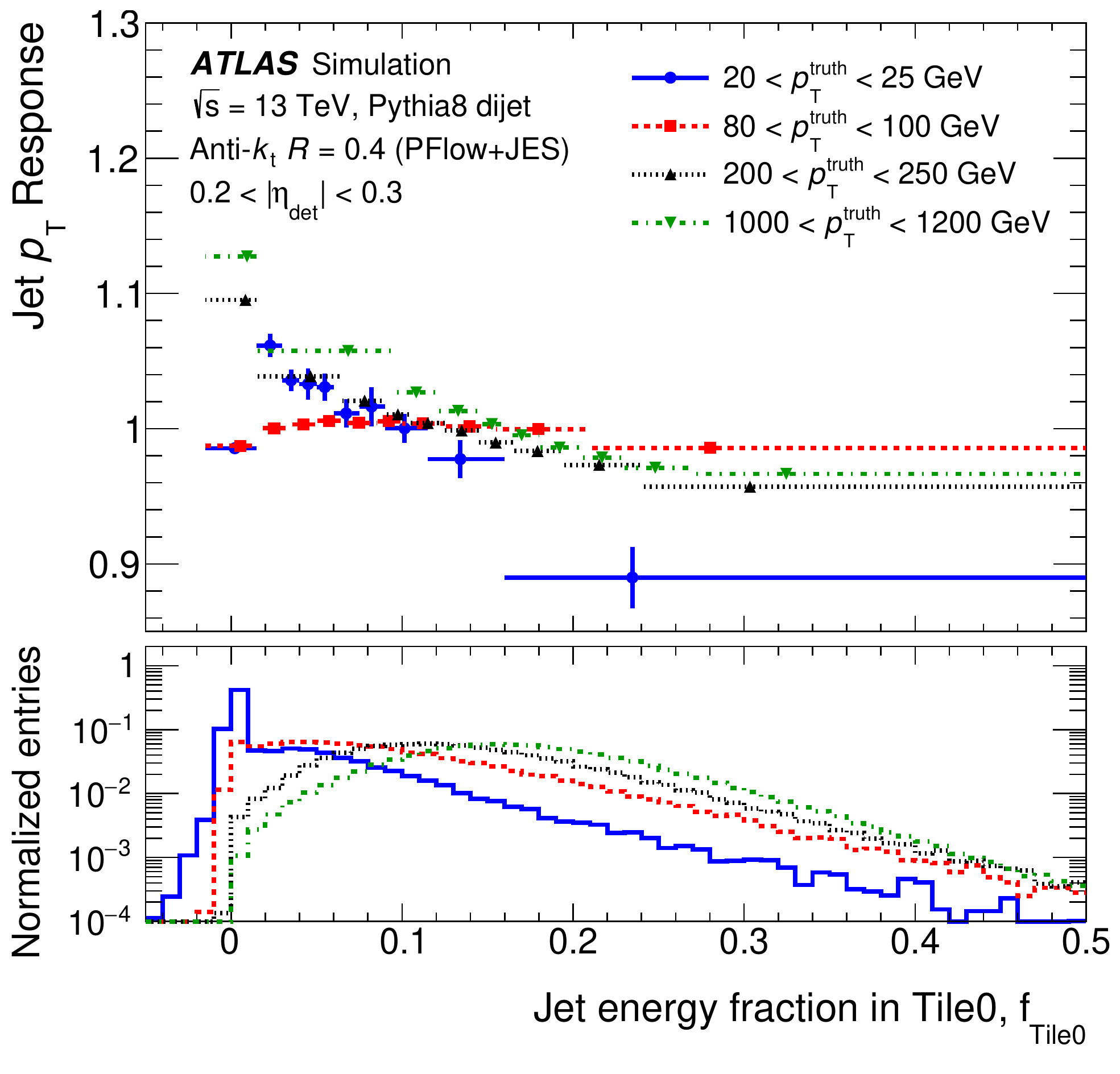}
\caption{\ftile}
\label{subfig:gsc_pflow_responses_tile0}
\end{subfigure}
 
\begin{subfigure}[b]{0.49\linewidth}
\centering
\includegraphics[width=0.8\linewidth]{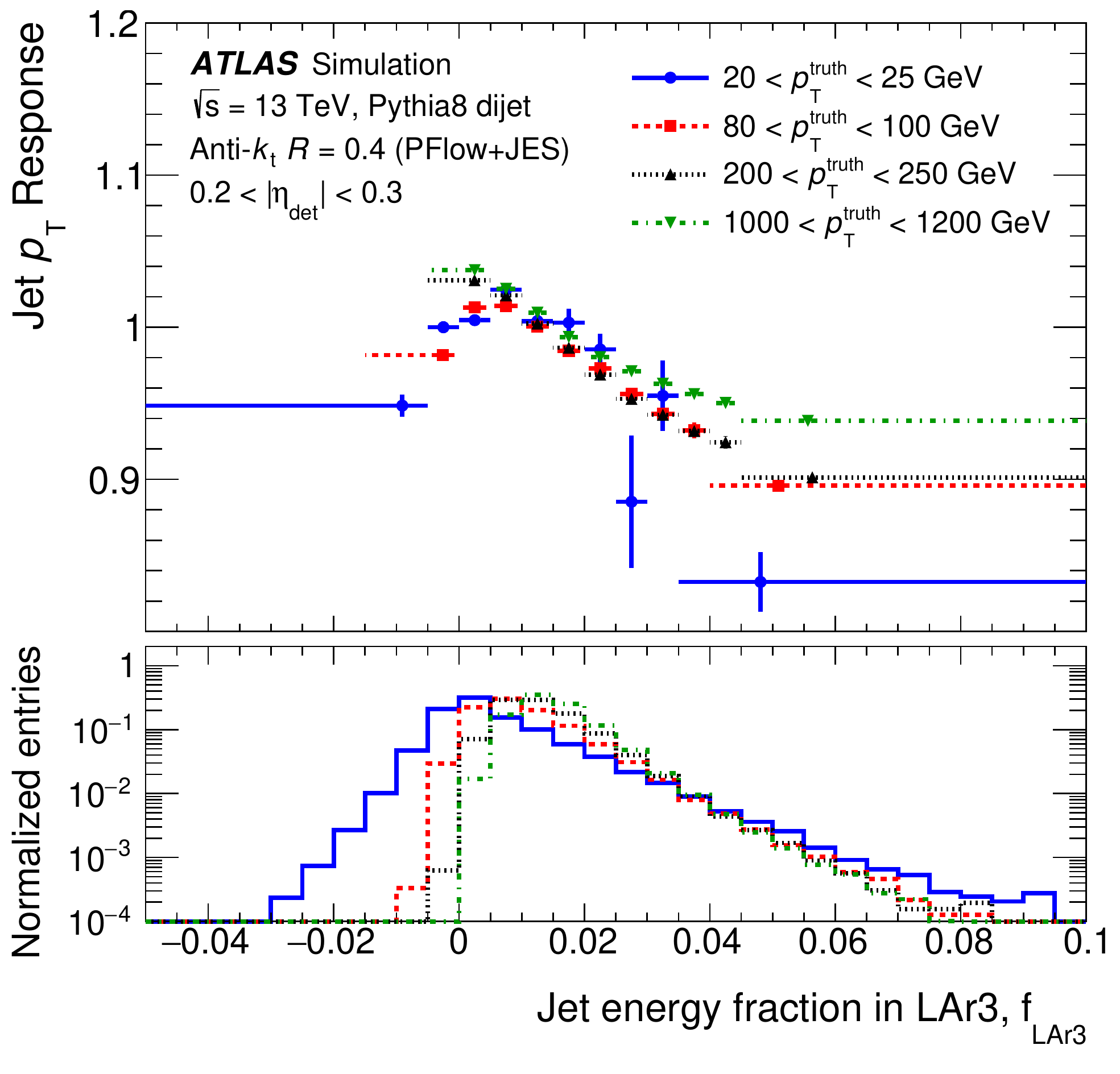}
\caption{\fem}
\label{subfig:gsc_pflow_responses_em3}
\end{subfigure}
\begin{subfigure}[b]{0.49\linewidth}
\centering
\includegraphics[width=0.8\linewidth]{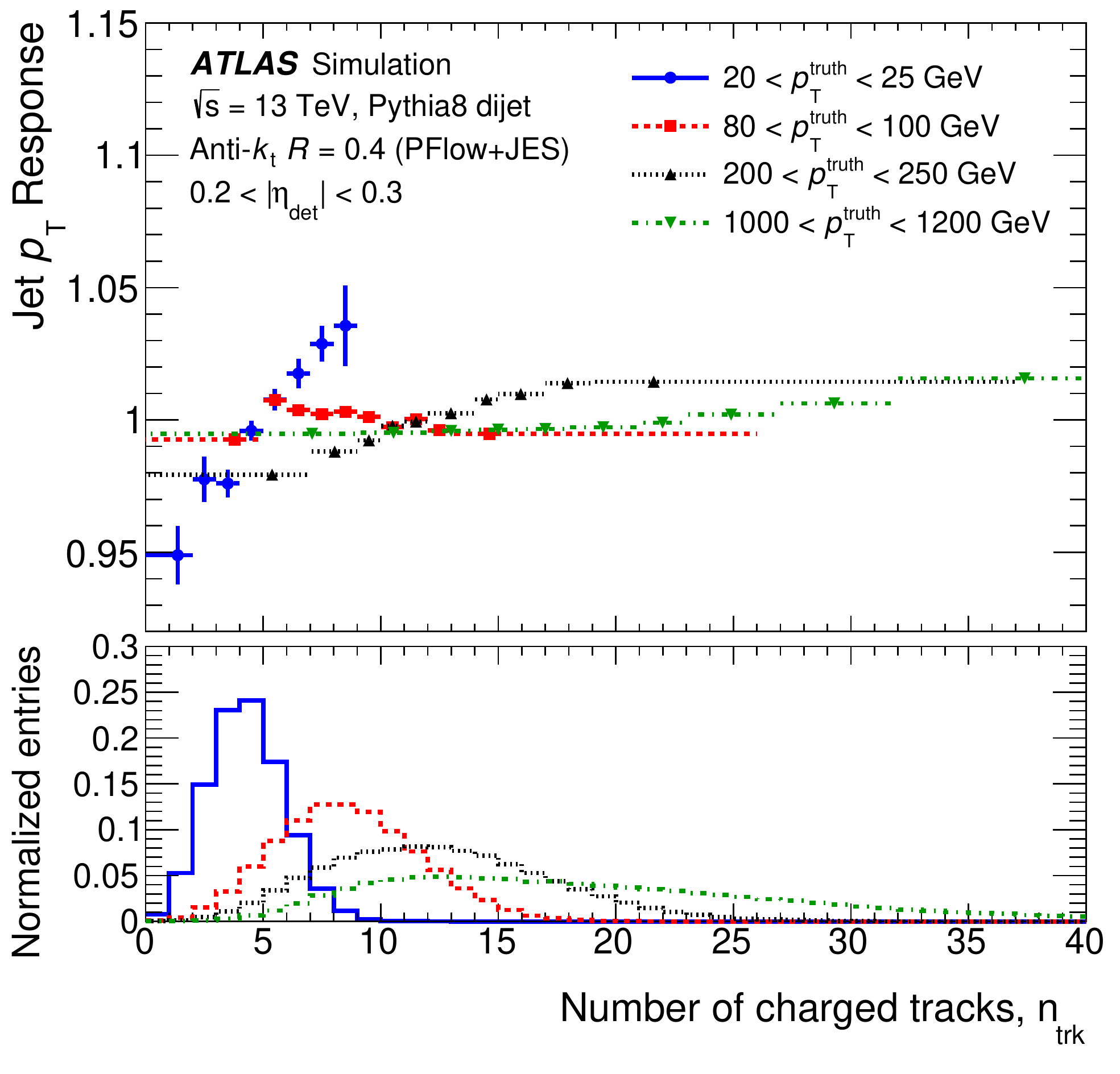}
\caption{\nTrk}
\label{subfig:gsc_pflow_responses_nTrk}
\end{subfigure}
 
\begin{subfigure}[b]{0.49\linewidth}
\centering
\includegraphics[width=0.8\linewidth]{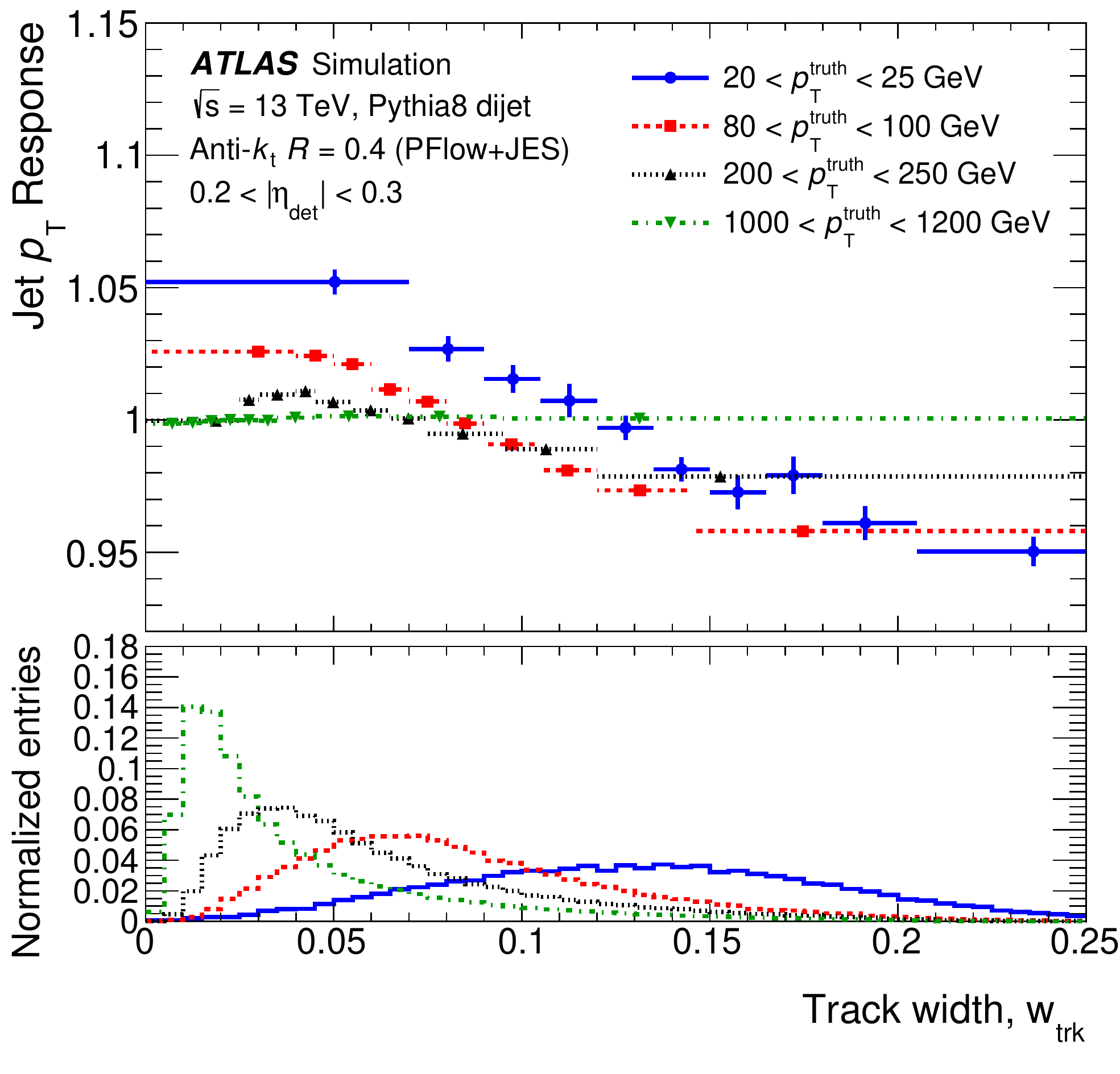}
\caption{\wTrk}
\label{subfig:gsc_pflow_responses_wTrk}
\end{subfigure}
\begin{subfigure}[b]{0.49\linewidth}
\centering
\includegraphics[width=0.8\linewidth]{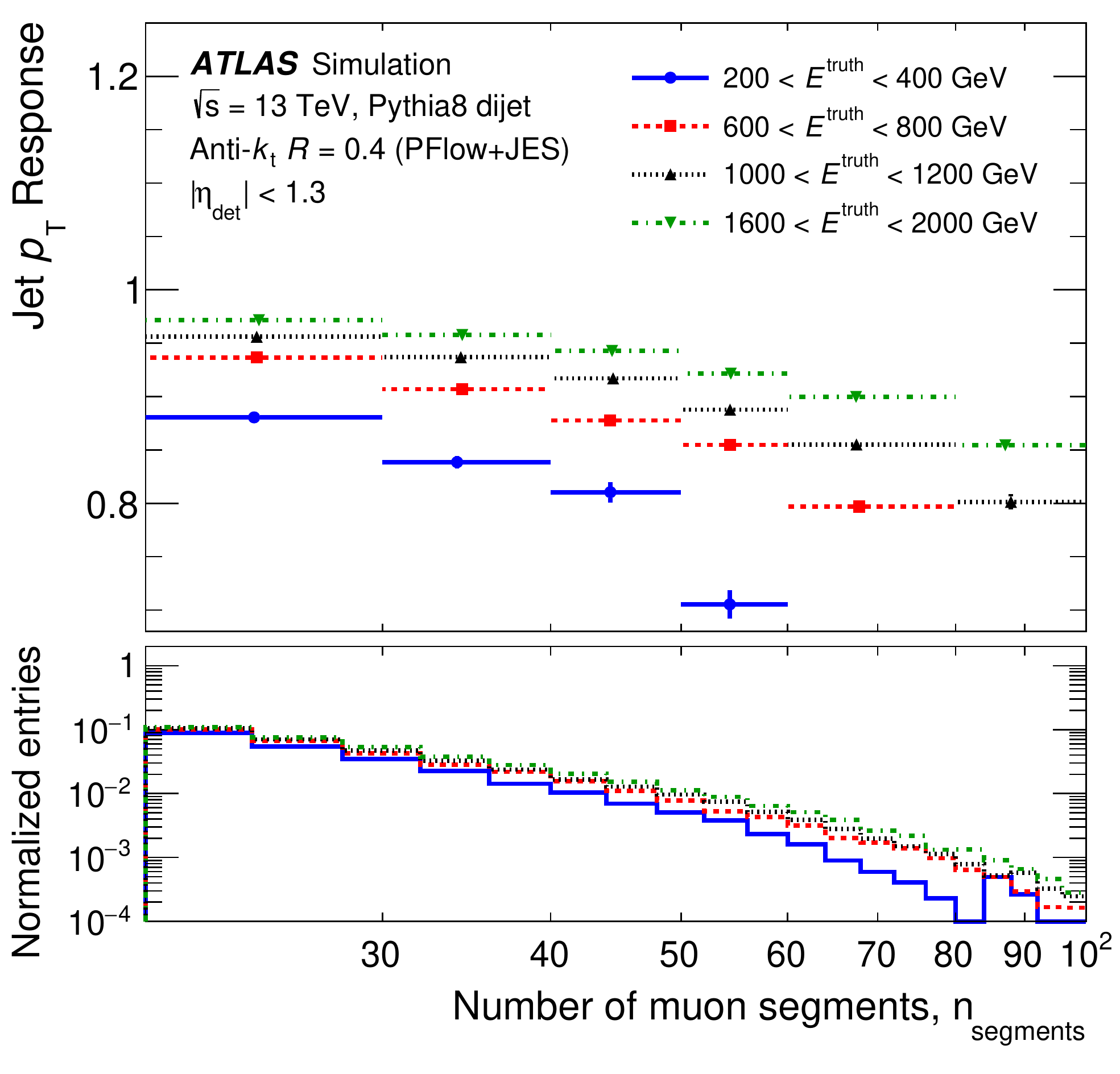}
\caption{\nseg}
\label{subfig:gsc_pflow_responses_nseg}
\end{subfigure}
\caption{Jet response for \pflow jets in four broad \pTtruth ranges as a function of each of the six observables used in the GSC \subref{subfig:gsc_pflow_responses_cf} the fraction of the jet \pt carried by charged particles, \subref{subfig:gsc_pflow_responses_tile0} the fraction of energy in the first layer of the Tile calorimeter, \subref{subfig:gsc_pflow_responses_em3} the fraction of energy in the third layer of the electromagnetic calorimeter, \subref{subfig:gsc_pflow_responses_nTrk} the number of tracks, \subref{subfig:gsc_pflow_responses_wTrk} the track width, and \subref{subfig:gsc_pflow_responses_nseg} the number of muon spectrometer track segments associated with the jet.
Jets at the \EMPFlowJES scale with $0.2 <|\etadet| < 0.3$ (except for \nseg which is shown for $|\etadet| < 1.3$ due to low statistics) are selected from a sample of \pythia{}8 dijet MC events and the corresponding preceding GSC steps have been applied accordingly.
The error bars show only the statistical uncertainty.
The bottom panels show the normalized distributions of the variables.
}
\label{fig:gsc_pflow_responses}
\end{center}
\end{figure}

The fractional jet resolution, defined as $\sigmaR/\mathcal{R}$, is used to determine the size of the fluctuations in the jet energy reconstruction and is shown for \pflow jets with $0.2<|\etadet|<0.3$ in MC simulation in Figure~\ref{fig:gsc_resolution}.
As more corrections are applied, the fractional jet resolution improves and the jet response dependence on the jet flavour is reduced.
No improvement is observed in \Fig{\ref{fig:gsc_resolution}} from the punch-through correction since only a small fraction of jets received this calibration, but there are analyses where their region of interest has a large fraction of jets that would receive this correction~\cite{EXOT-2019-03,STDM-2016-03}.
 
\begin{figure}[!htb]
\begin{center}
\includegraphics[width=0.8\linewidth]{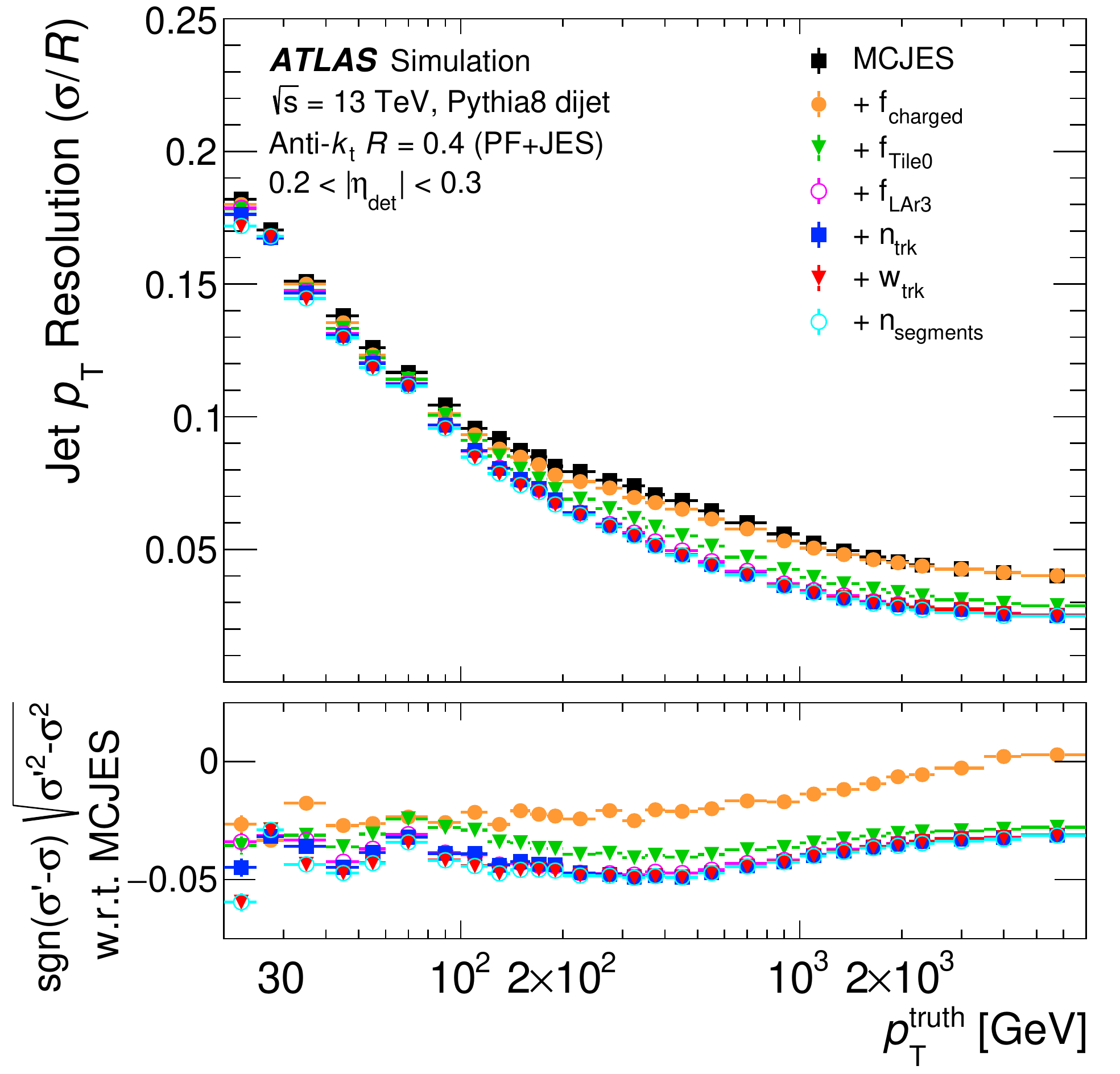}
\end{center}
\caption{Resolution of jets at the \EMPFlowJES scale with $0.2 < |\etadet| < 0.3$ measured in \pythia{}8 dijet MC simulation after each stage of the global sequential calibration (GSC).
All jet flavours, including \bjets, are considered.
The lower panel shows the difference in quadrature between the resolution before any GSC correction is applied ($\sigma$) and after the corresponding GSC step is applied ($\sigma'$).
}
\label{fig:gsc_resolution}
\end{figure}
 
 
\FloatBarrier
 
\subsection{\Insitu jet calibrations}
\label{sec:jes-insitu}
Once jets are corrected to the particle level using the MCJES and GSC, they require one final calibration step to account for differences between the jet response in data and simulation. These differences are caused by imperfect simulation of both the detector materials and the physics processes involved: the hard scatter and underlying event, jet formation, pile-up, and particle interactions with the detector. The final \insitu calibration measures the jet response in data and MC simulation separately and uses the ratio as an additional correction in data.
 
Jet response is calculated by balancing the \pt of a jet against that of a well-calibrated reference object or system. The response $\mathcal{R}_{\insitu}$ is defined as the average ratio of the jet \pt to the reference object \pt in bins of reference object \pt, where that average is taken from the peak location found by fitting the distribution with a Gaussian function. $\mathcal{R}_{\insitu}$ is sensitive to effects such as the presence of additional radiative jets or the transition of energy into or out of the jet cone, although these effects can be mitigated through careful event selection.\footnote{Requirements on the angle between the leading jet and the well-calibrated reference object, as well as cuts on the maximum \pt{} of the second jet in the event, help suppress additional radiation to ensure the events are as clean as possible.}
A better method is to form the double ratio from the response in data and MC simulation:
\begin{equation}\nonumber
c = \frac{\cal{R}^{\text{data}}_{\insitu}}{\cal{R}^{\text{MC}}_{\insitu}}\,,
\end{equation}
which is robust to secondary effects so long as they are well modelled in simulation and is therefore a reliable measure of the jet energy scale difference between data and MC simulation. The double ratio $c$ is transformed via numerical inversion from a function of reference object \pt to a function of jet \pt (and jet \etajet where applicable). This is the final \insitu calibration.
 
There are three stages of \insitu analyses. First, the \etajet intercalibration analysis corrects the energy scale of forward ($0.8 \leq |\etadet| < 4.5$) jets to match those of central ($|\etadet| < 0.8$) jets using the \pt balance in dijet events. Second, the \zjet and \gammajet analyses balance the hadronic recoil in an event against the \pt of a calibrated $Z$ boson or photon. The missing-$E_{\text{T}}$ projection fraction (MPF) method uses the full hadronic recoil instead of a jet to compute the balance to help mitigate effects of \pileup and jet reconstruction threshold which otherwise make low-\pT measurements challenging~\cite{ATLAS-CONF-2015-057}. Finally, the multijet balance (MJB) analysis uses a system of well-calibrated low-\pt jets to calibrate a single high-\pt jet~\cite{ATLAS-CONF-2013-003}. The \vjet and MJB analyses are computed only for central jets, but are also applicable to forward jets due to the effect of the $\eta$ intercalibration. Each measurement is translated from a function of reference object \pt into jet \pt. A statistical combination of the \vjet and MJB analyses provides a single smooth calibration applicable across the full momentum range.
 
Since the three \insitu analyses ($\eta$ intercalibration, \vjet MPF, and MJB) are performed sequentially, systematic uncertainties are propagated from each to the next. Within each analysis, systematic uncertainties arise from three sources: modelling of physics processes in simulation, uncertainties in the measurement of the reference object, and uncertainties in the expected \pt balance due to the event's topology. Mis-modelling is accounted for by comparing the predictions of two MC generators and taking their difference as the uncertainty. Systematic uncertainties in the measurement of the reference object are taken from the $\pm 1 \sigma$ uncertainties in each object's calibration and propagated through the analysis. Event topology uncertainties are estimated by varying the event selections used and observing the impact on the final MC simulation to data ratio.
 
A rebinning procedure is applied to each systematic uncertainty to ensure that the features represented in the final result are statistically significant and not the result of fluctuations in small numbers of simulated or data events.  This is only performed where the response does not vary sharply with \pt, ensuring it does not obscure real physics effects. The rebinning procedure follows a bootstrapping method: pseudo-experiment datasets are created by sampling from a Poisson distribution with a mean of one for each event in the data or MC simulation~\cite{bohm2010introduction}. The pseudo-experiments are therefore statistically correlated yet unique, and the root mean square of the response distribution across the pseudo-experiments provides a measure of the statistical uncertainty of the analysis. The measured result for each systematic uncertainty is then rebinned as appropriate for each analysis to ensure that the final shapes are statistically significant.
 
The \vjet and MJB calibrations and uncertainties are derived from the full 2015--2017 combined datasets with a total luminosity of 80~\ifb.  The $\eta$ intercalibration analysis uses a dataset of total size 81~\ifb, but since this analysis is more sensitive than the others to year-by-year fluctuations, the dataset is split into two blocks and a time-dependent result is computed instead. One $\eta$ intercalibration is derived from and applies to the $2015+2016$ dataset while a second independent calibration is derived from the $2017$ dataset and applies to $2017+2018$ data. These two data periods are treated separately due to a change in LAr calorimeter read-out that occurred between 2016 and 2017 data taking and affected jet reconstruction in the endcap regions. With no changes of similar scale made between 2017 and 2018 data taking, the 2017 calibration can be reasonably applied to 2018 as well. The post-calibration jet performance is consistent between these two different data periods and therefore a single set of uncertainties based on the $2015+2016$ dataset is used for the $\eta$ intercalibration in all years, with only a small localized additional uncertainty added for 2018 as described in Section~\ref{sec:jes-etaintercal}.
 
Certain common selection criteria are applied to all three \insitu analyses. Each event must have a reconstructed vertex with at least two associated tracks of $\pt>500~\MeV$. All jets must satisfy quality criteria to reject non-collision background, calorimeter noise, and cosmic rays~\cite{ATLAS-CONF-2015-029}. Furthermore, each jet with $20~\GeV < \pt < 60~\GeV$ and $|\etadet| < 2.4$ must pass jet vertex tagging, or JVT, requirements with selection criteria that are specific to the jet definition~\cite{ATLAS-CONF-2014-018}. These requirements match jets to the primary vertex and are 92\% efficient for \EMTopo jets and 97\% efficient for \pflow jets.

\subsubsection{Relative calibration measurement in $\eta$ using dijet events}
\label{sec:jes-etaintercal}
 
The $\eta$ intercalibration analysis produces a correction which is applied to forward jets ($0.8 \leq |\etadet| < 4.5$) to bring them to the same energy scale as central jets ($|\etadet| < 0.8$).  Jets in the central region of the detector are taken to be well-calibrated, while jets in the forward regions vary in response and must be corrected accordingly. Events are selected with exactly two jets in different $\eta$ regions of the detector. To maximize statistics, neither jet need be in the central region: instead, all regions will be calibrated relative to one another.
 
For these dijet events, momentum balance requires that the transverse momentum of the two jets must be equal and opposite.
Therefore, the momentum \textit{asymmetry} of the two jets is a metric for the response difference between the two detector regions (\textit{left} and \textit{right} for simplicity):
 
\begin{equation}\nonumber
\asym = \frac{\pt^{\text{left}}-\pt^{\text{right}}}{\pTavg}\,,
\label{eq:asym-etacal}
\end{equation}
where $\pTavg = (\pt^{\text{left}}+\pt^{\text{right}})/2$.  The response ratio \Response of the two jets defines the calibration factor $c$ for each jet and is then:
\begin{equation}\nonumber
\Response = \frac{c^{\text{left}}}{c^{\text{right}}} = \frac{2 + \avg{\asym}}{2 - \avg{\asym}} \cong \frac{\pt^{\text{left}}}{\pt^{\text{right}}}\,.
\end{equation}
 
The average response ratio $\avg{\Response_{ijx}}$ is measured in each bin $i$ of $\eta^{\text{left}}$, $j$ of $\eta^{\text{right}}$, and $x$ of $\pTavg$; $\Delta \avg{\Response_{ijx}}$ is the statistical uncertainty in each bin. All $\eta$ values are in detector coordinates rather than corrected jet coordinates (\etadet) since the properties of interest correlate to specific regions of detector hardware.
The following function relates the correction factors and responses in each of the $N$ bins:
\begin{equation}\nonumber
\label{etaintercal_function}
S(c_{1x},\ldots,c_{Nx}) = \sum_{j=2}^{N}\sum_{i=1}^{j-1} \left(\frac{1}{\Delta \avg{\Response_{ijx}}} (c_{ix}\avg{\Response_{ijx}} - c_{jx}) \right)^{2} + X(c_{ix})\,.
\end{equation}
Here, the function $X(c_{ix})$ quadratically imposes a penalty on correction factors deviating from 1.\footnote{This penalty function takes the form $X(c_{i})= \lambda \left( \frac{1}{N} \sum_{i=1}^{N} c_{i} -1 \right)^2$, where $\lambda$ introduces the Lagrange multiplier visible in Eq.~(\ref{eq:etaintercal_bigboy}). The purpose of the penalty function is to ensure that the appropriate minimum is selected by suppressing local minima with large values of $c_{ix}$, and as such its exact form is somewhat arbitrary.} Minimizing this function produces the correction factors to be used in the calibration.
 
Previous iterations of the jet energy scale have used a fit in \textsc{Minuit} to minimize $S(c_{ix})$.  The current calibration instead minimizes the function analytically. Suppressing the $x$ indices for clarity and setting the derivative of $S$ with respect to some correction factor $c_\alpha$ equal to zero, the following equation defines the correction factor values which minimise S:
\begin{multline}
\label{eq:etaintercal_bigboy}
\sum_{i=1}^{\alpha -1}\left( \Big(\frac{-\avg{\Response_{i\alpha}}}{\Delta^2 \avg{\Response_{i\alpha}}} + \frac{\lambda}{N^2}\Big) c_i \right) +
\left( \sum_{i=1}^{\alpha -1}\frac{1}{\Delta^2 \avg{\Response_{i\alpha}}} + \sum_{i=\alpha +1}^{N}\frac{\avg{\Response_{\alpha i}}^2}{\Delta^2 \avg{\Response_{\alpha i}}} + \frac{\lambda}{N^2} \right) c_{\alpha} \\
+  \sum_{i=\alpha +1}^{N} \left( \Big( \frac{- \avg{\Response_{\alpha i}}}{\Delta^2 \avg{\Response_{\alpha i}}} + \frac{\lambda}{N^2} \Big) c_i \right)
- \frac{\lambda}{N}  = 0~.
\end{multline}
Here $\lambda$ is a Lagrange multiplier arising from the penalty term whose value has no effect on the minimization result but prevents the trivial solution where all the $c_i$ are null.
 
Equation~(\ref{eq:etaintercal_bigboy}) can then be expressed as a matrix system of linear equations. This matrix system is solved independently for each $\pTavg$ bin $x$ to obtain values for the correction factors $c_{ix}$ for each \etadet bin $i$ in this momentum range. Solving analytically for the $c_{ix}$ in this way allows the result to be found approximately a thousand times more quickly than using a fit.
This large reduction in computational requirements in turn allows the analysis to use a finer binning in \etadet, capturing more details of the detector structure. The two methods agree well and each shows good closure when tested in simulation. Finally, the full set of correction factors are normalized such that the average correction factor in the central region $|\etadet| < 0.8$ is unity. 
 
Events are selected using a combination of single-jet triggers, with each trigger only considered in the jet \pt range for which it is at least 99\% efficient~\cite{TRIG-2012-01,TRIG-2016-01}.
Events may pass either a central jet trigger or a forward jet trigger, or both.
In the case that a trigger is prescaled, the passing event is weighted by the appropriate amount.
Jets with $|\etadet|<2.4$ are also required to satisfy JVT criteria to minimize contributions from \pileup and must pass basic cleaning requirements~\cite{PERF-2014-03,ATLAS-CONF-2015-029}.
Each selected event must have two jets with $\pt > 25$~\GeV\ and $|\eta| < 4.5$.
To ensure a clean dijet topology, events are further required to have no third jet with significant \pt: $\pt^\text{third}/\pTavg<0.25$, where $\pTavg$ is the average momentum of the two leading jets. The two leading jets are required to be back-to-back in the azimuthal plane such that $\Delta \phi > 2.5$~rad.
 
Like the other \textit{\insitu} analyses, the goal of the $\eta$ intercalibration is to correct for data--simulation differences, so the quantity of interest is the ratio of the measured calorimeter response in MC simulation to the response in data.  The nominal calibration is derived by comparison with \POWPYTHIA{}8 simulated events.
The analysis binning in \pTavg and \etadet is selected to balance the requirements of both sufficient statistics in sparse regions and resolution of narrow detector features. As such, it varies for different values of \etadet.
Remaining statistical fluctuations in the final calibration are smoothed using a two-dimensional Gaussian kernel with parameters selected to preserve significant structures.
 
\begin{figure}[htp!]
\begin{center}
\begin{subfigure}[b]{0.49\textwidth}
\includegraphics[width=\textwidth]{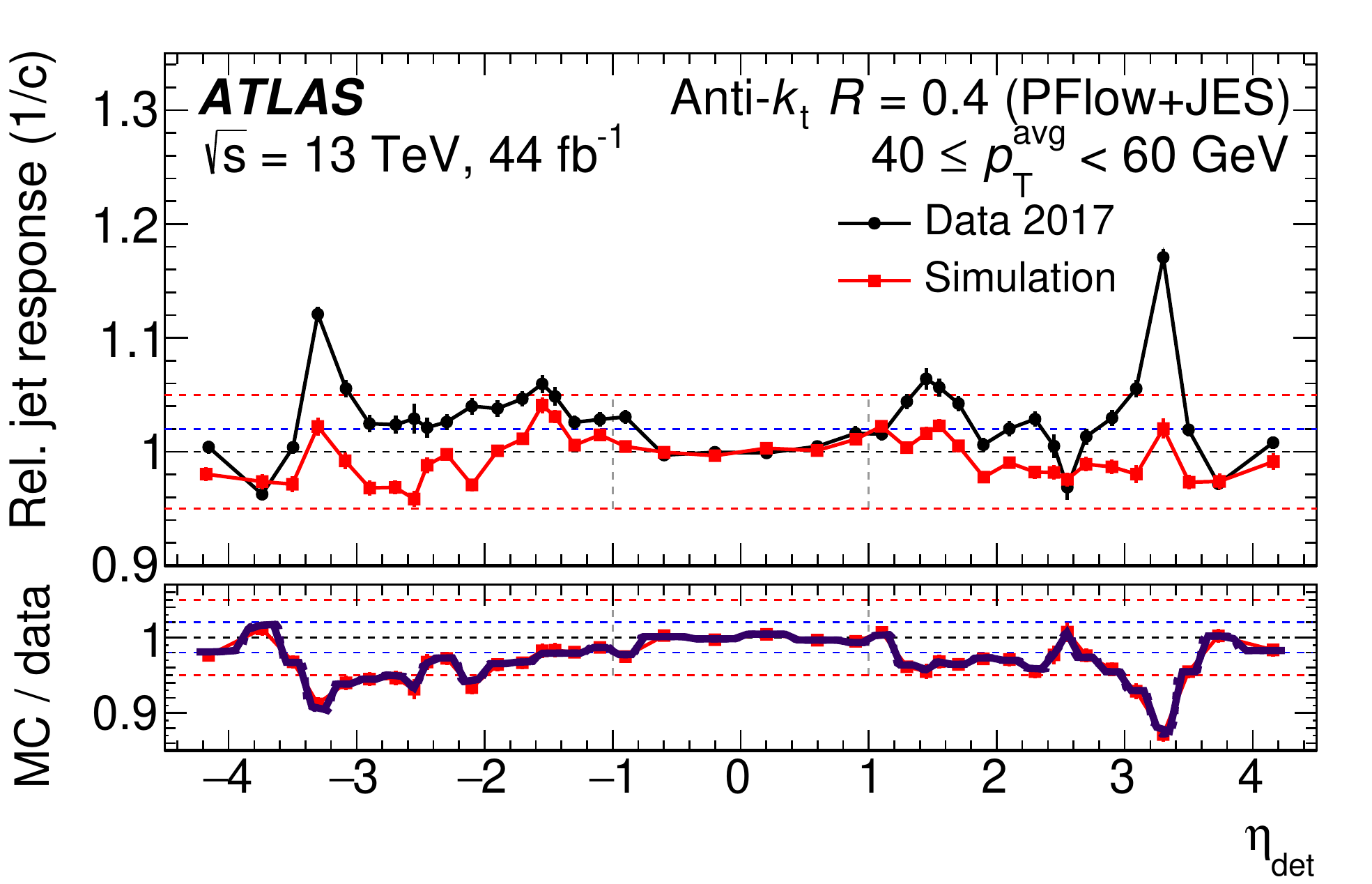}
\caption{}
\label{subfig:etaintercal_vseta_lowpt}
\end{subfigure}
\begin{subfigure}[b]{0.49\textwidth}
\includegraphics[width=\textwidth]{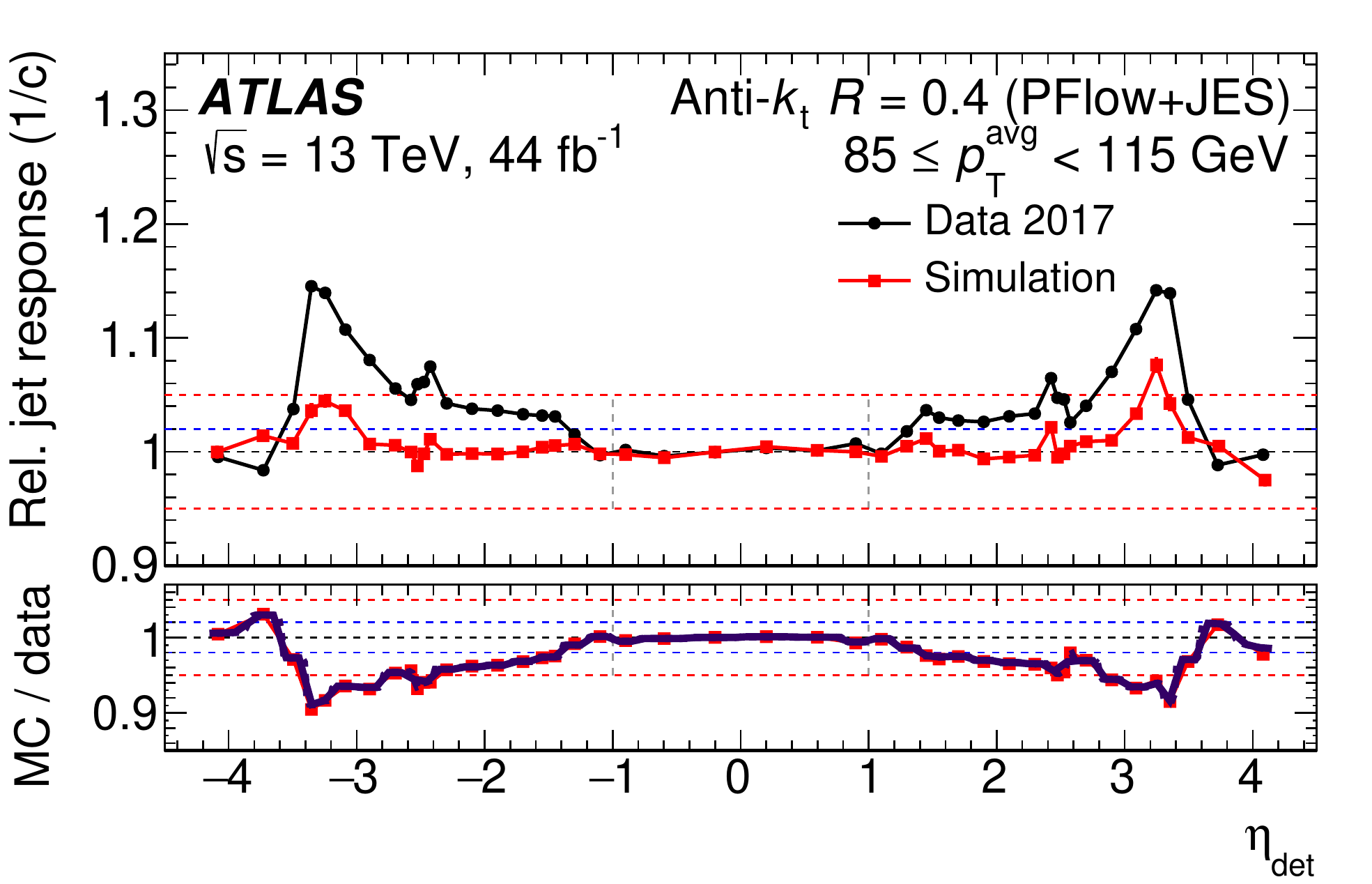}
\caption{}
\label{subfig:etaintercal_vseta_medpt}
\end{subfigure}
 
\begin{subfigure}[b]{0.49\textwidth}
\includegraphics[width=\textwidth]{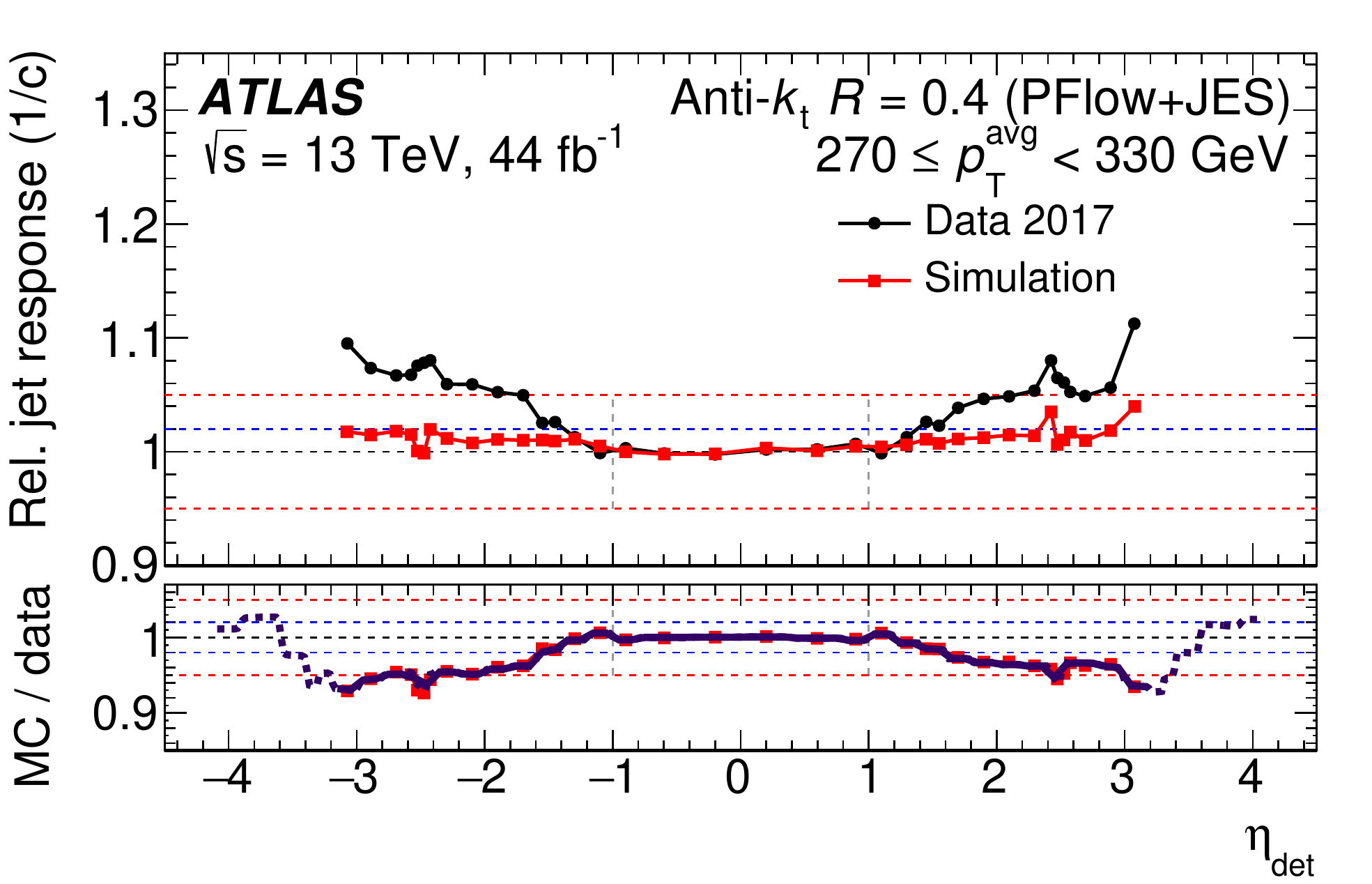}
\caption{}
\label{subfig:etaintercal_vseta_highpt}
\end{subfigure}
\begin{subfigure}[b]{0.49\textwidth}
\includegraphics[width=\textwidth]{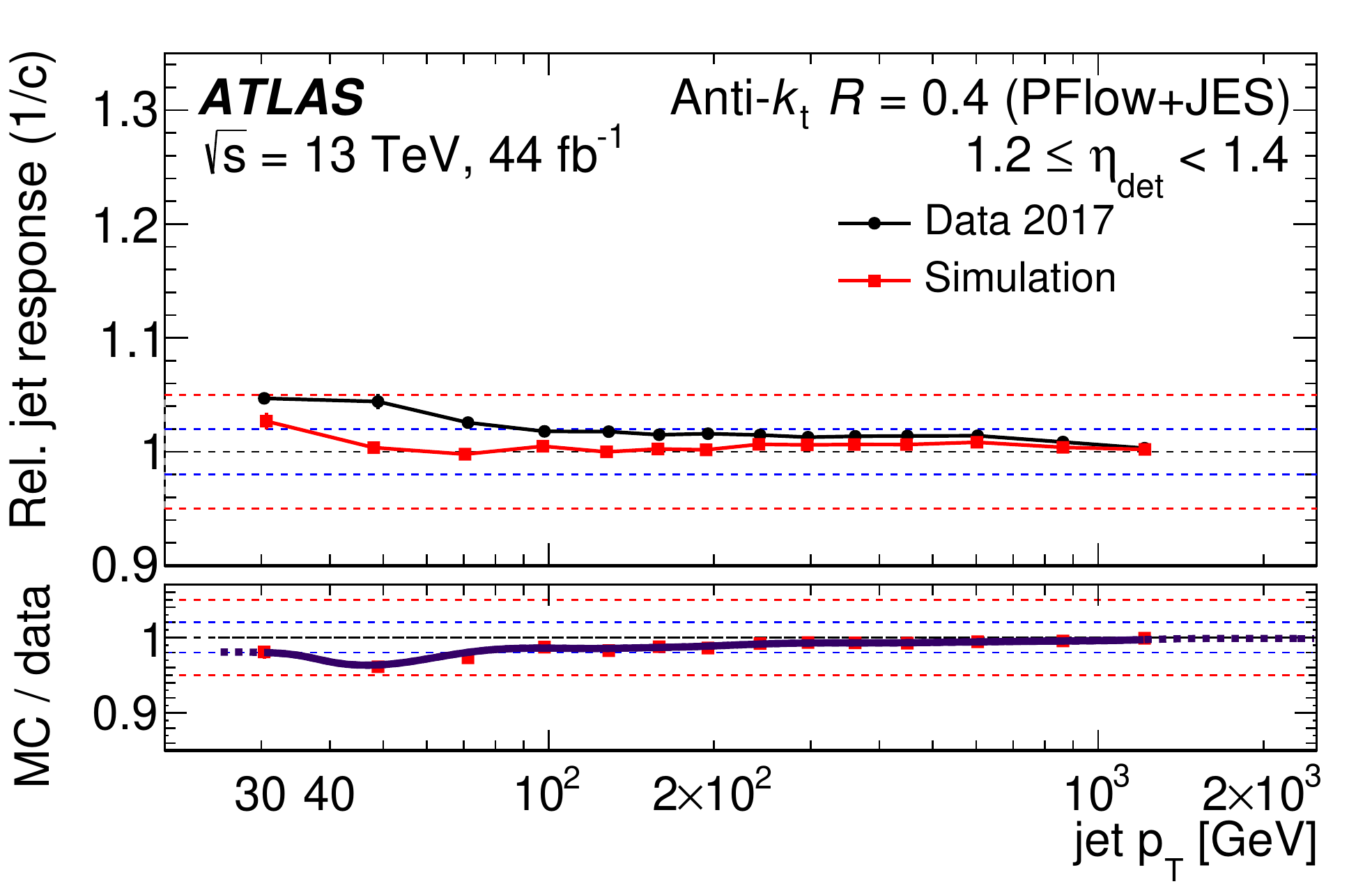}
\caption{}
\label{subfig:etaintercal_vspt_central}
\end{subfigure}
 
\begin{subfigure}[b]{0.49\textwidth}
\includegraphics[width=\textwidth]{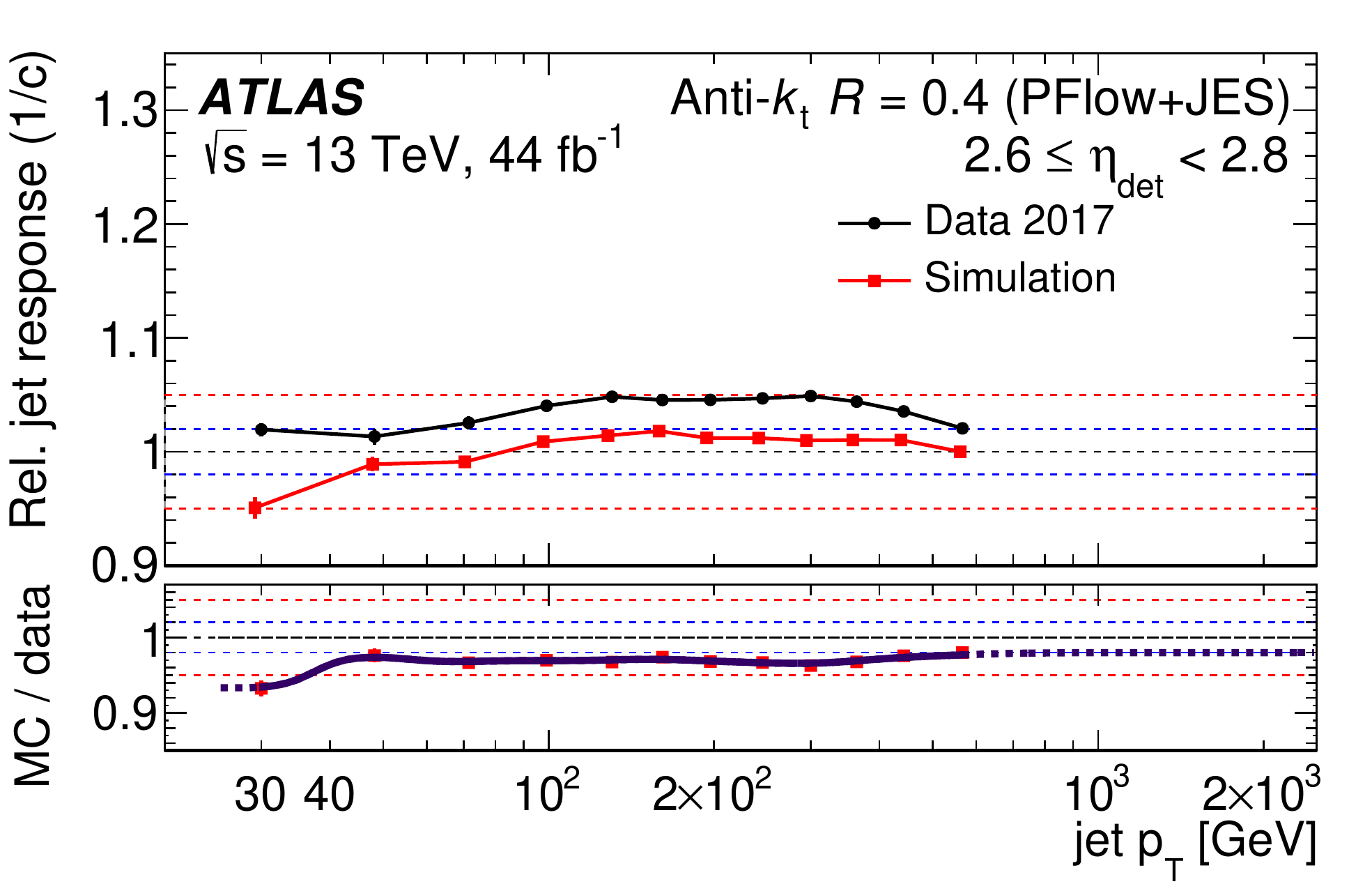}
\caption{}
\label{subfig:etaintercal_vspt_forward}
\end{subfigure}
\begin{subfigure}[b]{0.49\textwidth}
\includegraphics[width=\textwidth]{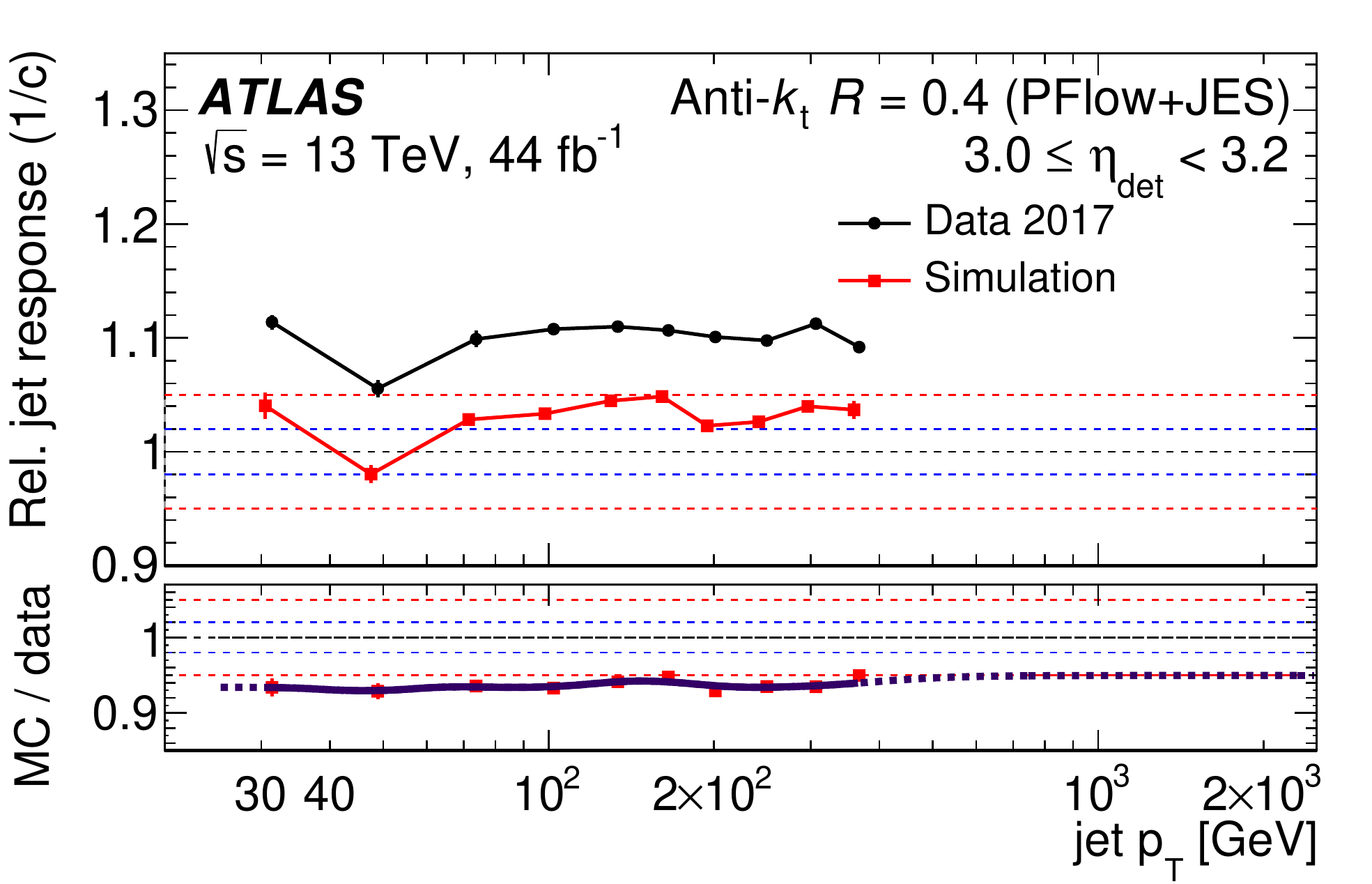}
\caption{}
\label{subfig:etaintercal_vspt_veryforward}
\end{subfigure}
\caption{Relative response of jets calibrated with \EMPFlowJES in data (black circles) and \POWPYTHIA{}8 MC simulation (red squares). Response is shown as a function of \etadet for jets of \subref{subfig:etaintercal_vseta_lowpt} $40~\GeV < \ptjet < 60~\GeV$, \subref{subfig:etaintercal_vseta_medpt} $85~\GeV < \ptjet < 115~\GeV$, and \subref{subfig:etaintercal_vseta_highpt} $270~\GeV < \ptjet < 330~\GeV$, and as a function of \pt for jets of \subref{subfig:etaintercal_vspt_central} $1.2 < \etadet < 1.4$, \subref{subfig:etaintercal_vspt_forward} $2.6 < \etadet < 2.8$, and \subref{subfig:etaintercal_vspt_veryforward} $3.0 < \etadet < 3.2$.  The lower panel shows the response ratio of simulation to data (red squares) as well as the smoothed \insitu calibration factor derived from the ratio (solid curve) which is used to perform the $\eta$ intercalibration. Dotted lines show the extrapolation of the \insitu calibration to the regions without data points. The dashed red and blue horizontal lines provide reference points for the viewer.}
\label{fig:etaintercal_responses}
\end{center}
\end{figure}
 
\Fig{\ref{fig:etaintercal_responses}} shows the measured response in data and \POWPYTHIA{}8 MC simulation for the 2017 dataset as a function of \etadet for three different \pTavg ranges (Figures~\ref{subfig:etaintercal_vseta_lowpt}, \ref{subfig:etaintercal_vseta_medpt}, \ref{subfig:etaintercal_vseta_highpt}) and as a function of \pTavg for three different \etadet ranges (Figures~\ref{subfig:etaintercal_vspt_central}, \ref{subfig:etaintercal_vspt_forward}, \ref{subfig:etaintercal_vspt_veryforward}).
The simulation can be seen to approximately reproduce the \etadet-dependent features of the response observed in data, although the response in data is consistently higher than the response in simulation. The simulation/data response ratio as directly measured is shown in discrete points in the bottom panel, while the calibration derived from smoothing the response ratio is overlaid as the solid curve.
The dashed curve shows the extrapolation to \pt ranges beyond the available data, taken from the Gaussian smoothing results. Since the smoothing is stronger in the \pt direction and weaker in $\etadet$ to preserve detector features, this sets each extrapolated value to approximately the value of the last populated bin at lower \pt. Above $\pT=2~\TeV$ the value is kept constant.
 
Uncertainties are derived as a function of \etadet and \pt and account for mis-modelling of physics, detector, and event topology effects on the momentum balance of dijet events. The dominant uncertainty is in MC mis-modelling and is taken to be the difference between the smoothed calibration curves derived from the \POWPYTHIA{}8 and \SHERPA dijet samples.  Additional uncertainties in the physics and topology modelling are assessed by varying the $\pt^\text{third}$, $\Delta \phi$, and \pileup suppression cuts and using a bootstrapping method to ensure observed shapes are statistically significant as discussed in \Sect{\ref{sec:jes-insitu}}. Similarly, the JVT uncertainty is determined by comparison with tighter and looser working points.
These uncertainties can take positive or negative values.
The statistical uncertainty is strictly positive and is taken from the data and MC simulation sample sizes.
Finally, a non-closure uncertainty is assessed by comparing the response in data with that in \POWPYTHIA{}8 after applying the derived $\eta$ intercalibration. This uncertainty is largest for $|\etadet| \sim 2.1$--$2.6$, where detector transitions make modelling of the LAr pulse shape particularly difficult~\cite{PERF-2014-07}, and for jets near the kinematic limit, where they have the maximum possible \pT for a given $\etadet$ subject to the constraint of a 13~\TeV\ centre-of-mass energy. The non-closure uncertainty is treated as three independent nuisance parameters, two covering the regions around $\pm 2.4$ in $\etadet$ and one at the kinematic limit, since these two non-closure uncertainties are uncorrelated.
 
After being corrected each with their dedicated calibration, the 2015+2016 and 2017 datasets are in good agreement, and therefore a single set of uncertainties is sufficient to cover both cases.
The uncertainties calculated with the 2015+2016 dataset are selected for this role.
The only dataset-dependent uncertainty is an additional small non-closure uncertainty used for 2018 data only. It covers the region around $\eta = \pm 1.5$ to account for the difference in Tile calorimeter calibration during this year of data-taking and has a maximum size of 2\%.
 
The method uncertainties are shown in \Fig{\ref{fig:etaintercal_systs}}. Three illustrative \pt values are selected. The uncertainties decrease slightly as a function of \pt and increase significantly as a function of $\etadet$ outside of the central detector region, while in the central region they are zero by construction. For practical use the various systematic uncertainty terms are summed in quadrature to produce one single systematic uncertainty dominated by the modelling term. In the cases where up and down variations differ, the largest absolute value of the two is used at each point. The total systematic uncertainty and the statistical uncertainty are both symmetrized in $\etadet$. The non-closure uncertainties, not included in \Fig{\ref{fig:etaintercal_systs}} as they are not method uncertainties, are instead shown in \Fig{\ref{fig:combination_uncertainties}} where it can be seen that they are kept asymmetric to reflect real differences in the detector.
 
\begin{figure}[htp!]
\begin{center}
\begin{subfigure}[b]{0.49\textwidth}
\includegraphics[width=\textwidth]{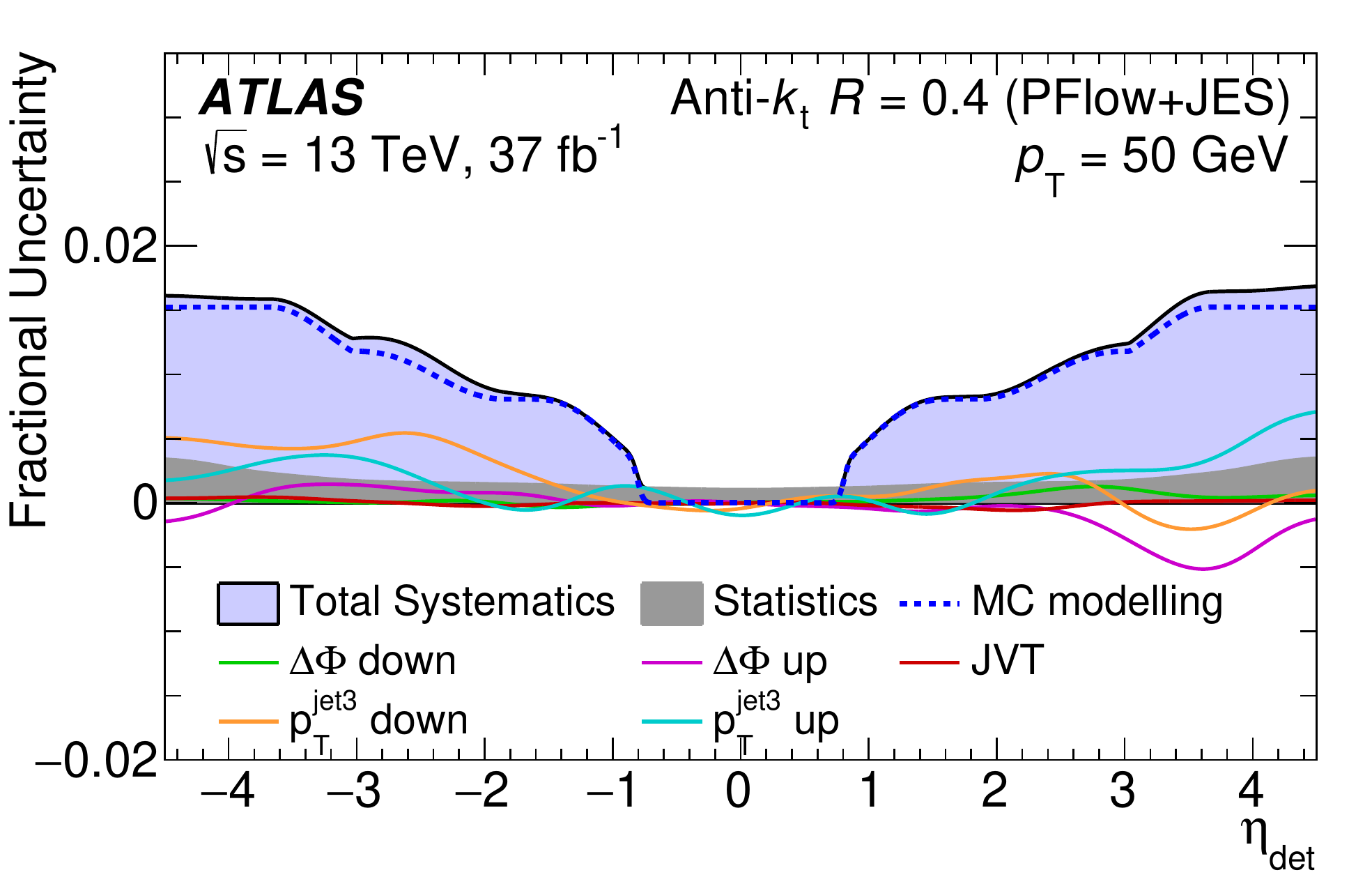}
\caption{}
\label{subfig:etaintercal_syst_50}
\end{subfigure}
\begin{subfigure}[b]{0.49\textwidth}
\includegraphics[width=\textwidth]{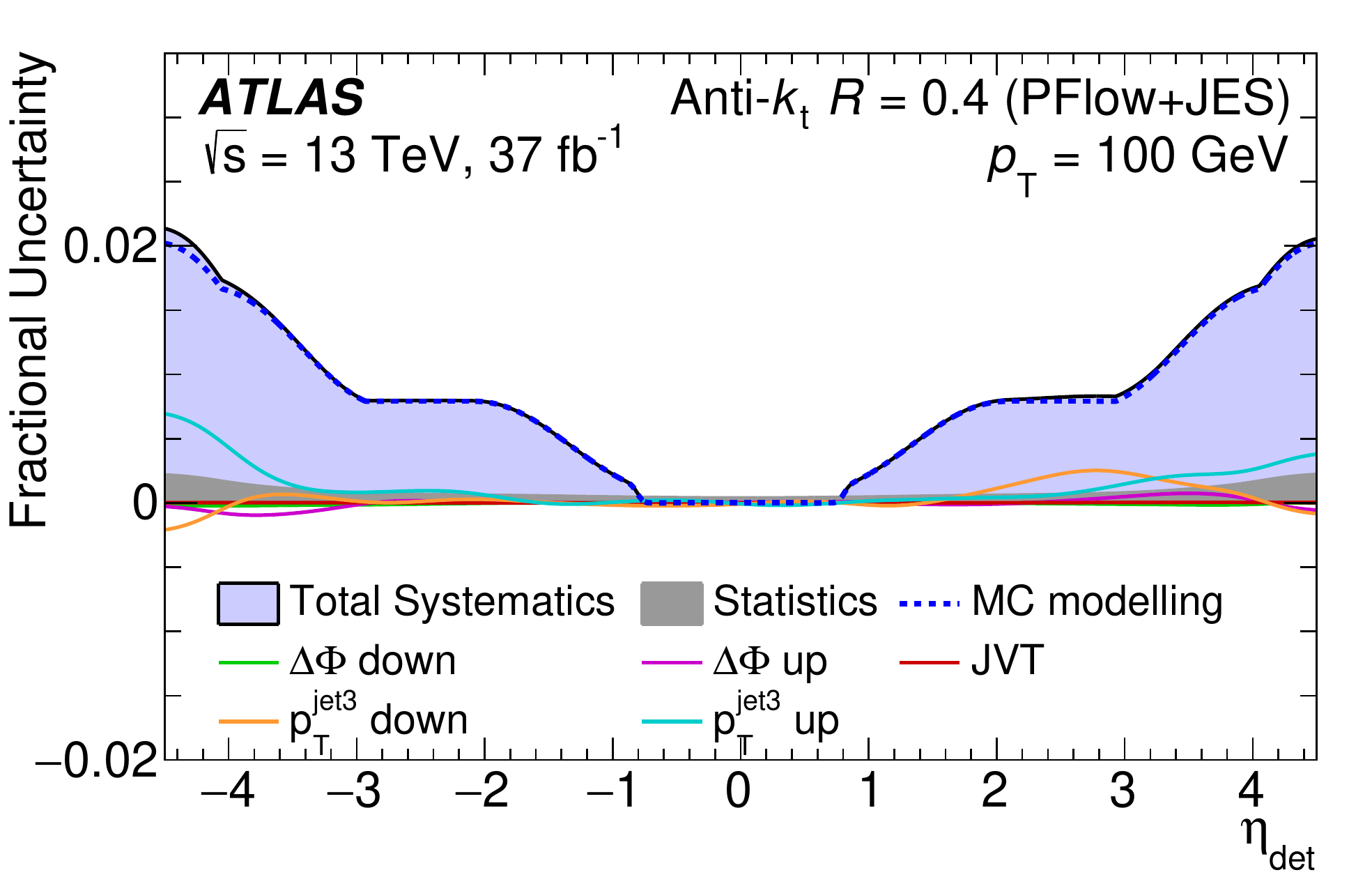}
\caption{}
\label{subfig:etaintercal_syst_100}
\end{subfigure}
 
\begin{subfigure}[b]{0.49\textwidth}
\includegraphics[width=\textwidth]{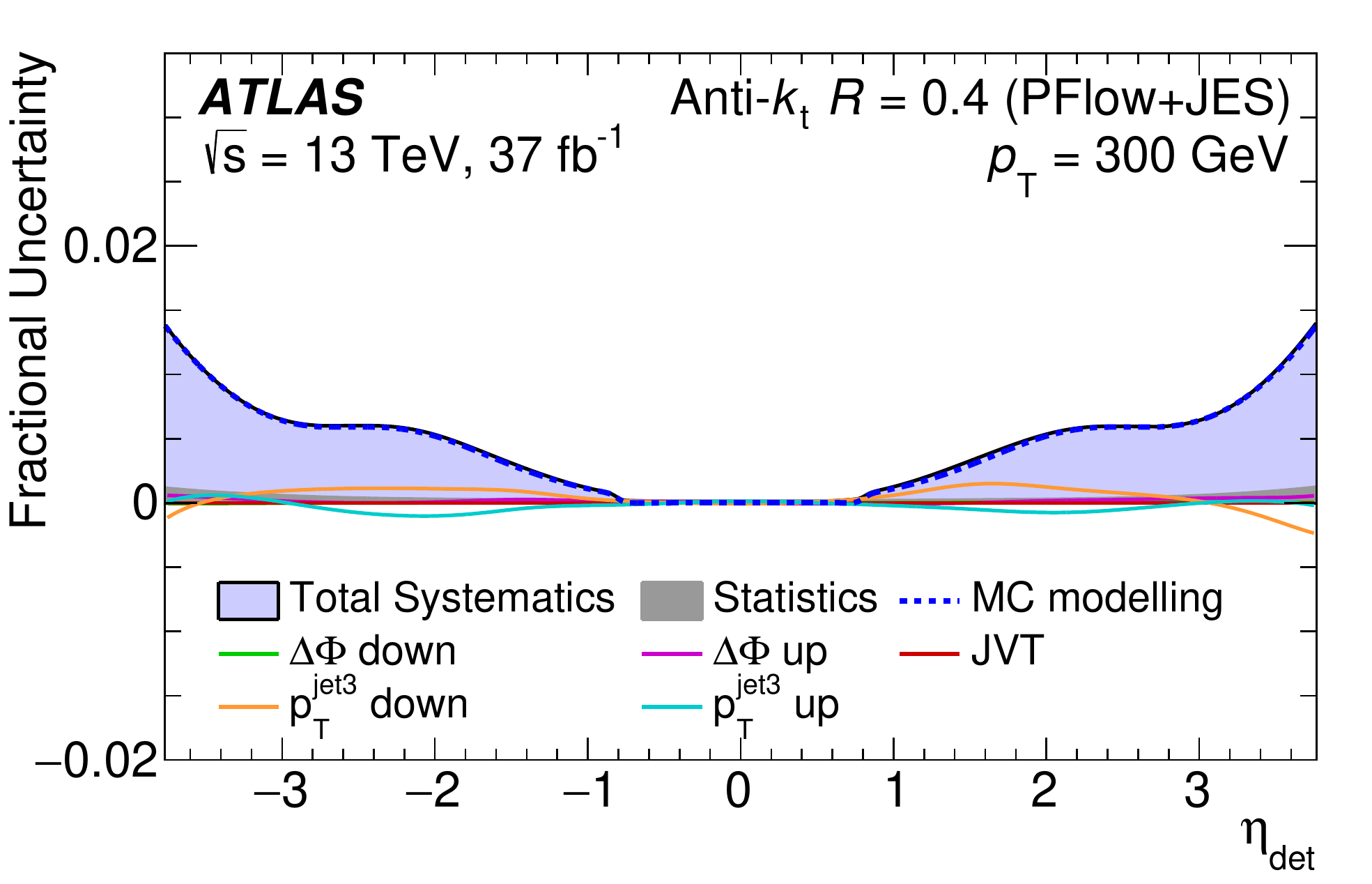}
\caption{}
\label{subfig:etaintercal_syst_300}
\end{subfigure}
\caption{Systematic uncertainties associated with the $\eta$ intercalibration procedure as a function of $\etadet$ for \EMPFlowJES jets of \subref{subfig:etaintercal_syst_50} $\pt=50~\GeV$, \subref{subfig:etaintercal_syst_100} $\pt=100~\GeV$, and \subref{subfig:etaintercal_syst_300} $\pt=300~\GeV$. The solid purple band shows the total systematic uncertainty, while the grey band shows the statistical uncertainty alone. Individual sources of uncertainty are marked by coloured lines. These have been smoothed to remove the impact of statistical fluctuations. Thus the visible shapes are statistically significant. The MC modelling term is the dominant source of uncertainty.}
\label{fig:etaintercal_systs}
\end{center}
\end{figure}
 
The calibrations are similar in size and shape between \pflow and \EMTopo jets. Systematic uncertainties are also similar in size and shape since the dominant MC modelling component does not differ meaningfully between the two jet collections.
 
\FloatBarrier
 
\subsubsection{Calibration measurement using \zjet and \gammajet events}
\label{sec:jes-vjet}
The next stage of the \insitu calibration corrects for the differences between data and MC simulation using the momentum balance between the measured hadronic activity in the event and the \pt of a well-calibrated photon or $Z$ boson. Only the central region of the detector ($|\eta| < 0.8$) is used for this analysis: the $\eta$ intercalibration ensures that a correction derived centrally translates directly to forward jets as well. 
 
The \vjet analyses rely on the energy scale of the photon or the electrons and muons from the $Z$ decay being well measured. All three objects are cleanly measured in the ATLAS detector and the uncertainties in their energy scales are small~\cite{EGAM-2018-01,PERF-2015-10}. The response is calculated separately in $Z\rightarrow e^{+}e^{-}$ and $Z\rightarrow\mu^{+}\mu^{-}$ events since the sources of uncertainties propagated from $e$ and $\mu$ calibration are independent, and the three channels are combined at a later stage. The \zjet response measurement is limited at moderate to high \pt by low statistics and thus covers a range in jet \pt from $17~\GeV$ to $1~\TeV$ with large uncertainties in the final bin. The \gammajet response measurement benefits from much higher statistics and extends to 1.2~\TeV{} with little loss in sensitivity. However, it is limited at low jet \pt by both the trigger prescales and the prevalence of soft jets misidentified as photons and so begins at 25~\GeV.
 
The missing-$E_{\text{T}}$ projection fraction technique is used for both of the \vjet analyses and balances the reference object \pt against the full hadronic recoil in an event. By doing so, it is possible to compute the calorimeter response to hadronic showers directly. This approach is robust to both \pileup and the underlying event, which each cancel out directionally on average over a large collection of events, and is not strongly affected by jet definitions since these become relevant only in the application of the calibration. The showering and topology effects in moving from a recoil-level quantity to a jet-level quantity are studied and found to be small, as discussed below.
Taking $\ptvec^{\text{\,recoil}}$ as the total transverse momentum of the hadronic activity in a clean \vjet event and \ptref as the transverse momentum of the photon or $Z$ boson, conservation of transverse momentum means that at the particle level:
 
\begin{equation}
\vec{p}_\text{T,truth}^\text{\,ref} + \vec{p}_\text{T,truth}^{\text{\,recoil}} = 0\,.
\label{eq:conservofmom}
\end{equation}
 
This balance could be altered by the presence of initial- or final-state radiation.\footnote{The balance can also be affected by leptonic processes in heavy flavour decays, though heavy flavour production is rare in \zjet events. Electrons are included in the balance calculation but muons and neutrinos are not. However, this effect is not a problem like additional radiation, since the balance in heavy flavour events actually contributes to measuring the energy loss and correctly calibrating their momenta.} To suppress the effects of such additional radiation, a cut is placed on the azimuthal angle $\Delta \phi$ between the jet and the reconstructed photon or $Z$ boson in the event and an uncertainty due to the topology is evaluated by varying the event selection requirements.
If the calorimeter response to the hadronic activity in this event is $r_{\text{MPF}}$ and the response for the calibrated reference object is 1, and assuming any missing energy in the event is due to the low response to the hadronic recoil ($r_{\text{MPF}} < 1$), then at the detector level Eq.~(\ref{eq:conservofmom}) becomes:
 
\begin{equation}\nonumber
\ptrefvec + r_{\text{MPF}}\,\, \ptrecvec = - \vmet
\end{equation}
 
After taking the projection of each term in the direction of the reference object, defined by a unit vector $\hat{n}_\text{ref}$, the response to the hadronic recoil is then seen to depend only on the missing energy in the event and the momentum of the reference object. The MPF response $\Response_\text{MPF}$ is defined by measuring the average of $r_{\text{MPF}}$ across events binned in the reference object \pt. Thus,
\begin{equation}\nonumber
\Response_{\text{MPF}} = \left< 1 + \frac{\hat{n}_\text{ref} \cdot \vmet}{\ptref } \right>\,.
\end{equation}
 
This peak location is taken to be the average response in that bin, and the response is mapped from reference to jet \pT by finding the average jet \pT in the events in each bin after $\eta$ intercalibration but before the application of any other \insitu steps.
 
Missing energy in each event is reconstructed from calorimeter \topos\ in the case of \EMTopo jet calibration and from particle-flow objects in the case of \pflow jet calibration, ensuring that the energy scale is consistent.  The $Z\rightarrow ee$ events are required to pass a dielectron trigger with $\pt^{e1,e2} > 15~\GeV$; $Z\rightarrow \mu\mu$ events must pass a similar dimuon trigger with $\pt^{\mu1,\mu2} > 14~\GeV$~\cite{TRIG-2018-01,TRIG-2018-05}. Electrons entering the analysis must have $\pt > 20~\GeV$, ensuring that the trigger is fully efficient, must be contained within the tracker such that $|\eta_{e}| < 2.47$, and must not fall in the calorimeter transition region ($1.37 < |\eta| < 1.52$). Muons entering the analysis are required to have $\pt > 20~\GeV$ and to fall within $|\eta| < 2.4$. Both electron and muon candidates must also pass loose identification and isolation requirements~\cite{EGAM-2018-01,PERF-2015-10}. All \zjet events are selected such that the reconstructed mass calculated from the electron or muon pair must be close to the $Z$ boson mass: $66~\GeV < m_{ee/\mu\mu} < 116~\GeV$.
A combination of single-photon triggers are used for the \gammajet analysis, with the lowest trigger threshold corresponding to $\ET^{\gamma} > 15~\GeV$. Offline photons must have $\ET^{\gamma} > 25~\GeV$ and $|\eta^{\gamma}| < 1.37$ and must satisfy tight identification and isolation criteria~\cite{EGAM-2018-01}.
 
Both the \zjet and \gammajet analyses have further selection requirements on the jets and event topology to suppress \pileup and initial- and final-state radiation. All jets within $\Delta R = 0.2$ of a photon or $\Delta R = 0.35$ of a lepton are removed.
Jets must satisfy basic cleaning requirements and pass the JVT selection to suppress \pileup.
Selected events must have one jet with $\pT > 10~\GeV$ and $|\eta| < 0.8$. Additional event activity is suppressed by requiring that any subleading jet must have $\pt < \text{max}(0.3 \times \pt^{\text{ref}},12)~\GeV$ and that the leading jet and reference object must be relatively back-to-back with $\Delta \phi^{\text{ref, jet}} > 2.9$. The relatively loose \pt cut on subleading jets is shown to be acceptable for the MPF analysis due to its intrinsic robustness to pile-up effects.
 
Figures~\ref{fig:MPF_response_zjet} and~\ref{fig:MPF_response_gammajet} show the MPF response calculated in \zjet and \gammajet events for data and for two MC samples using different generators. The lower panels show the MC simulation to data ratio for both generators. The results using \POWPYTHIA{}8 (\zjet) and \Pythia{}8 (\gammajet) constitute the nominal calibration while \textsc{Sherpa} is used to define an uncertainty due to the generator choice. In the lowest \pT bin of the \gammajet measurement, the discrepancy between the MC predictions is caused by a generator-level cut at 35~\GeV\ present in the \Sherpa sample.  This point is included in the final \insitu combination, but due to its large generator uncertainty it contributes very little to the overall weighted-average-based result (see Section~\ref{sec:jes-combination} and Figure~\ref{subfig:weights_comb}) and the total effect is negligible. The \gammajet generator uncertainty at this point has therefore been set to its value in the second-lowest bin for display purposes in Figure~\ref{fig:mpf_uncert_gammajet} to better reflect its actual contribution to the total systematic uncertainties. The apparent dip near the lowest \pT range of each measurement is due to the interplay of two factors: an asymmetry in the $\Response_{\text{MPF}}$ distribution near the low \pT reconstruction threshold which causes the measured response to increase for the lowest \pT values, and the natural increase in response with higher jet \pT. One motivation for the use of the MPF technique is increased resilience to this threshold effect.
 
\begin{figure}[htp!]
\begin{center}
\begin{subfigure}[b]{0.49\textwidth}
\includegraphics[width=\textwidth]{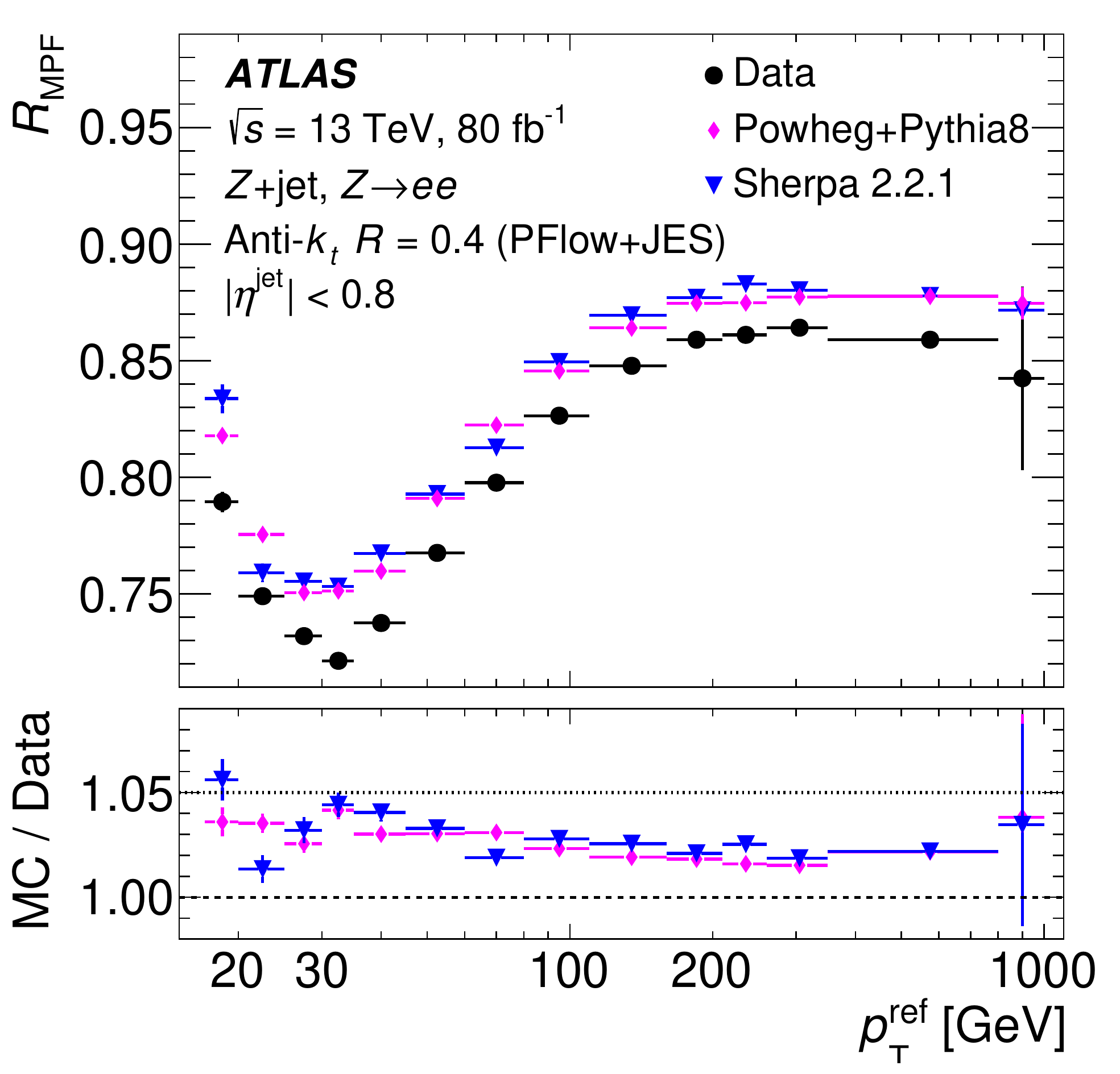}
\caption{$Z\rightarrow ee+$jet}
\label{subfig:mpf_response_zjet_e}
\end{subfigure}
\begin{subfigure}[b]{0.49\textwidth}
\includegraphics[width=\textwidth]{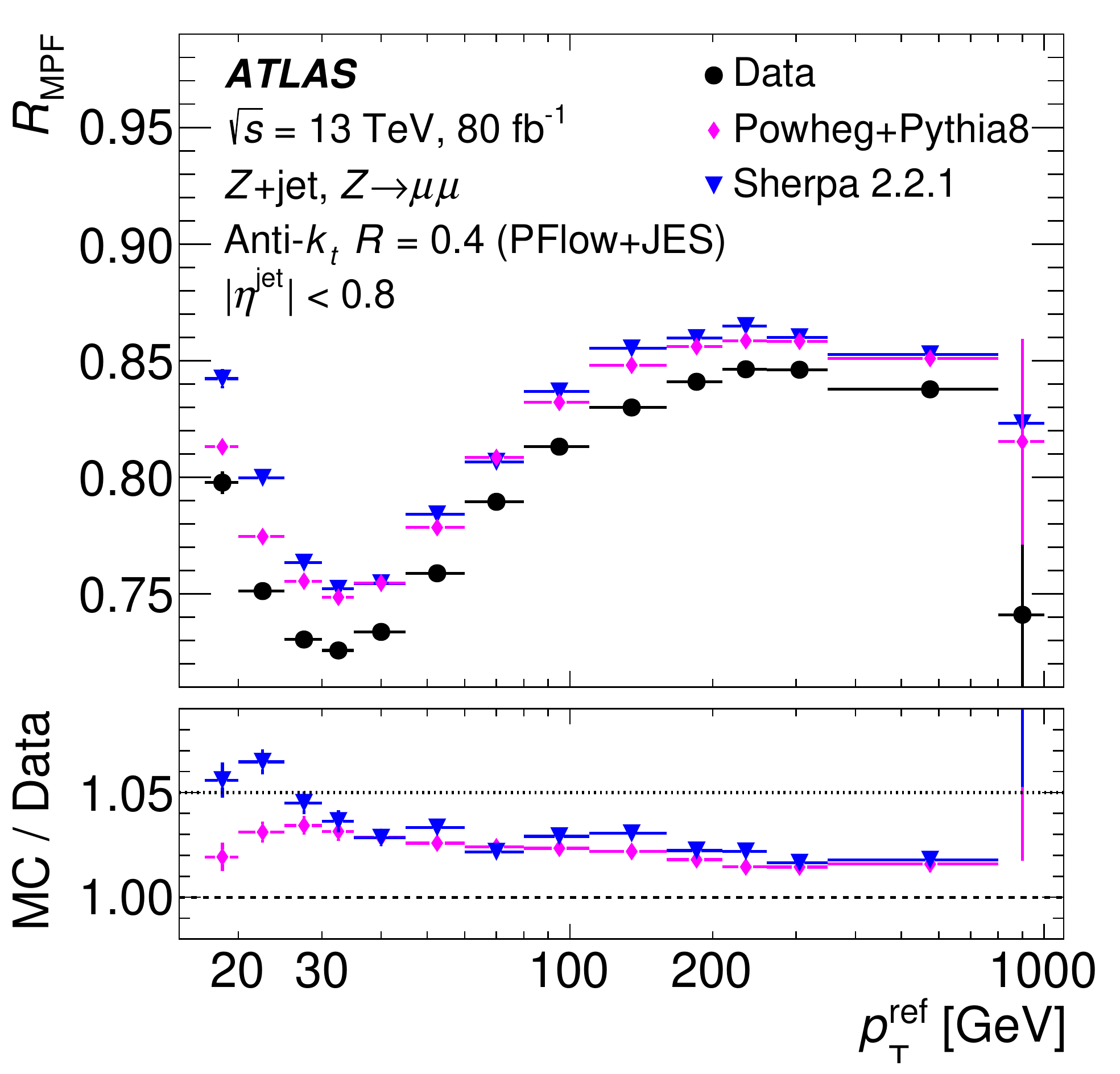}
\caption{$Z\rightarrow \mu\mu+$jet}
\label{subfig:mpf_response_zjet_mu}
\end{subfigure}
\caption{Average \pflow jet response as a function of reference \pt for \zjet events where the $Z$ boson decays into \subref{subfig:mpf_response_zjet_e} electrons and into \subref{subfig:mpf_response_zjet_mu} muons calculated using the MPF technique. $Z\rightarrow ee$ and $Z \rightarrow \mu \mu$ events are combined at a later stage. The black points correspond to 2015--2017 data while the pink diamonds and blue triangles correspond to independent Monte Carlo samples from two different generators, and their error bars show the statistical uncertainties. The ratio of MC simulation to data for both generators is shown in the bottom panel and defines the \insitu correction to be applied. The dotted lines at 1 and 1.05 serve as a reference.}
\label{fig:MPF_response_zjet}
\end{center}
\end{figure}
 
\begin{figure}[htp!]
\begin{center}
\includegraphics[width=0.7\textwidth]{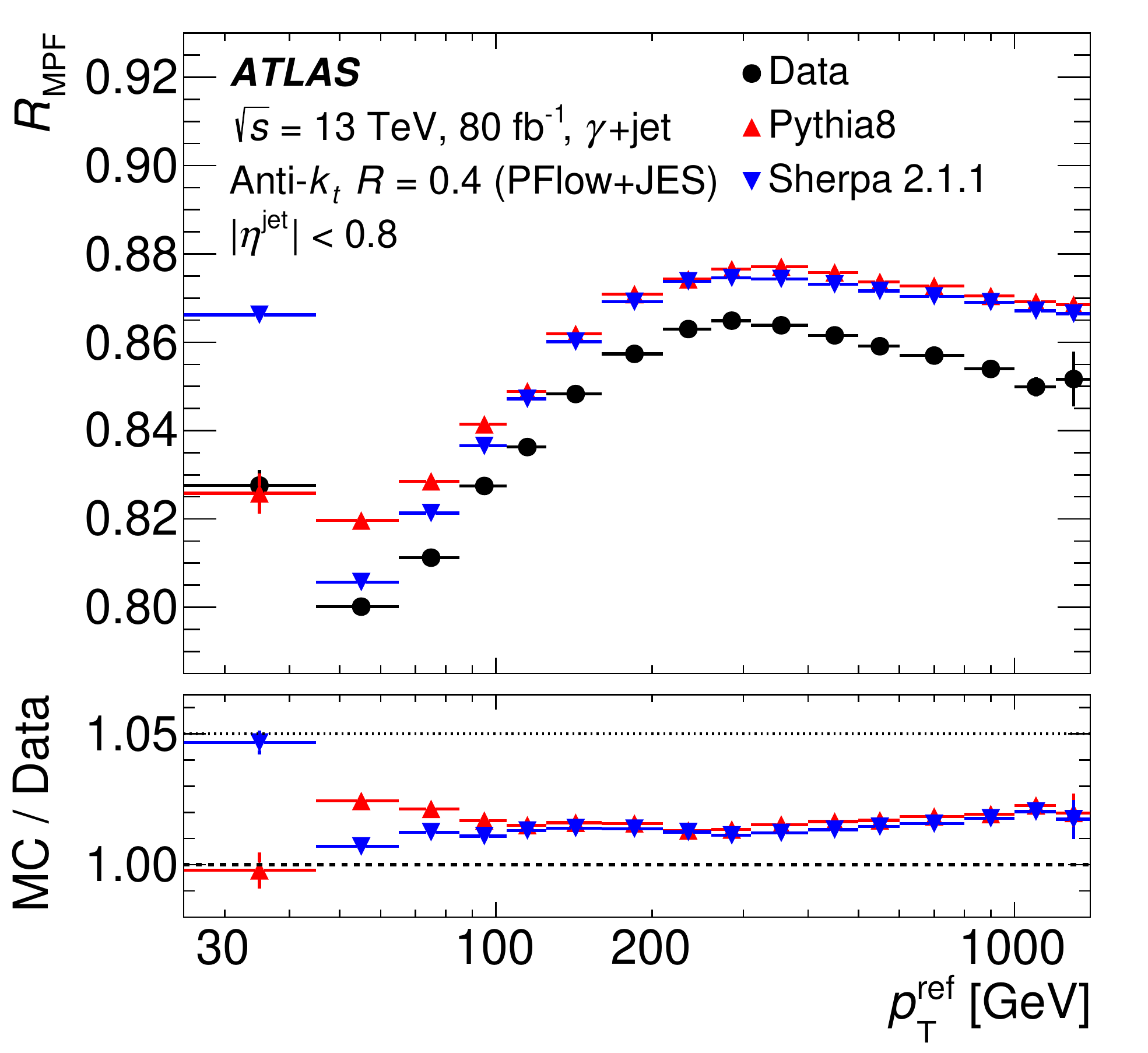}
\caption{Average \pflow jet response as a function of reference \pt for \gammajet events calculated using the MPF technique. The black points correspond to 2015--2017 data. The red and blue triangles correspond to independent Monte Carlo samples from two different generators. Error bars show the statistical uncertainties. The ratio of MC simulation to data for both generators is shown in the bottom panel and defines the \insitu correction to be applied. The dotted lines at 1 and 1.05 serve as a reference.}
\label{fig:MPF_response_gammajet}
\end{center}
\end{figure}
 
Two small correction factors are derived in simulation and use the \textit{true calorimeter response}, defined as the ratio of measured energy in the calorimeter deposited by particles belonging to a particle-level jet to the total energy of the particle-level jet.
The \textit{topology correction} accounts for the differences in calorimeter response for sparse energy depositions versus those in the dense cores of jets, and is found by taking the average of the ratio of $\Response_{\text{MPF}}$ to the true calorimeter response in each \pt bin.
The \textit{showering correction} accounts for the flow of particles entering or exiting across the boundaries of the jet definition and is calculated from the ratio of the true calorimeter response to the measured response of the reconstructed jet, therefore varying with the jet algorithm and size.
The total correction factor is the product of the two and is found to be less than $2\%$ for jets of $\pt < 50~\GeV$ and negligible above that. This correction factor would in principle be applied identically to $\Response_{\text{MPF}}$ in both data and simulation to better estimate jet response, but since the ratio of $\Response_{\text{MPF}}$ in data and simulation is the quantity of interest for the \insitu calibration, the correction would cancel out in the ratio and only the uncertainty in its derivation is relevant. This uncertainty is taken from a comparison of two different physics lists (FTFP BERT~\cite{PERF-2015-05} and QGSP BIC~\cite{Folger:2003sb}) in the simulation of the particle/detector interactions and is found to be $\sim2\%$ for jets with $\pt < 20~\GeV$, $\sim0.5\%$ for jets with $20~\GeV < \pt < 40~\GeV$ and zero for jets with $\pt > 40~\GeV$.
 
The full set of uncertainties is shown for the $Z \rightarrow ee + \text{jet}$ and $Z \rightarrow \mu\mu + \text{jet}$ analyses in Figure~\ref{fig:mpf_uncertainties_zjet} and for the \gammajet analysis in Figure~\ref{fig:mpf_uncert_gammajet}. The dominant systematic uncertainties are due to generator differences at lower \pT and to the photon energy scale at higher \pT. Uncertainties in the $e,~\mu,~\text{and}~\gamma$ energy scales and resolutions are taken from the calibrations provided for each physics object and are propagated through the analysis \cite{EGAM-2018-01,PERF-2015-10}. The $\Delta \phi$ and second-jet veto uncertainties are estimated by varying the cuts and comparing the resulting response measurements. As in the $\eta$ intercalibration, the JVT uncertainty is determined by comparison with tighter and looser working points. A photon purity uncertainty is estimated for the \gammajet analysis using control regions dominated by dijet events where one of the jets can be misidentified as a photon. The uncertainty on the final state modelling is taken, as discussed, from the generator comparison. Limited data and MC statistics contribute to the statistical uncertainty, which is largest for the lowest and highest bins of each analysis. A bootstrapping procedure is applied to the uncertainties to suppress statistical fluctuations as previously described.
 
\begin{figure}[htp!]
\begin{center}
\begin{subfigure}[b]{0.49\textwidth}
\includegraphics[width=\textwidth]{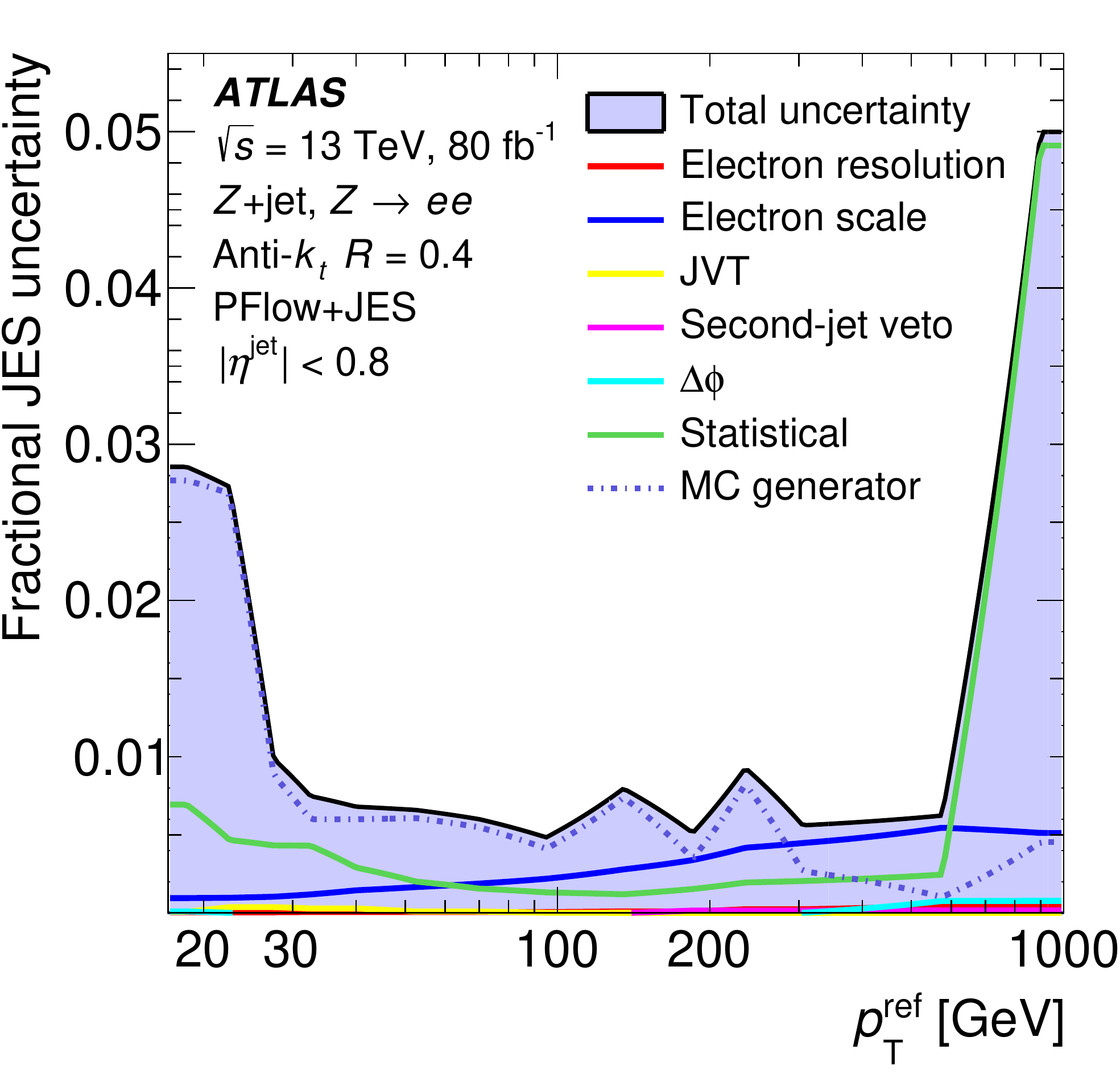}
\caption{}
\label{subfig:mpf_uncert_zjet_ee}
\end{subfigure}
\begin{subfigure}[b]{0.49\textwidth}
\includegraphics[width=\textwidth]{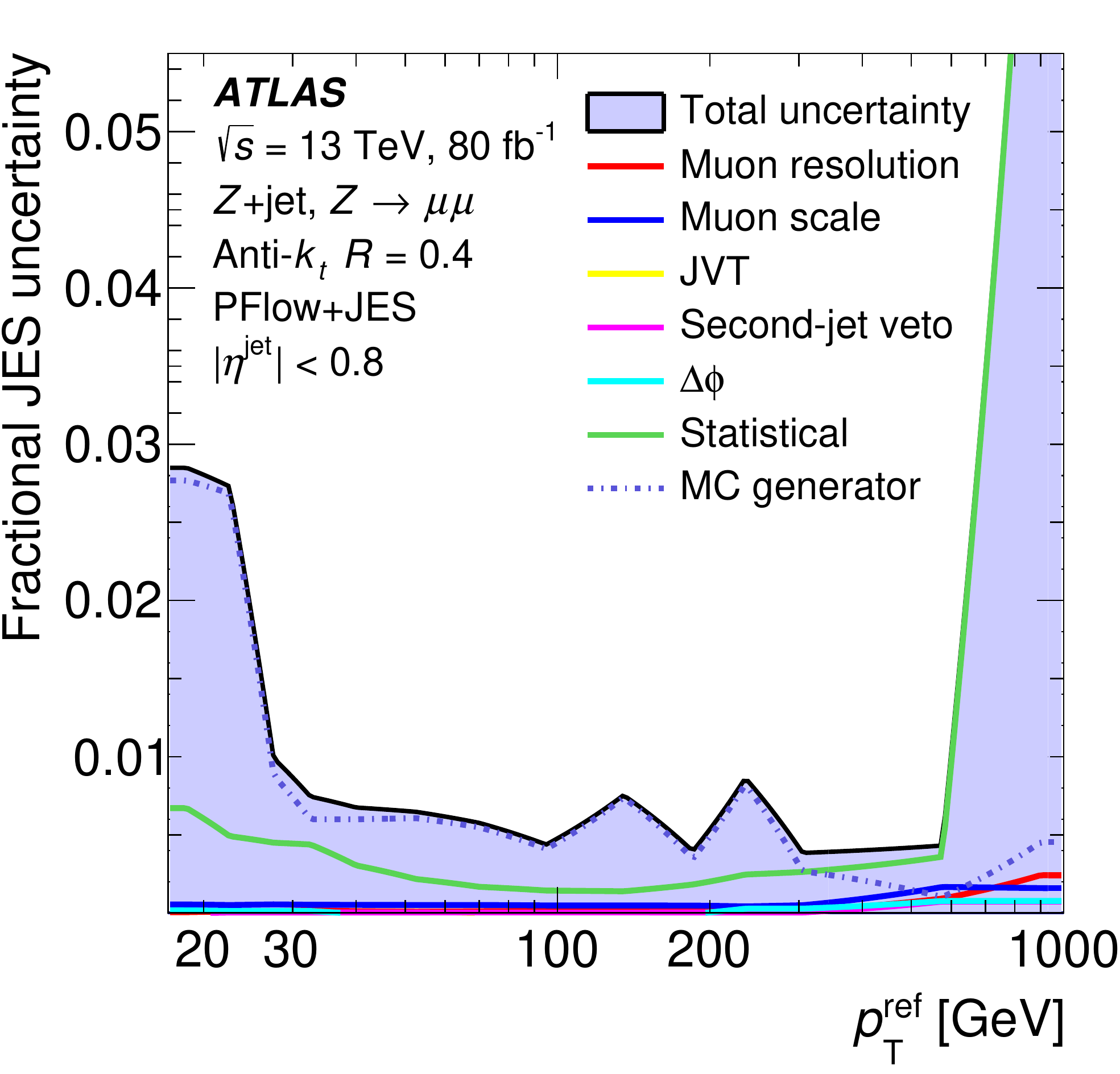}
\caption{}
\label{subfig:mpf_uncert_zjet_mumu}
\end{subfigure}
\caption{Systematic uncertainties for \pflow jets as a function of reference \pt for \subref{subfig:mpf_uncert_zjet_ee} $Z\rightarrow ee+$jet events and \subref{subfig:mpf_uncert_zjet_mumu} $Z\rightarrow \mu\mu+$jet events calculated using the MPF technique. Uncertainties due to the JVT, second-jet veto, and $\Delta \phi$ cuts derive from the analysis technique. Electron or muon (as appropriate) scale and resolution uncertainties are propagated through the analysis from the uncertainties associated with the individual objects. The statistical uncertainties come from the MC simulation/data ratio and reach a maximum value of 0.083 in \subref{subfig:mpf_uncert_zjet_mumu} while the difference between the \Pythia{}8 and \Sherpa samples defines the MC generator uncertainty. All uncertainties are smoothed to ensure that the visible fluctuations are statistically significant.}
\label{fig:mpf_uncertainties_zjet}
\end{center}
\end{figure}
 
\begin{figure}[htp!]
\begin{center}
\includegraphics[width=0.7\textwidth]{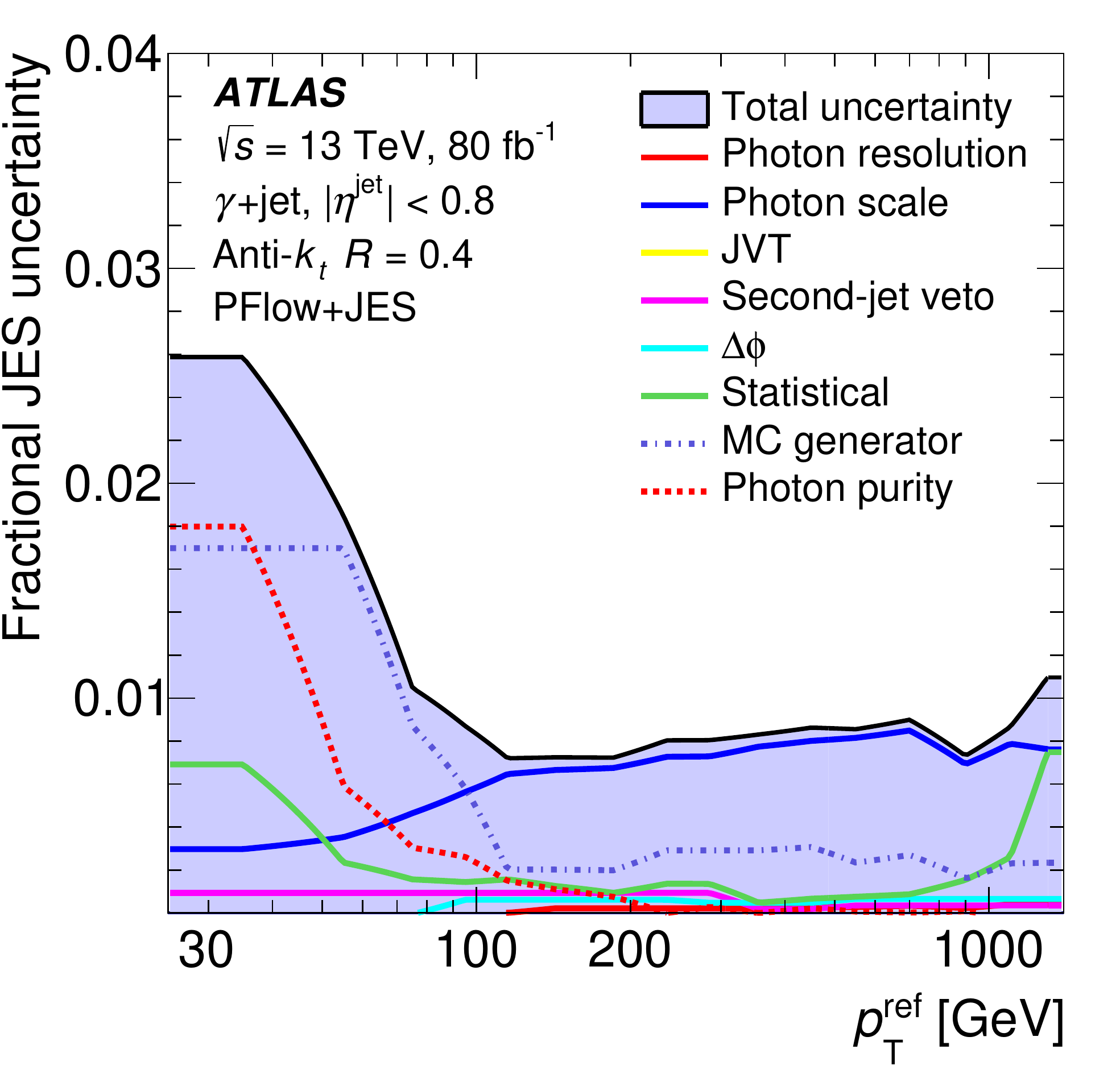}
\caption{Systematic uncertainties on \pflow jets as a function of reference \pt for \gammajet events calculated using the MPF technique. Uncertainties due to the JVT, second-jet veto, and $\Delta \phi$ cuts derive from the analysis technique. Photon scale and resolution uncertainties are propagated through the analysis from the uncertainties associated with the individual objects. The statistical uncertainties come from the MC simulation/data ratio while the difference between the \Pythia{}8 and \Sherpa samples defines the MC generator uncertainty. All uncertainties are smoothed to ensure that the visible fluctuations are statistically significant.}
\label{fig:mpf_uncert_gammajet}
\end{center}
\end{figure}
 
Similar analyses in the \vjet final states but explicitly balancing the reference \pt against the \pt of a reconstructed jet (\textit{direct balance}) are used to cross-check the jet energy scale calibration. The JES results computed using direct balance showed good agreement with those derived via MPF.
 
The innate difference in response between \EMTopo and \pflow jets can be seen by comparing their measured MPF responses. Since the MPF method uses topo-clusters and \pflow objects in computing the missing energy, the measured responses are independent of the MCJES calibration and reflect the precalibration response for each jet input type. The MPF responses measured in the \gammajet analysis for \EMTopo and \pflow jets are shown in Figure~\ref{fig:gammajet_emtopo_pflow}. The shape of the \EMTopo measurement follows the form of the Groom's function, which corresponds to the response expected from a hadronic calorimeter~\cite{GROOM2007633}. The \pflow measurement does not follow the same shape but instead shows an improvement over the baseline calorimeter response at low \pT thanks to the inclusion of information from tracks.
 
\begin{figure}[htp!]
\begin{center}
\includegraphics[width=0.7\textwidth]{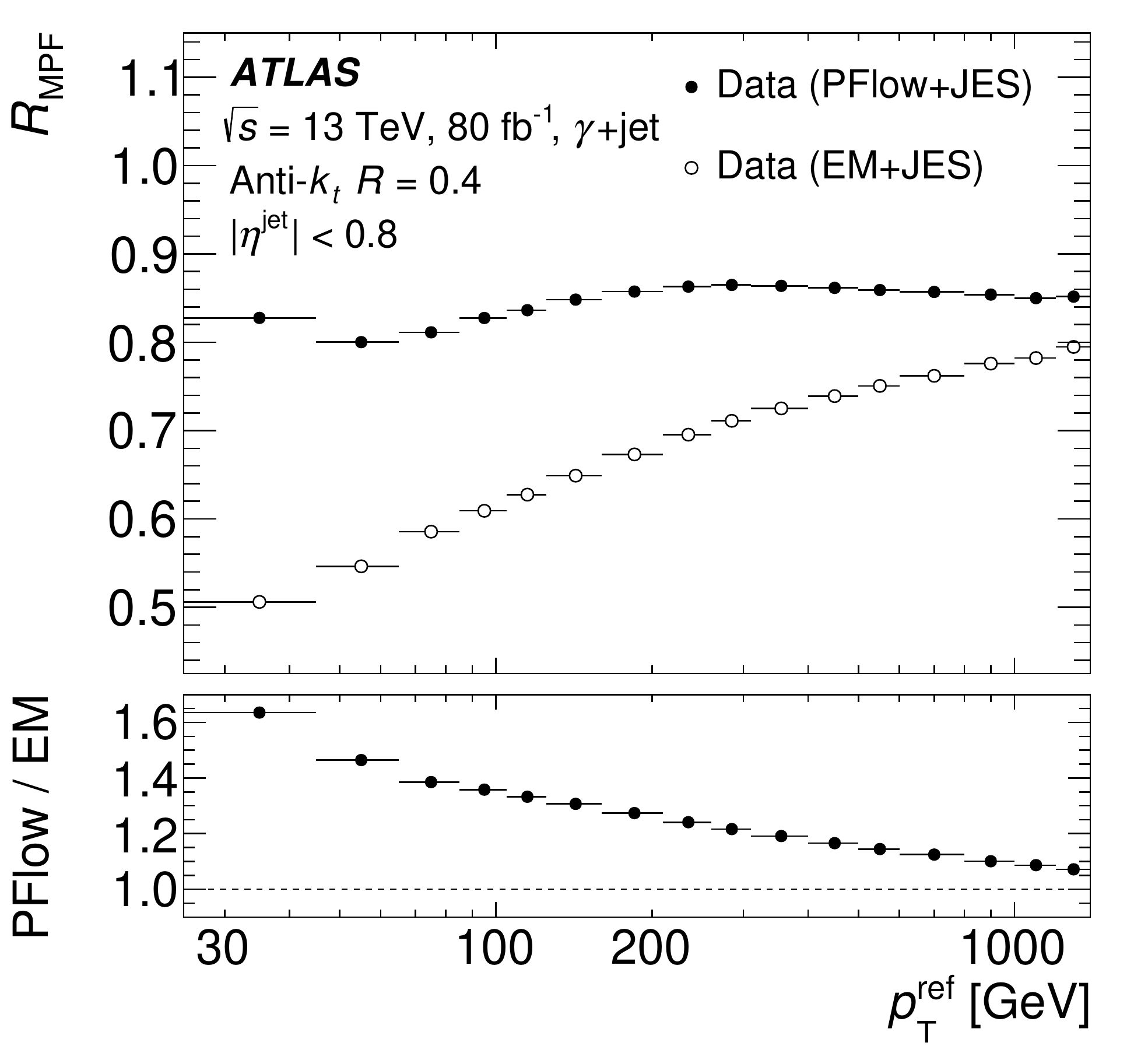}
\caption{Average jet response as a function of reference \pt for \gammajet events calculated using the MPF technique in 2015--2017 data. The solid points correspond to \pflow jets while the hollow points correspond to \EMTopo jets. The ratio of \pflow response to \EMTopo response is shown in the bottom panel.}
\label{fig:gammajet_emtopo_pflow}
\end{center}
\end{figure}
 
\FloatBarrier
 
\subsubsection{High-\pt jet calibration using multijet balance}
\label{sec:jes-multijets}
The final stage of \textit{in~situ} calibration derives a correction for jets with \pt above the range of the \vjet analyses using the multijet balance (MJB) technique. Events are selected with a single high-\pt jet balanced against a system of lower-\pt jets (the \textit{recoil system}). The jets of the recoil system are selected to ensure they are well calibrated using a combination of the \vjet results (Section~\ref{sec:jes-vjet}), while the leading jet is left at the scale of the $\eta$ intercalibration.
The response of the system is defined as:
\begin{equation}\nonumber
\mathcal{R}_{\text{MJB}} = \left\langle \frac{\pt^{\text{lead}}}{\pt^{\text{ref}}} \right\rangle\,,
\end{equation}
where $\pt^{\text{ref}}$ is taken from the vector sum of all jets in the recoil system. In a procedure parallel to that used for the \vjet analyses, the response is measured in bins of $\pt^{\text{ref}}$ and the correction is then mapped to the uncalibrated leading jet by finding the average $\pt^{\text{lead}}$ of the events in each bin.
 
Since the MJB analysis can only include events where all jets of the recoil system can already be well-calibrated, events with very high $\pt^{\text{lead}}$ are often excluded as their second and third leading jets can have momenta outside the range of calibration by the \vjet analyses. To address this, MJB proceeds via two iterations. In the first iteration, a combination of the \vjet results is used to calibrate the recoil system, so only events with subleading jets of $\pt < 1.2~\TeV$ are included. In the second iteration, events with subleading jets up to $\pt = 1.8~\TeV$ are included and calibrated using the MJB results from the first iteration. This extends the range of the calibration to $\pt^{\text{lead}}=2.4~\TeV$.
 
Events are selected for the MJB analysis using a variety of single-jet triggers with each corresponding to a unique range of $\pt^{\text{lead}}$. To suppress dijet topologies and ensure that only true multijet events are used, events must have at least three jets with $\pt > 25~\GeV$ and $|\eta| < 2.8$ and the subleading jet must not have a momentum above $0.8\pt^{\text{lead}}$. Jets are as usual required to pass JVT selections, limiting the effects of \pileup. Isolation of the leading jet from contamination by the recoil system is ensured by requiring that the azimuthal angle $\Delta \phi$ between the leading jet and the direction of the recoil system is at least 0.3~radians and that the $\Delta \phi$ between the leading jet and any individual jet in the recoil system with a $\pt > 0.05 \pt^{\text{lead}}$ is at least 1.0~radians.
 
The MJB response in data and in four MC samples with different generators is shown in Figure~\ref{subfig:mjb_response}. In both data and MC simulation, the response decreases at lower~\pt due to the intrinsic bias in $\mathcal{R}_{\text{MJB}}$ from the combined effects of the leading jet isolation and \pt asymmetry requirements. This bias is greater for lower-\pt leading jets, but is well modelled in simulation, leaving the calibration unbiased. The lower panel shows the ratio of the response of each MC sample to data. Here, the ratio of the \textsc{Sherpa} sample to data defines the nominal correction while the ratio based on \Pythia defines an uncertainty on the generator choice. This response ratio is constant and approximately 2\% for jets above 1~\TeV; below this point the calculated correction is slightly smaller.
 
\begin{figure}[htp!]
\begin{center}
\begin{subfigure}[b]{0.49\textwidth}
\includegraphics[width=\textwidth]{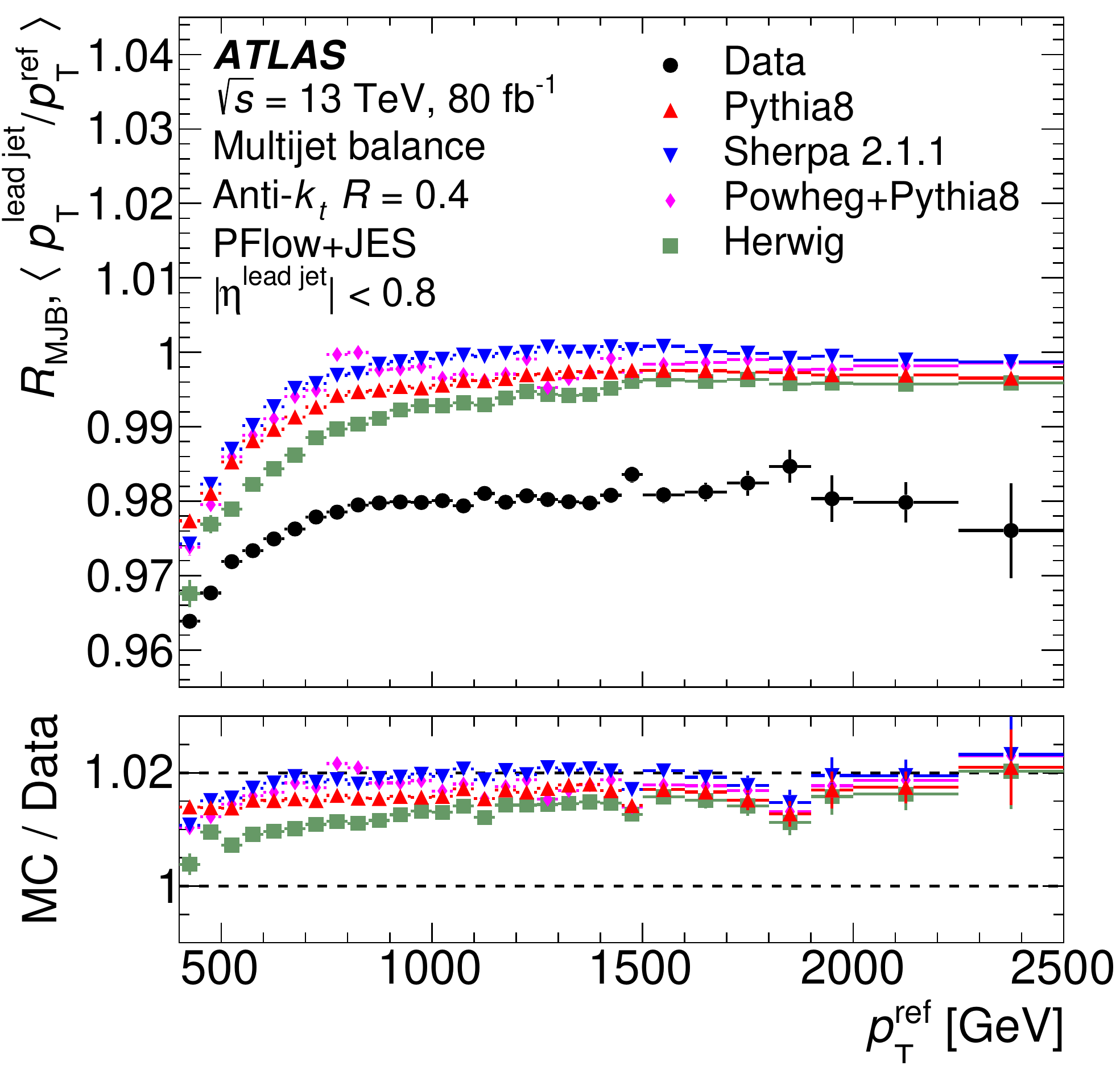}
\caption{Response}
\label{subfig:mjb_response}
\end{subfigure}
\begin{subfigure}[b]{0.49\textwidth}
\includegraphics[width=\textwidth]{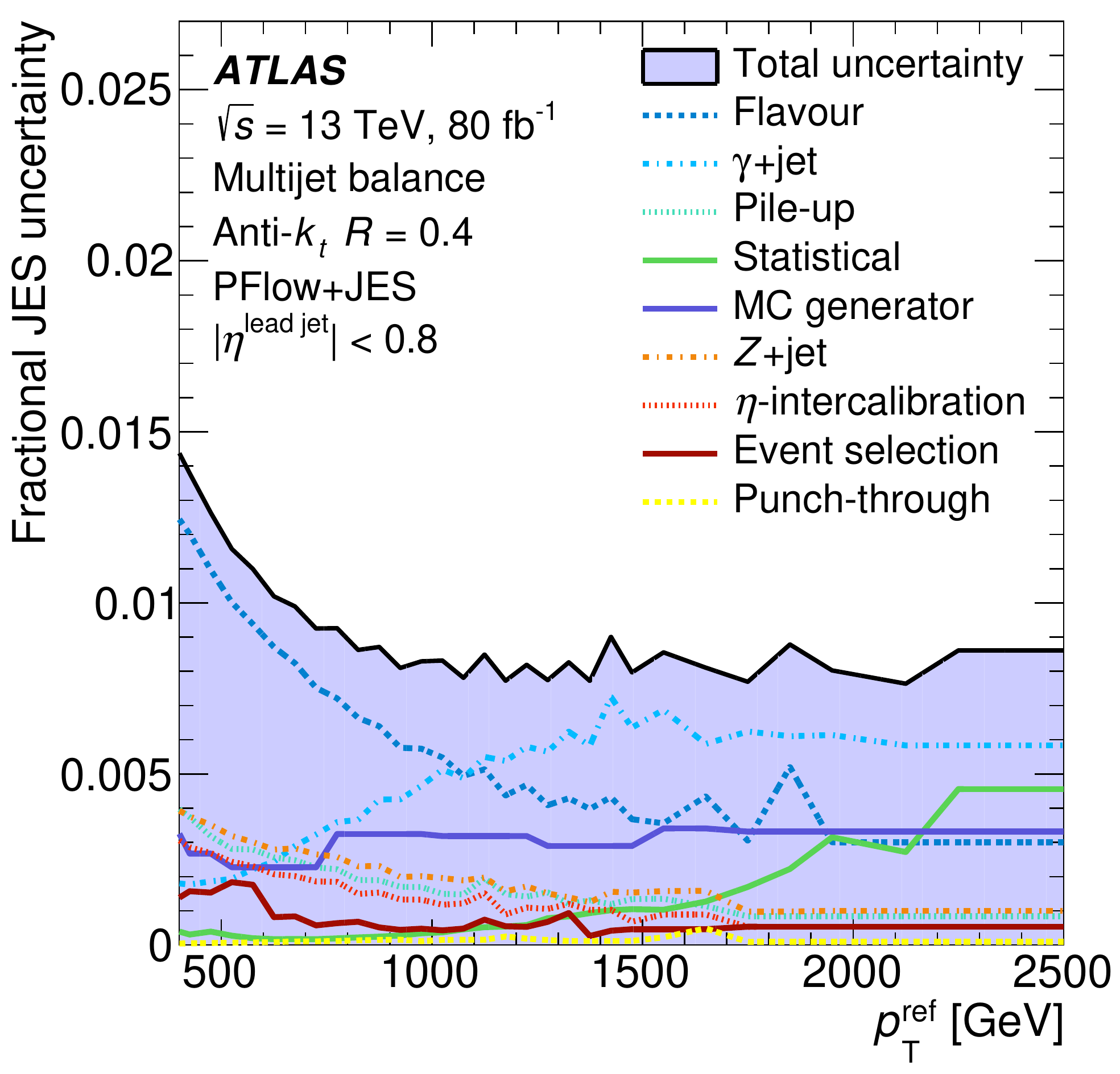}
\caption{Uncertainty}
\label{subfig:mjb_uncertainty}
\end{subfigure}
\caption{\subref{subfig:mjb_response} Response for the leading \pflow jet in multijet events as a function of \ptref and \subref{subfig:mjb_uncertainty} the systematic uncertainties on the response. Subleading jets in the event are calibrated using the \vjet MPF corrections, while the leading jet is calibrated only up to the $\eta$ intercalibration. The response is shown for data and for simulation using four different MC generators, and the MC simulation-to-data response ratios in the bottom panel correspond to the derived $in~situ$ calibration. The error bars show the statistical uncertainties. The nominal calibration is defined by the comparison with \textsc{Sherpa}; its difference from the \Pythia result defines the `MC generator' uncertainty in~\subref{subfig:mjb_uncertainty}. This uncertainty is defined in a single-sided way by the measured response difference and therefore it is not symmetrised for display in~\subref{subfig:mjb_uncertainty} but instead its full one-sided value is shown. Other uncertainties come from the event selection and MC simulation/data statistics or are propagated from the \zjet, \gammajet, flavour, \pileup, $\eta$ intercalibration, and punch-through studies.}
\label{fig:mjb}
\end{center}
\end{figure}
 
All uncertainties in the MJB analysis are shown in Figure~\ref{subfig:mjb_uncertainty}. The dominant term at low $\pt^{\text{lead}}$ is the uncertainty from jet flavour, derived in simulation and reflecting the difference in jet response for quark-initiated and gluon-initiated jets. Two terms contribute, one reflecting the uncertainty in the fraction of gluon-initiated jets in the sample, the other based on the difference in MC simulation-derived gluon response between generators. Other independently derived uncertainties correspond to \pileup and punch-through effects and are propagated through the MJB analysis via the recoil system.  The \zjet, \gammajet, and $\eta$ intercalibration uncertainties are propagated from the previous stages of \insitu analysis. Event selection uncertainties are determined by varying each of the analysis cuts and determining the effects on the measured response ratio. Finally, the MC generator uncertainty is derived as described above by comparing the response ratio of \textsc{Sherpa} with \Pythia as an alternative. Results using \herwig and \POWPYTHIA{}8 are shown for reference but are not included in the uncertainty definition as they are less reliable for this measurement.  All uncertainties are smoothed via the bootstrapping procedure to ensure statistical significance, and the total uncertainty is found to be below $1.5\%$ for all considered values of $\pt^{\text{lead}}$. The MC generator uncertainty, which is defined in a one-sided fashion from the response ratios, is symmetrised by the \insitu combination process along with the other uncertainties. However, its full one-sided size is shown in~\Fig{\ref{subfig:mjb_uncertainty}} for easier comparison with~\Fig{\ref{subfig:mjb_response}}.
 
For \EMTopo jets the intrinsic bias at low \pt is slightly smaller and more closely tracked by simulation, leading in turn to slightly reduced systematic uncertainties for jets below $\pt \sim 700~\GeV$. Above $\pt > 1~\TeV$, \insitu uncertainties propagated from lower-\pt jets have a greater impact, and therefore the uncertainty is smaller for \pflow jets than for \EMTopo jets.
 
\FloatBarrier
 
\subsubsection{\Pileup and the \textit{\insitu} analyses}
\label{subsubsec:pileup-insitu}
 
One of the primary changes in LHC run conditions over the course of \RunTwo was an increase in \pileup. The average number of interactions per crossing ($\mu$) during 2015+2016 data taking was 23.7, which increased to 37.8 in 2017. The data taken during 2018 and to which the calibrations in this paper are also applied has an average of 36.1 interactions per crossing~\cite{ATLAS-CONF-2019-021}. The consistency of the calibrations for events with different \pileup conditions is therefore an important feature of the methods.
 
Figure~\ref{fig:vjet_stability} shows individual bins in the response ratios of the \zjet and \gammajet analyses separated out as a function of number of primary vertices in the event. The \zjet results are shown for $25~\GeV < \ptref < 30~\GeV$ and the \gammajet results for $45~\GeV < \ptref < 65~\GeV$, in the regions where each has appropriate statistical significance. The multijet balance analysis is not shown: due to the higher \pt regime in which it operates it is more robust to \pileup effects.  A linear fit to the data/simulation ratio has a slope compatible with zero within the fit uncertainties in each plot, demonstrating the stability of the \insitu calibration as a function of \Npv. This in turn illustrates the efficacy of the \pileup corrections described in Section~\ref{sec:jes-pileupcorr} and shows that the inclusively derived calibration is applicable to events with a range of \pileup conditions.
 
\begin{figure}[htp!]
\begin{center}
\begin{subfigure}[b]{0.49\textwidth}
\includegraphics[width=\textwidth]{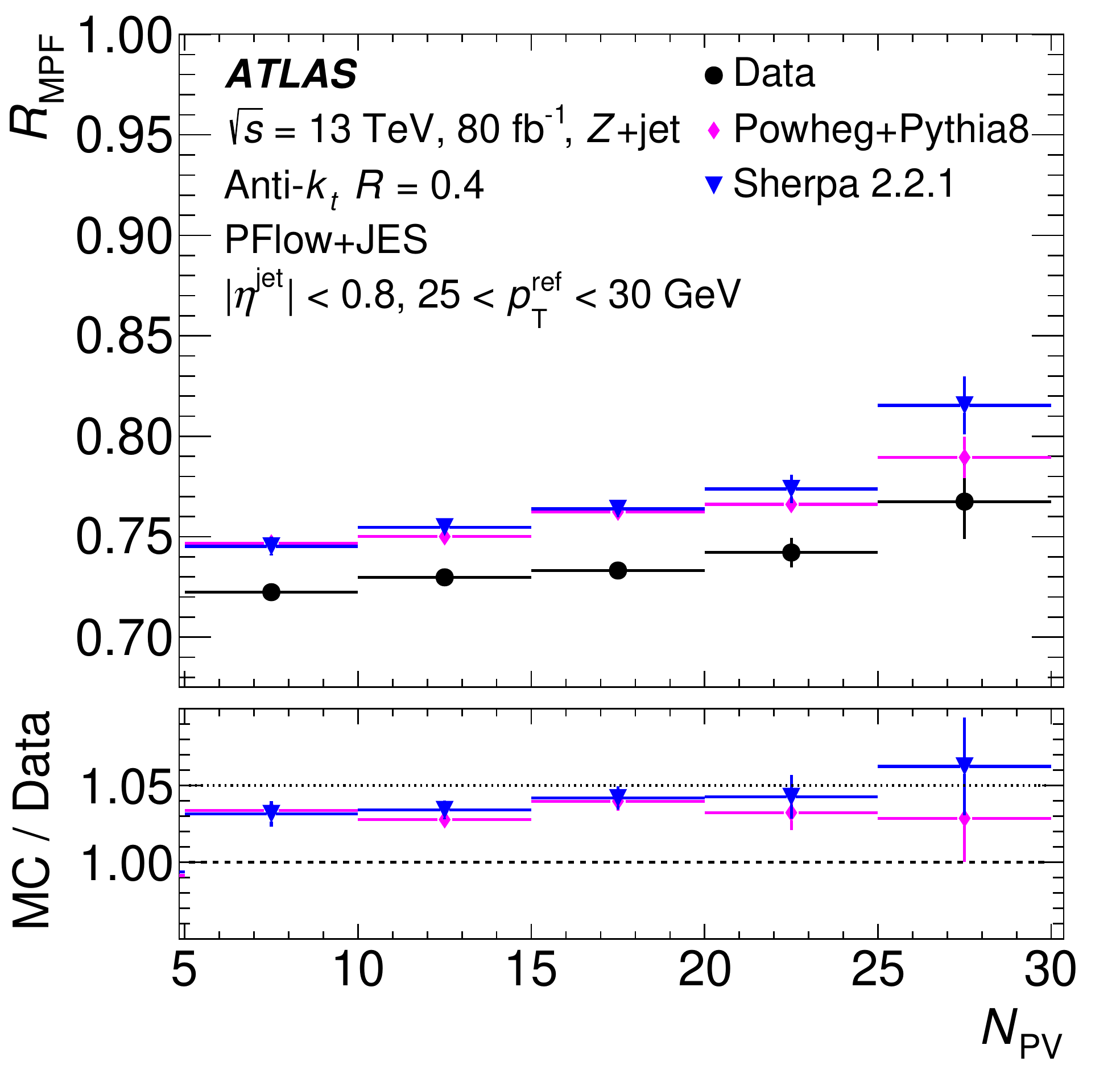}
\caption{}
\label{subfig:zjet_stability}
\end{subfigure}
\begin{subfigure}[b]{0.49\textwidth}
\includegraphics[width=\textwidth]{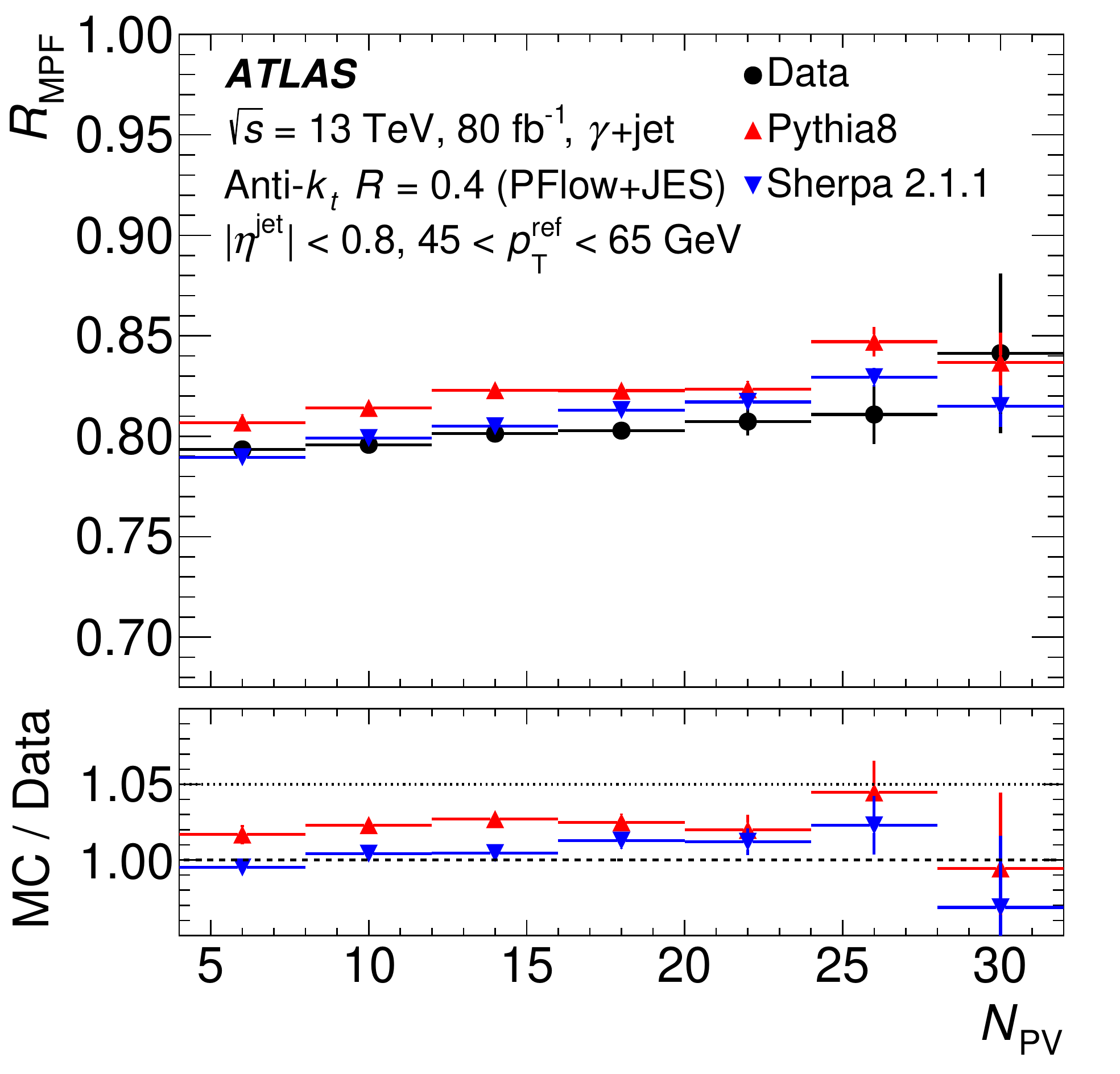}
\caption{}
\label{subfig:gammajet_stability}
\end{subfigure}
 
\caption{Average \pflow MPF jet response as a function of $\Npv$ for \subref{subfig:zjet_stability} \zjet events with reference \pt derived from the reconstructed $Z$ boson in the range $25~\GeV < \ptref < 30~\GeV$ and for \subref{subfig:gammajet_stability} \gammajet events with reference \pt defined from the photon in the range $45~\GeV < \ptref < 65~\GeV$.  For the \zjet analysis, results from the $Z\rightarrow ee$ and $Z\rightarrow \mu\mu$ channels are combined to reduce statistical fluctuations. The black points correspond to 2015--2017 data while the pink and blue points correspond to Monte Carlo samples from two different generators. The error bars reflect the statistical uncertainties. The ratio of MC simulation to data for both generators is shown in the bottom panel.}
\label{fig:vjet_stability}
\end{center}
\end{figure}

The \insitu JES measurements can be used to calculate the dependence of the measured median \pt density $\rho$ on the event topology in simulation and data and to derive an uncertainty, as mentioned in Section~\ref{sec:jes-pileupcorr}.  The density $\rho$ is computed as a function of $\mu$ for each of the \zjet, \gammajet, and dijet topologies as shown in the top panels of \Fig{\ref{fig:rhotop}}.  Taking $\rho_{2017}$ as the value of $\rho$ for the average \pileup conditions during 2017 data taking and  t1 and t2 as any two \insitu measurement topologies out of \zjet, \gammajet, and dijet, then the following metric of consistency can be defined:
\begin{equation}\nonumber
\Delta = \left(\rho_{2017}^{\text{t1}} - \rho_{2017}^{\text{t2}} \right)_{\text{MC}} - \left(\rho_{2017}^{\text{t1}} - \rho_{2017}^{\text{t2}} \right)_{\text{data}}\,.
\end{equation}
The quantity $\max(|\Delta|)$ is then the largest value of $\Delta$ across the various topology comparisons. The total $\rho$ topology systematic uncertainty is given by
\begin{equation}\nonumber
\Delta_{\pt} = \max(|\Delta|) \times C^{\text{JES}}_{\pt} \times \pi R^2\,,
\end{equation}
where $C^{\text{JES}}$ is the size of the MCJES correction for a jet with the relevant \pt. The second panels in \Fig{\ref{fig:rhotop}} show $\rho_{2017}^{\text{t1}} - \rho_{2017}^{\text{t2}}$ for the comparisons (\zjet, dijet) and (\gammajet, dijet) in both MC simulation and data. The lower panels show the difference of these two quantities between data and MC simulation, that is, $\Delta_{\text{\zjet, dijet}}$ and $\Delta_{\text{\gammajet, dijet}}$. The input to the systematic uncertainty $\max(|\Delta|)$ is the most discrepant of the two lines in the lower panel evaluated at $\mu=37.8$, the value in 2017 data. As \Fig{\ref{fig:rhotop}} illustrates, this uncertainty is larger for \pflow jets than for \EMTopo jets. This is understood to be due to a greater sensitivity to the underlying event when tracking information is included, which leads to greater differences among the simulated samples.
 
\begin{figure}[htp!]
\begin{center}
\begin{subfigure}[b]{0.49\textwidth}
\includegraphics[width=\textwidth]{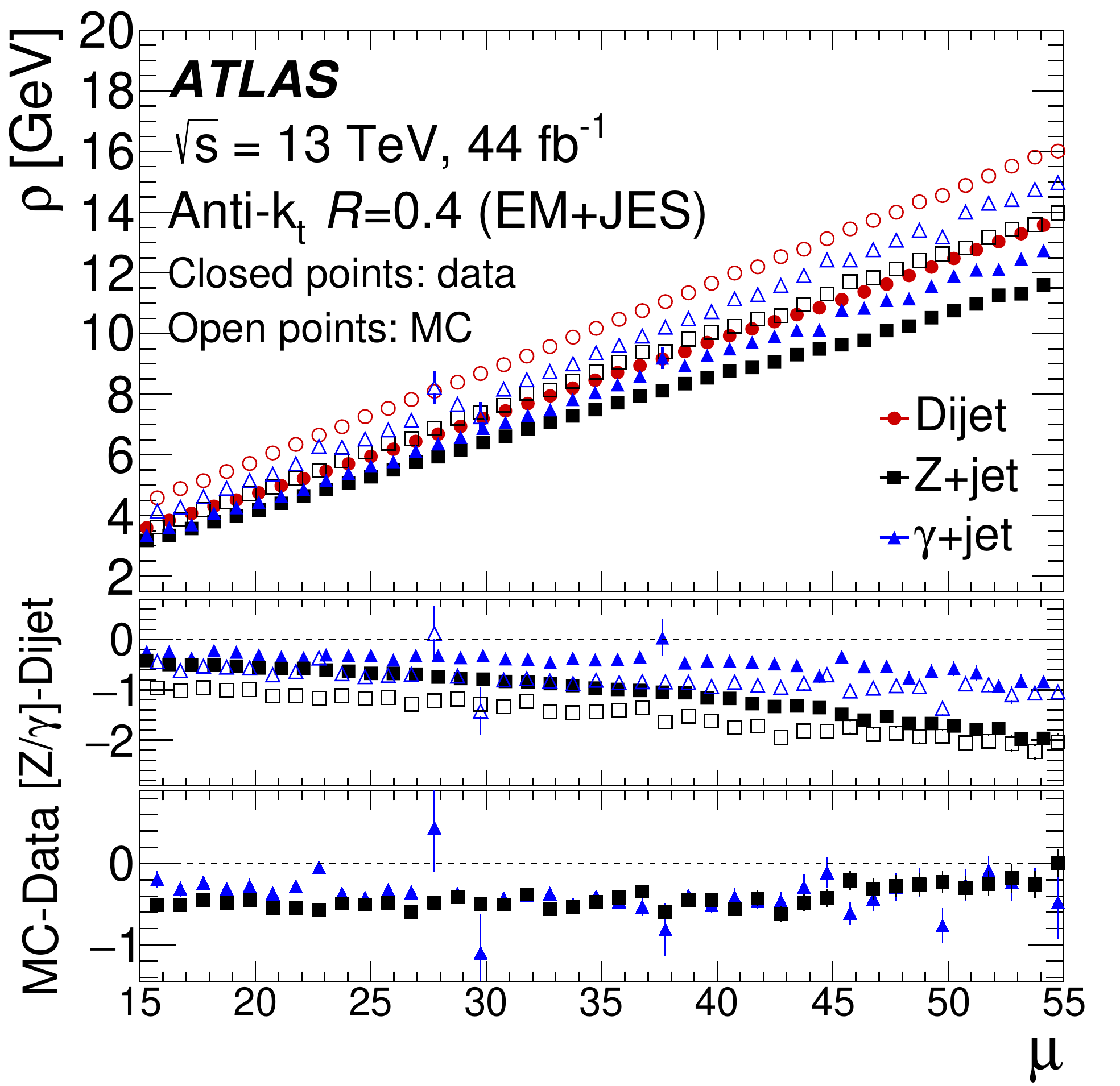}
\caption{}
\label{subfig:rhotop_EM}
\end{subfigure}
\begin{subfigure}[b]{0.49\textwidth}
\includegraphics[width=\textwidth]{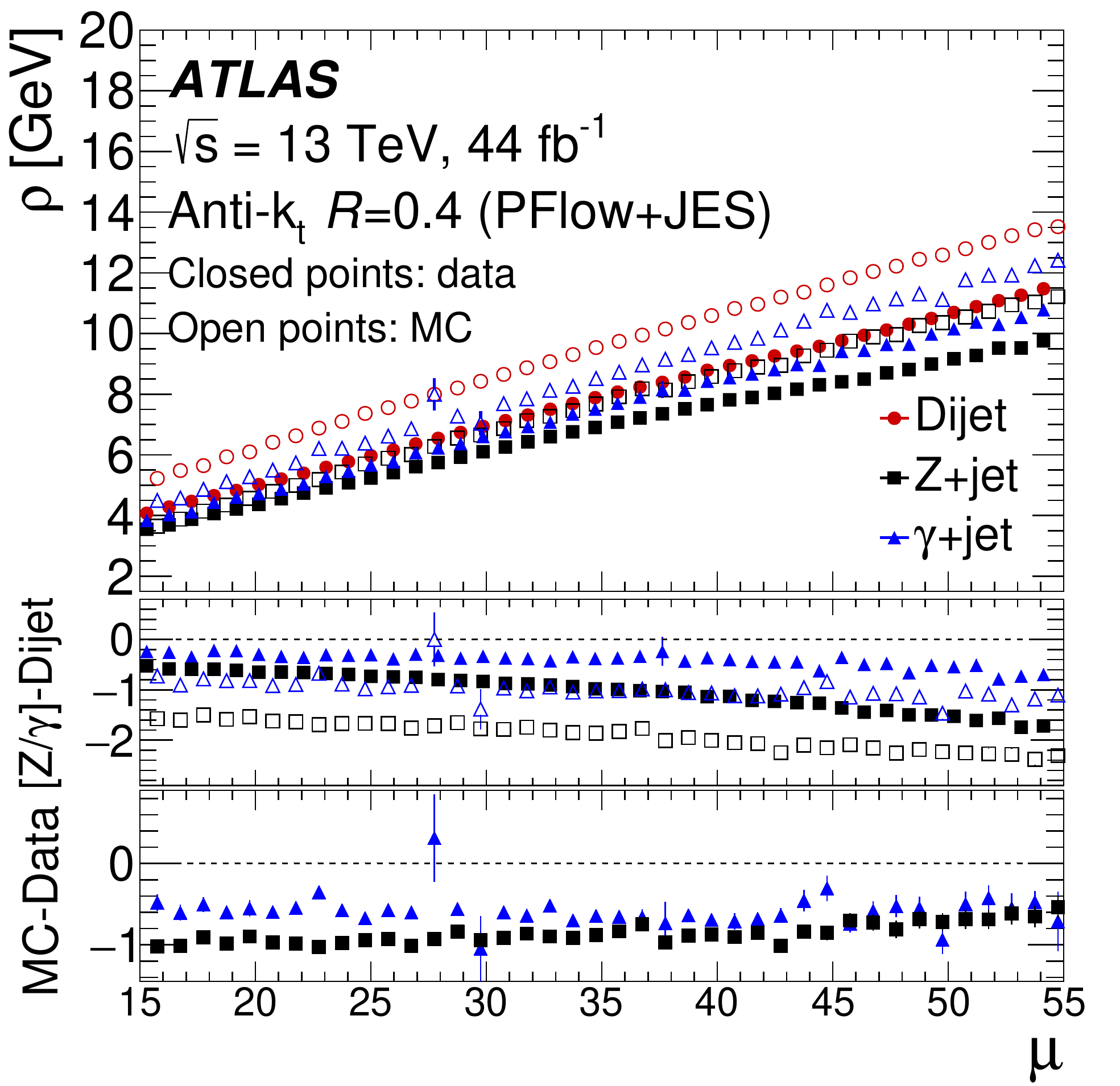}
\caption{}
\label{subfig:rhotop_PF}
\end{subfigure}
\caption{Inputs to the $\rho$ topology uncertainty derived in the \zjet, \gammajet, and dijet \insitu analyses. The error bars show the statistical uncertainties. The top panels relate the \pt density $\rho$ to the mean number of interactions per bunch crossing $\mu$ in data and MC simulation for the three input analyses. The second panels show the difference between the \zjet and dijet and between the \gammajet and dijet measurements. The lowermost panels show the difference between the data and MC simulation lines in the second panels: this defines the size of the topology uncertainty. The two plots show \subref{subfig:rhotop_EM} \EMTopo and \subref{subfig:rhotop_PF} \pflow jets, illustrating why this uncertainty is larger for \pflow jets than for \EMTopo jets.}
\label{fig:rhotop}
\end{center}
\end{figure}
 
\FloatBarrier
 
\subsubsection{\Insitu combination}
\label{sec:jes-combination}
The data/MC simulation response ratios,
\begin{equation}\nonumber
\left\langle \pTjet/\pTref \right\rangle_{\text{data}} \Big/ \left\langle \pTjet/\pTref\right \rangle_{\text{MC}}\,,
\end{equation}
from the four different `absolute' \insitu measurements of $Z(\rightarrow ee)$+jet, $Z(\rightarrow \mu\mu)$+jet, \gammajet, and the multijet balance must be combined to produce a single calibration covering the full range of jet \pt from $17~\GeV$ to $2.4~\TeV$. The four measurements overlap one another in various \pt ranges, so this procedure must account for their relative statistical power as well as the tension between different response ratio measurements in the same $\pt$ range. The $Z(\rightarrow ee)$+jet and $Z(\rightarrow \mu\mu)$+jet channels, though compatible within uncertainties, are treated as separate measurements for the sake of the combination since they are affected by different systematic uncertainties.
 
The combination procedure is briefly summarized here; for a detailed description see Ref.~\cite{PERF-2012-01}. Each of the absolute \insitu measurements is converted from a parameterisation in terms of reference object \pt into jet \pt and divided into finer bins of 1~\GeV{} using second-order polynomial splines. A $\chi^2$ minimization is performed in each bin, taking as inputs the measurements available in that \pt range and their uncertainties. This minimisation functions as a weighted average, with the weight given to each input measurement decreasing as its uncertainty grows. In this way, the measurement with the smallest statistical and systematic uncertainties dominates the estimate of the response ratio in that bin.
 
The weights of each input measurement in this combination are shown in Figure~\ref{subfig:weights_comb} as a function of jet \pt. The \zjet measurements dominate for jet \pT below $\sim500$~\GeV\ where the statistical uncertainties on these measurements grow dramatically; the $Z(\rightarrow \mu\mu)$+jet is the more powerful of the two in the upper half of this range due to the size of the electron scale and resolution uncertainties affecting the $Z(\rightarrow ee)+$jet channel. The combination is then dominated by \gammajet until jet \pT of above 1~\TeV, where the lower statistics in this channel and the decreased flavour uncertainties in the multijet balance analysis allow the latter to dominate. The final calibration curve is determined by smoothing the outputs from the minimization with a Gaussian kernel.
 
\begin{figure}[htp!]
\begin{center}
\begin{subfigure}[b]{0.49\textwidth}
\includegraphics[width=\textwidth]{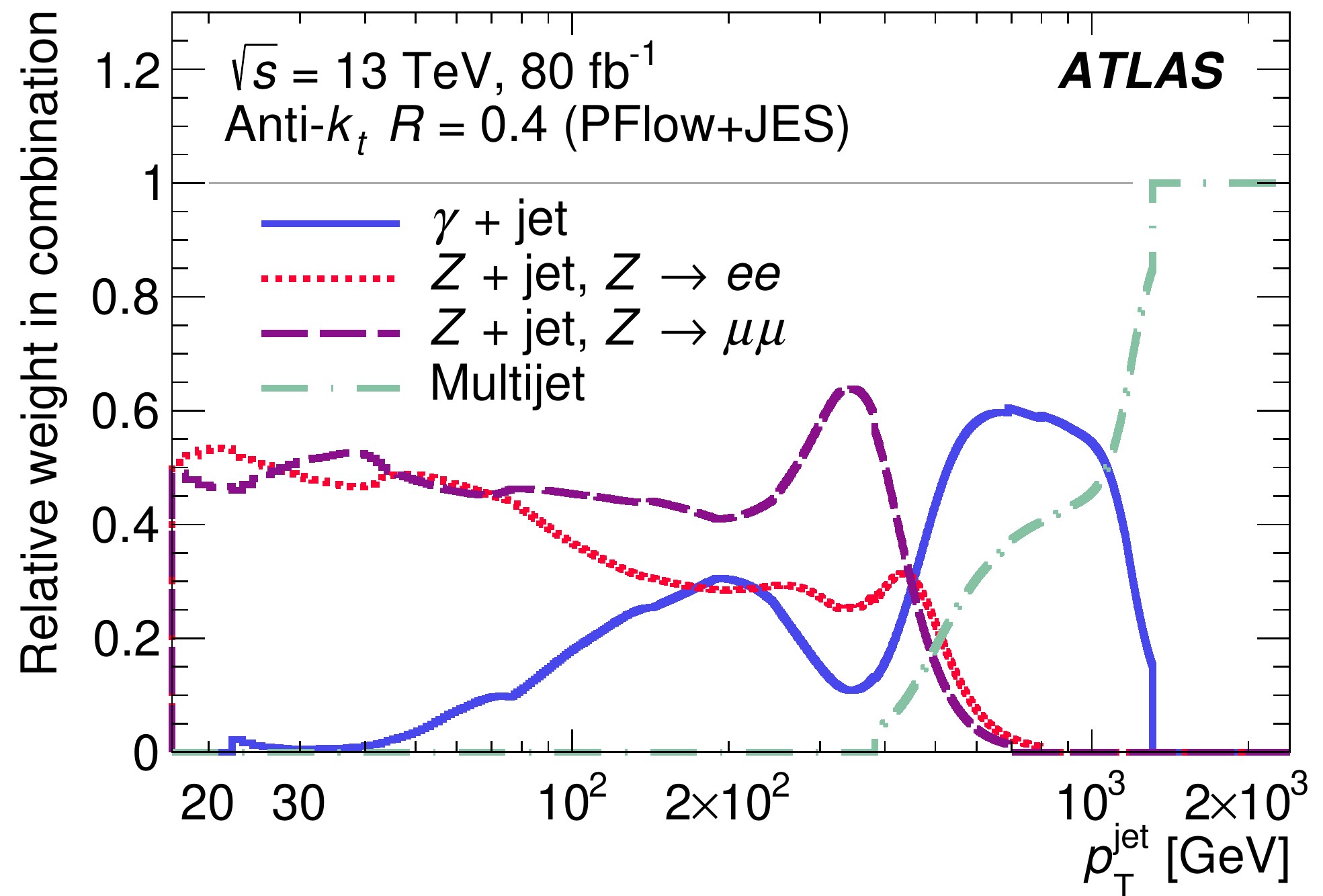}
\caption{}
\label{subfig:weights_comb}
\end{subfigure}
\begin{subfigure}[b]{0.49\textwidth}
\includegraphics[width=\textwidth]{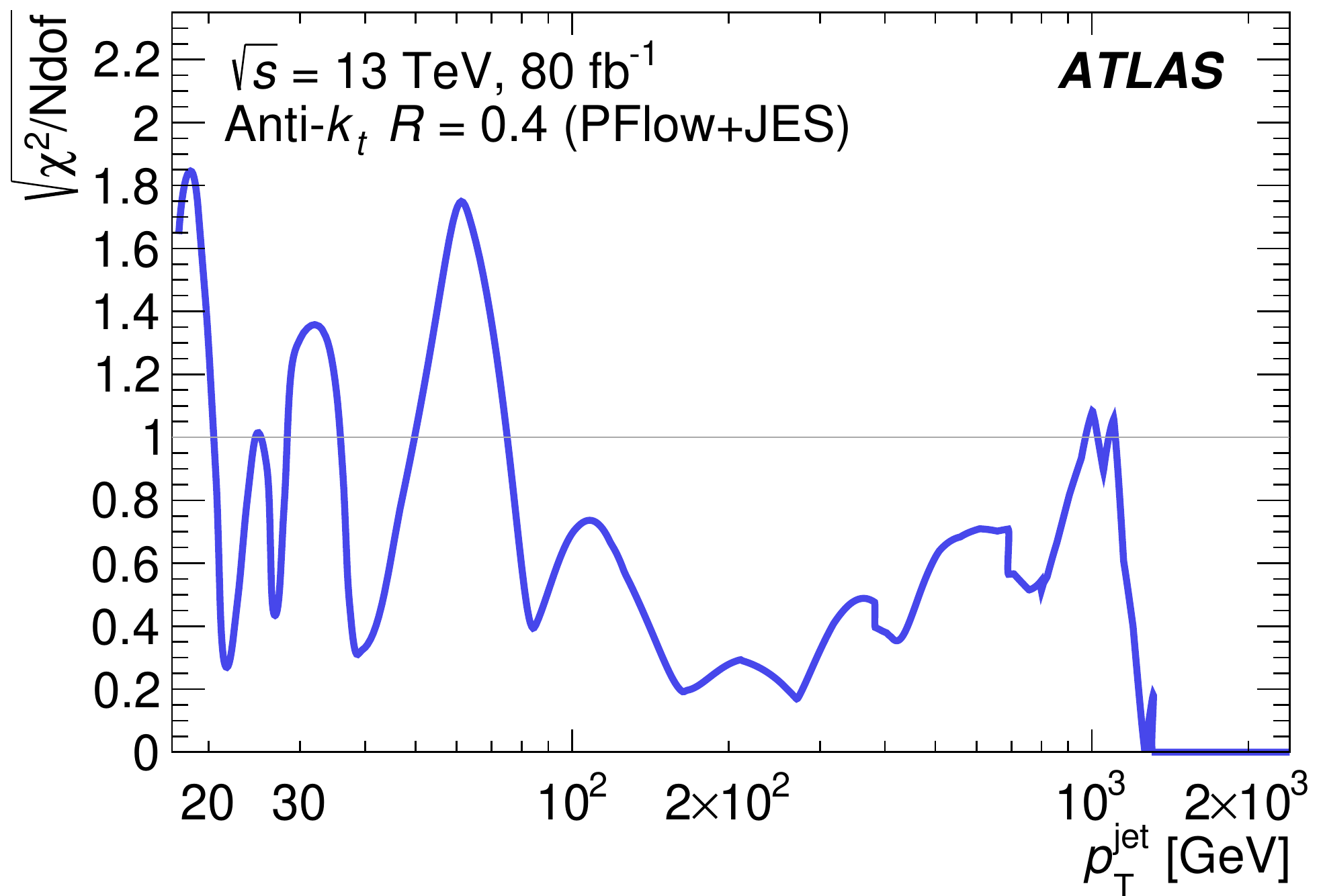}
\caption{}
\label{subfig:chi2_comb}
\end{subfigure}
\caption{\subref{subfig:weights_comb} The weight assigned to different techniques in the combination of \insitu measurements of the relative \pT response of \akt $R=0.4$ particle-flow jets in data and simulation, as a function of the jet \pT. For each \pT bin, the weights of the \zjet, \gammajet, and multijet balance methods are shown. \subref{subfig:chi2_comb} The $\chi^2/N_{\text{dof}}$ metric, illustrating the compatibility of the \insitu measurements being combined, as a function of jet \pT.  In the low \pT range, the combination is between three measurements ($Z(\rightarrow ee)+$jet, $Z(\rightarrow\mu\mu)+$jet, and \gammajet) of which the two \zjet measurements have several correlated uncertainties, resulting in increased tension compared to previous calibrations.}
\label{fig:weights_combination}
\end{center}
\end{figure}
 
The $\sqrt{\chi^{2}/N_{\text{dof}}}$ across the measurements, before any scaling is applied, is shown in Figure~\ref{subfig:chi2_comb}. This metric shows the degree of tension between the input measurements at each point: when they are in agreement well within uncertainties the value will be below 1, while when they differ relative to their uncertainties it will be above 1. Following PDG guidelines, in bins where tension between the input measurements, quantified by $\sqrt{\chi^{2}/N_{\text{dof}}}$, is found to be greater than 1, the uncertainties in the measurements in that bin are scaled by the same tension factor to ensure that the overall level of agreement between methods is acceptable within uncertainties for all $\pt$ values~\cite{PDG}.  However, since the tensions visible at low~\pT are primarily between the two \zjet measurements, and since the MC generator and showering and topology uncertainties are fully correlated between the two channels and therefore cannot contribute to this tension, these two components are excluded from the scaling procedure. The components which are not scaled are the dominant uncertainties.
 
\Fig{\ref{fig:jes_combination}} shows the final \insitu combination as a function of jet \pt. To complete the calibration, the inverse of the curve ($R_{\text{MC}}/R_{\text{data}}$) is taken as the scaling factor and applied to data. The combined measurement (solid line) for \EMPFlowJES jets is compared with each of the four absolute \insitu analyses (empty shapes) in \Fig{\ref{subfig:jes_combination_plain}}. The total size of the correction is approximately 3\% at low \pt and decreases to around 2\% for jets above 200~\GeV. A comparison between the results for \EMJES and \EMPFlowJES jets is shown in \Fig{\ref{subfig:jes_combination_comparison}}, where the overall size of both the correction and its uncertainty is seen to be slightly larger for \EMJES jets.
 
\begin{figure}[htp!]
\begin{center}
\begin{subfigure}[b]{0.49\textwidth}
\includegraphics[width=\textwidth]{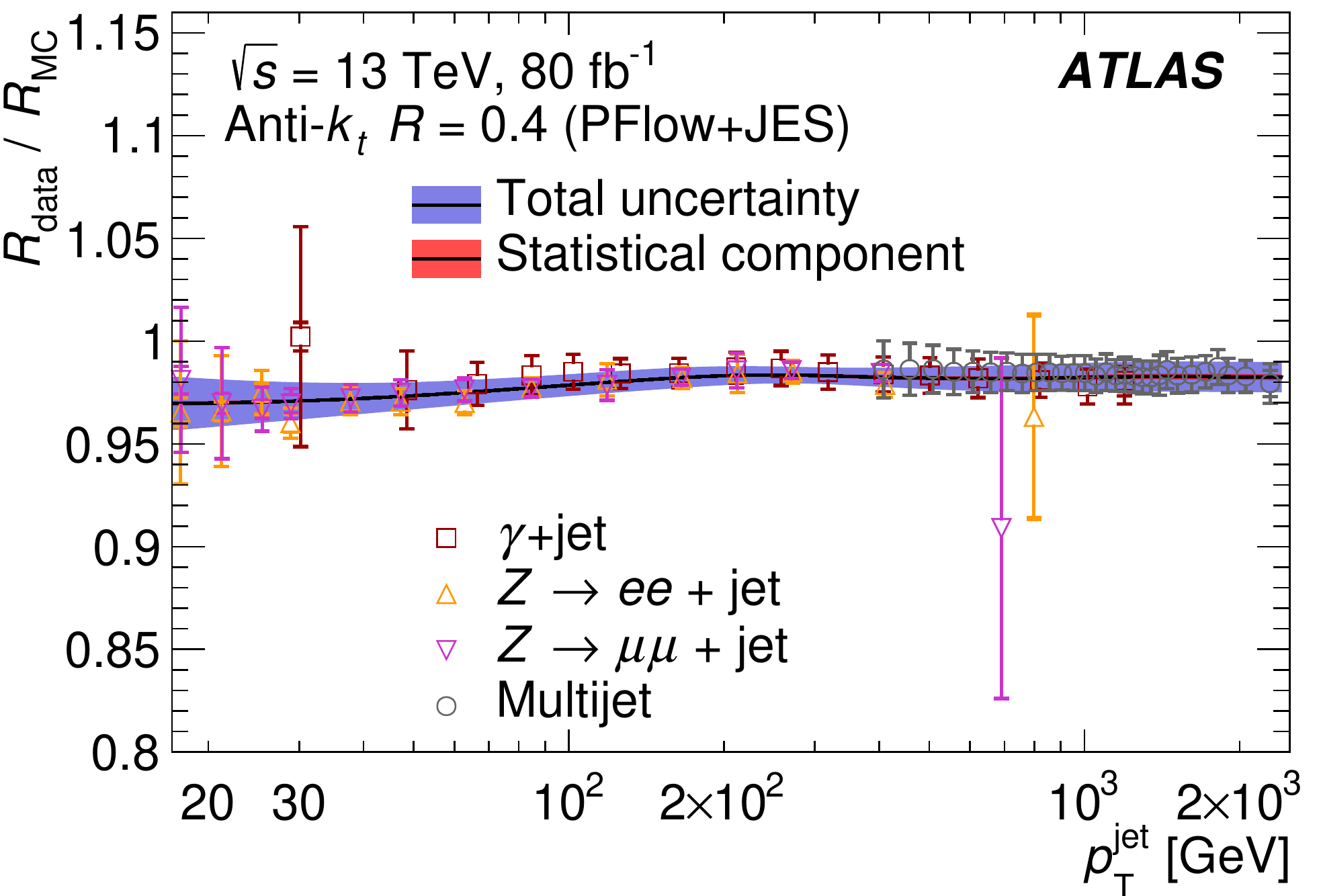}
\caption{}
\label{subfig:jes_combination_plain}
\end{subfigure}
\begin{subfigure}[b]{0.49\textwidth}
\includegraphics[width=\textwidth]{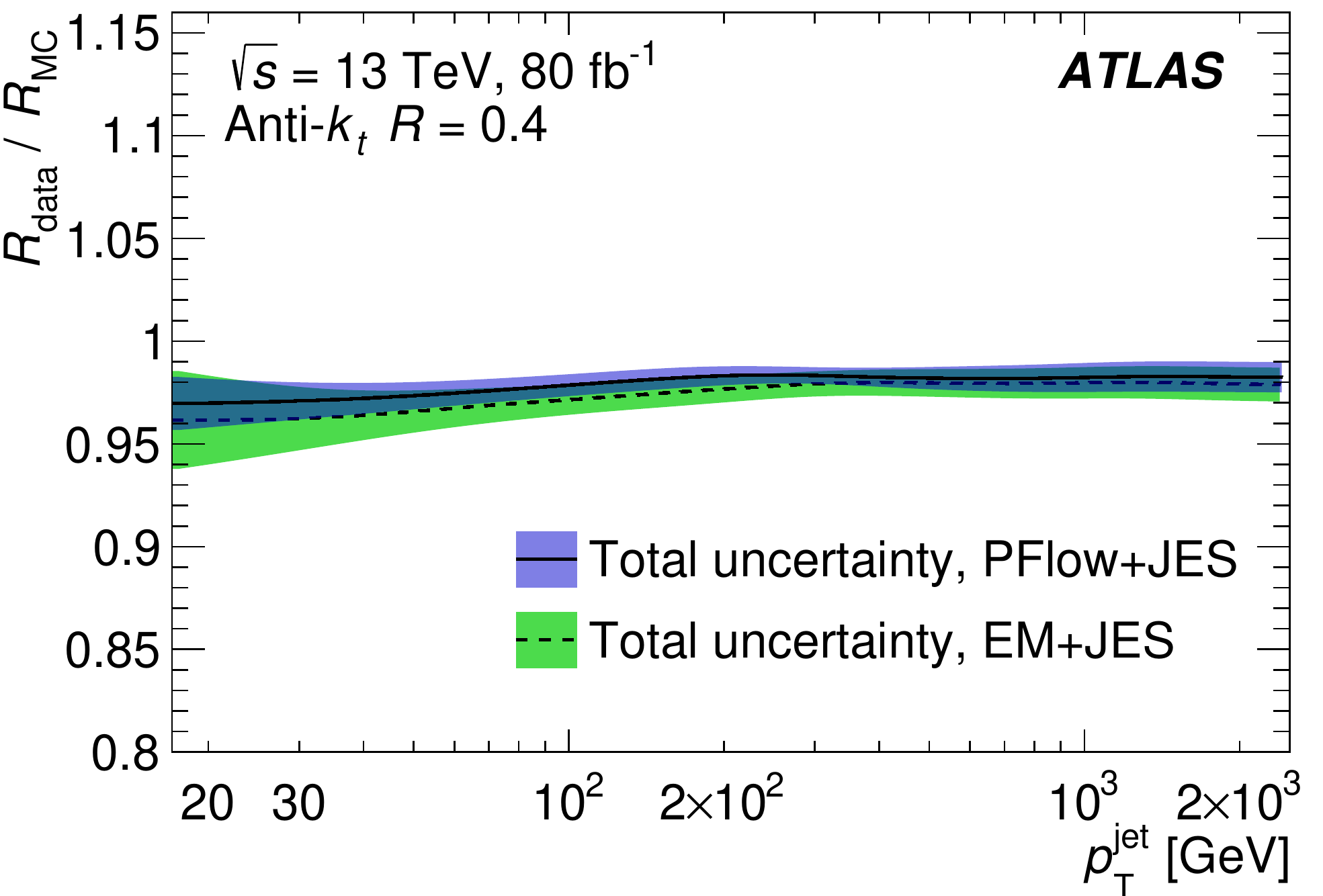}
\caption{}
\label{subfig:jes_combination_comparison}
\end{subfigure}
\caption{\subref{subfig:jes_combination_plain} Ratio of the \EMPFlowJES jet response in data to that in the nominal MC event generators as a function of jet \pt for \zjet, \gammajet, and multijet \insitu calibrations. The inner horizontal ticks in the error bars give the size of the statistical uncertainty while the outer horizontal ticks indicate the total uncertainty (statistical and systematic uncertainties added in quadrature). The final correction and its statistical and total uncertainty bands are also shown, although the statistical uncertainty is too small to be visible in most regions. \subref{subfig:jes_combination_comparison} A comparison of the combined correction and its uncertainty for \EMPFlowJES and \EMJES jets.}
\label{fig:jes_combination}
\end{center}
\end{figure}
 
Each uncertainty component from the \insitu analyses is individually propagated through the combination procedure. First, the relevant measured response is varied by $1\sigma$ in the uncertainty component within its standard binning. The finer rebinning, $\chi^{2}$ minimization, and combination procedure is repeated, although using the weights as determined for the nominal result to prevent the varied uncertainty from decreasing the contribution of the measurement. The difference between the combined calibration curve with the systematically shifted input and the nominal calibration curve is taken as $1\sigma$ in the varied uncertainty. Throughout this process, each individual uncertainty source is treated as fully correlated across $\eta$ and $\pt$ but entirely uncorrelated with all other uncertainty sources. After this step, the uncertainties from the \zjet analyses are taken to be fully correlated with the same uncertainties propagated through the multijet balance. Other assumptions of correlation between components can similarly be made and altered after their propagation, allowing multiple different assumptions.
 
\FloatBarrier
 
\subsection{Systematic uncertainties}
\label{sec:jes-systematics}
The full uncertainty in the jet energy scale consists of 125 individual terms derived from the \textit{in~situ} measurements, \pileup effects, flavour dependence, and estimates of additional effects as summarized in \Tab{\ref{tab:jes_uncertainties}}. The majority of the individual terms stem from the \insitu measurements and cover the effects of analysis selection cuts, event topology dependence, and MC mis-modelling and statistical limitations, as well as the uncertainties associated with the calibration of the electrons, muons, and photons.
 
The $\eta$ intercalibration analysis results in five nuisance parameters, with a sixth for 2018 data only, as discussed in \Sect{\ref{sec:jes-etaintercal}}: one covers systematic effects, one covers statistical uncertainty, and three (four in 2018) are used to parameterize the non-closure. \Pileup effects are described by four nuisance parameters which account for offsets and \pt dependence in \avgmu and \Npv as well as event topology dependence of the density metric $\rho$. The offset and \pt dependence terms are derived in data using a combination of \zjet measurements and measurements comparing reconstructed jets with track-jets. The $\rho$ topology term is the largest of the \pileup uncertainties and is determined by the maximum deviation in measured density between different \insitu measurements under the same \pileup conditions.
 
The two flavour dependence uncertainties are derived from simulation and account for relative flavour fractions and differing responses to quark- and gluon-initiated jets~\cite{PERF-2012-01,PERF-2014-02}. The flavour composition uncertainty accounts for the differing response of quark- and gluon-initiated jets in a sample with some uncertainty on the fraction of gluon-initiated jets $f_g$.  Where $\Response_q$ and $\Response_g$ are the responses measured in \Pythia and $\sigma_g^f$ is the uncertainty on $f_g$ in the sample, the flavour composition uncertainty is defined as:
\begin{equation*}
\sigma_\text{composition} = \sigma_g^f \frac{|\Response_q - \Response_g|}{f_g \Response_g + (1 - f_g)\Response_q}\,.
\end{equation*}
The flavour response uncertainty accounts for the fact that, unlike quark-initiated jet response, gluon-initiated jet response is found to differ significantly between generators. This uncertainty is defined by comparison between the nominal \Pythia sample and an alternative \herwig sample:
\begin{equation*}
\sigma_\text{response} = f_g (\Response_g^{\Pythia} - \Response_g^{\herwig})\,.
\end{equation*}
Figure~\ref{fig:flavour_uncertainties} shows the gluon-jet response and the difference between quark-jet and gluon-jet responses using both \Pythia and \herwig for \pflow jets. The samples are the same as those used for the multijet balance analysis and are dominated by gluon jets at low \pt.
For \herwig, $\Response_q - \Response_g$ becomes negative in the 90--600 \pt{} region (which appears as a bump in the $|\Response_q - \Response_g|$ curve).
 
\begin{figure}[htp!]
\begin{center}
\begin{subfigure}[b]{0.49\textwidth}
\includegraphics[width=\textwidth]{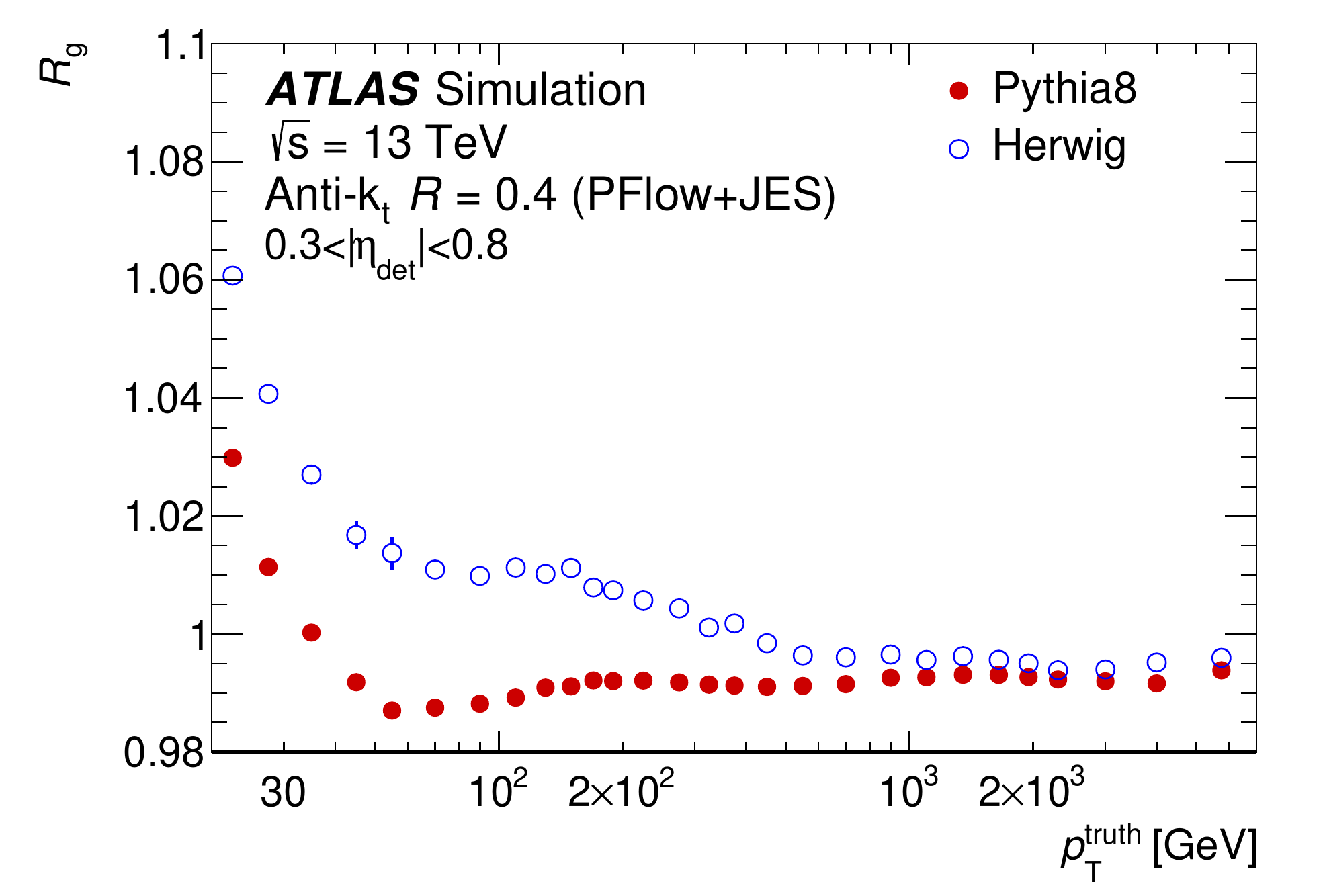}
\caption{}
\label{subfig:flavours_first_pf}
\end{subfigure}
\begin{subfigure}[b]{0.49\textwidth}
\includegraphics[width=\textwidth]{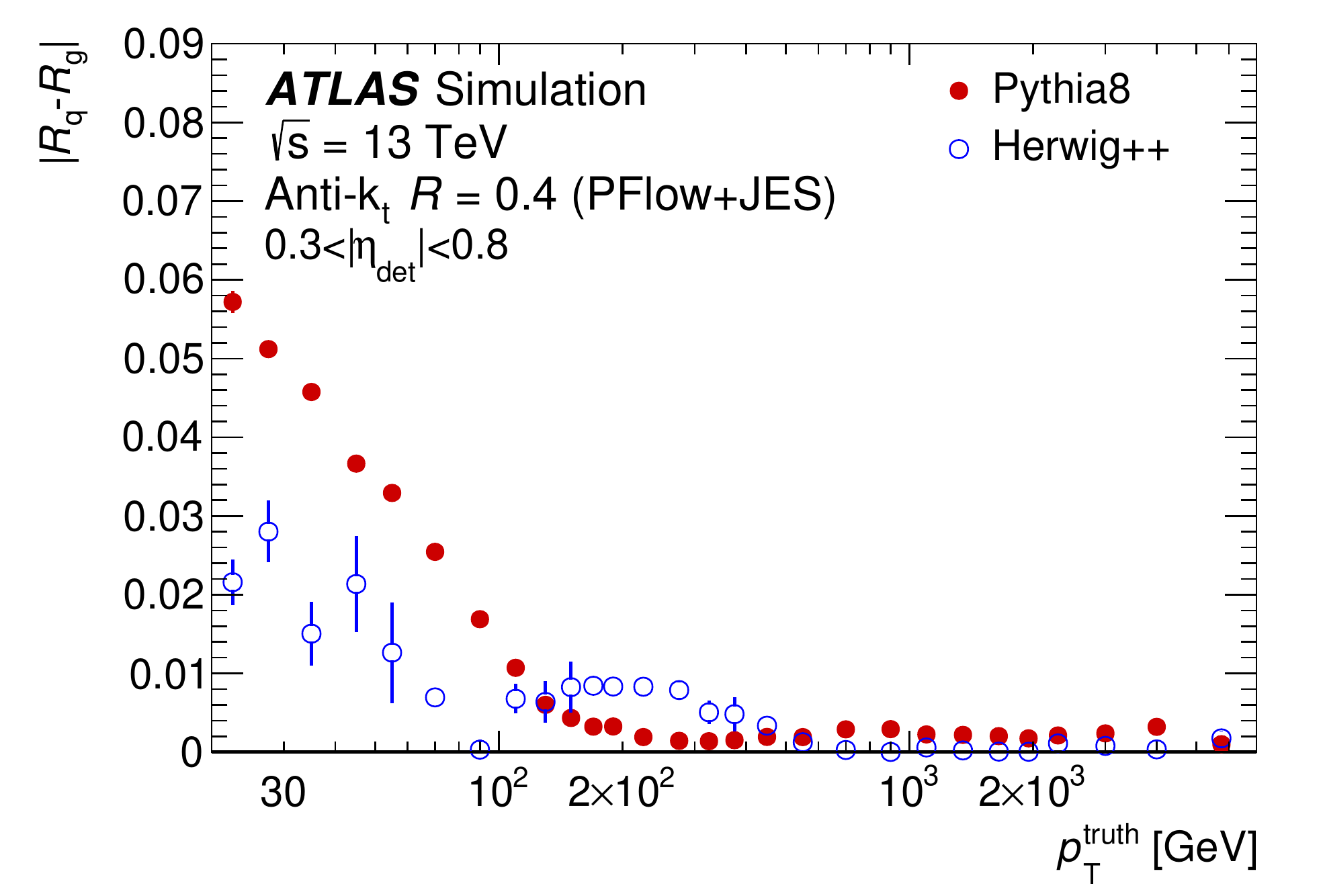}
\caption{}
\label{subfig:flavours_second_pf}
\end{subfigure}
\caption{\subref{subfig:flavours_first_pf} Measured gluon-initiated jet response and~\subref{subfig:flavours_second_pf} difference between quark- and gluon-initiated jet responses for \pflow jets using two different generators. These define the flavour response and composition uncertainties respectively.}
\label{fig:flavour_uncertainties}
\end{center}
\end{figure}
 
An additional uncertainty applied only to $b$-initiated jets covers the difference in response between jets from light- versus heavy-flavour quarks and replaces the flavour composition and response uncertainties for these heavy-flavour jets. The punch-through uncertainty accounts for mis-modelling of the GSC correction to jets which pass through the calorimeter and into the muon system, taking the difference in jet response between data and MC simulation in bins of muon detector activity as the systematic uncertainty. Both are discussed in more detail in Ref.~\cite{PERF-2014-02}. Finally, the high-\pt `single particle' uncertainty is derived from studies of the response to individual hadrons and is used to cover the region beyond 2.4~\TeV, where the MJB analysis no longer has statistical power~\cite{PERF-2015-05}. When calibrating MC samples simulated using AFII, an additional non-closure uncertainty is applied to account for the difference in jet response between these samples and those which used full detector simulation.
 
\begin{table}
\vspace{-0.5cm}
\small
\centering
\caption[]{Sources of uncertainty in the jet energy scale.}
\begin{tabular}{ l l}
\toprule
Component & \multicolumn{1}{c}{Description} \\
\midrule
\multicolumn{2}{c}{$\eta$ intercalibration} \\
\midrule
Systematic mis-modelling     & Envelope of the generator, \pileup, and event topology variations \\
Statistical component          & Statistical uncertainty (single component)\\
Non-closure 			 & Three components describing non-closure at high energy and at $\eta \sim \pm 2.4$ \\
Non-closure, 2018 only      & Single component describing non-closure at $\eta \sim \pm 1.5$ due to Tile calibration \\
\midrule
\multicolumn{2}{c}{$Z+\text{jet}$} \\
\midrule
Electron scale & 		Uncertainty in the electron energy scale \\
Electron resolution & 	Uncertainty in the electron energy resolution \\
Muon scale & 			Uncertainty in the muon momentum scale \\
Muon resolution (ID) & 	Uncertainty in muon momentum resolution in the ID \\
Muon resolution (MS) & 	Uncertainty in muon momentum resolution in the MS \\
MC generator & 		Difference between MC event generators \\
JVT cut & 				Jet vertex tagger uncertainty \\
$\Delta \phi$ cut &		Variation of $\Delta \phi$ between the jet and $Z$ boson \\
Subleading jet veto &		Radiation suppression through second-jet veto \\
Showering \& topology &	Modelling energy flow and distribution in and around a jet \\
Statistical & 			Statistical uncertainty in 28 discrete \pt terms \\
\midrule
\multicolumn{2}{c}{$\gamma+\text{jet}$} \\
\midrule
Photon scale &			Uncertainty in the photon energy scale \\
Photon resolution & 		Uncertainty in the photon energy resolution \\
MC generator &			Difference between MC event generators \\
JVT cut & 				Jet vertex tagger uncertainty \\
$\Delta \phi$ cut &		Variation of $\Delta \phi$ between the jet and photon \\
Subleading jet veto &		Radiation suppression through second-jet veto \\
Showering \& topology &	Modelling energy flow and distribution in and around a jet \\
Photon purity & 			Purity of sample used for $\gamma+\text{jet}$ balance \\
Statistical & 			Statistical uncertainty in 16 discrete \pt terms \\
\midrule
\multicolumn{2}{c}{Multijet balance} \\
\midrule
$\Delta \phi$ (lead, recoil system) & Angle between leading jet and recoil system \\
$\Delta \phi$ (lead, any sublead) & Angle between leading jet and closest subleading jet \\
MC generator & 		Difference between MC event generators \\
$\pt^{\text{asym}}$ selection & Second jet's \pt contribution to the recoil system \\
Jet \pt &				Jet \pt threshold \\
Statistical &			Statistical uncertainty in 28 discrete \pt terms \\
\midrule
\multicolumn{2}{c}{\Pileup} \\
\midrule
$\mu$ offset  	&	 	Uncertainty in the $\mu$ modelling in MC simulation \\
\Npv offset     & 	Uncertainty in the \Npv modelling in MC simulation \\
$\rho$ topology &		Uncertainty in the per-event \pt density modelling in MC simulation \\
\pt dependence &		Uncertainty in the residual \pt dependence \\
\midrule
\multicolumn{2}{c}{Jet flavour} \\
\midrule
Flavour composition &	Uncertainty in the proportional sample composition of quarks and gluons \\
Flavour response &		Uncertainty in the response of gluon-initiated jets \\
$b$-jets &				Uncertainty in the response of $b$-quark-initiated jets \\
\midrule
Punch-through &		Uncertainty in GSC punch-through correction \\
\midrule
Single-particle response & High-\pt jet uncertainty from single-particle and test-beam measurements \\
\midrule
AFII non-closure & 		Difference in the absolute JES calibration for simulations in AFII \\
\bottomrule
\end{tabular}
\label{tab:jes_uncertainties}
\end{table}
 
The total jet energy scale uncertainty is shown in \Fig{\ref{subfig:vs_pt}} as a function of jet \pt for fixed $\eta_{\text{jet}}=0$ and in \Fig{\ref{subfig:vs_eta}} as a function of jet $\eta$ for fixed $\ptjet = 60~\GeV$. A dijet-like composition of the sample (that is, predominantly gluons) is assumed in computing the flavour uncertainties. The uncertainties in the $\eta$ intercalibration analysis are labelled `relative \insitu JES' with the non-closure uncertainty creating the asymmetric peaks around $\eta=\pm 2.5$. Uncertainties in all other \insitu measurements are combined into the `absolute \insitu JES' term, which also includes the single-particle uncertainty.
 
\begin{figure}[htp!]
\begin{center}
\begin{subfigure}[b]{0.49\textwidth}
\includegraphics[width=\textwidth]{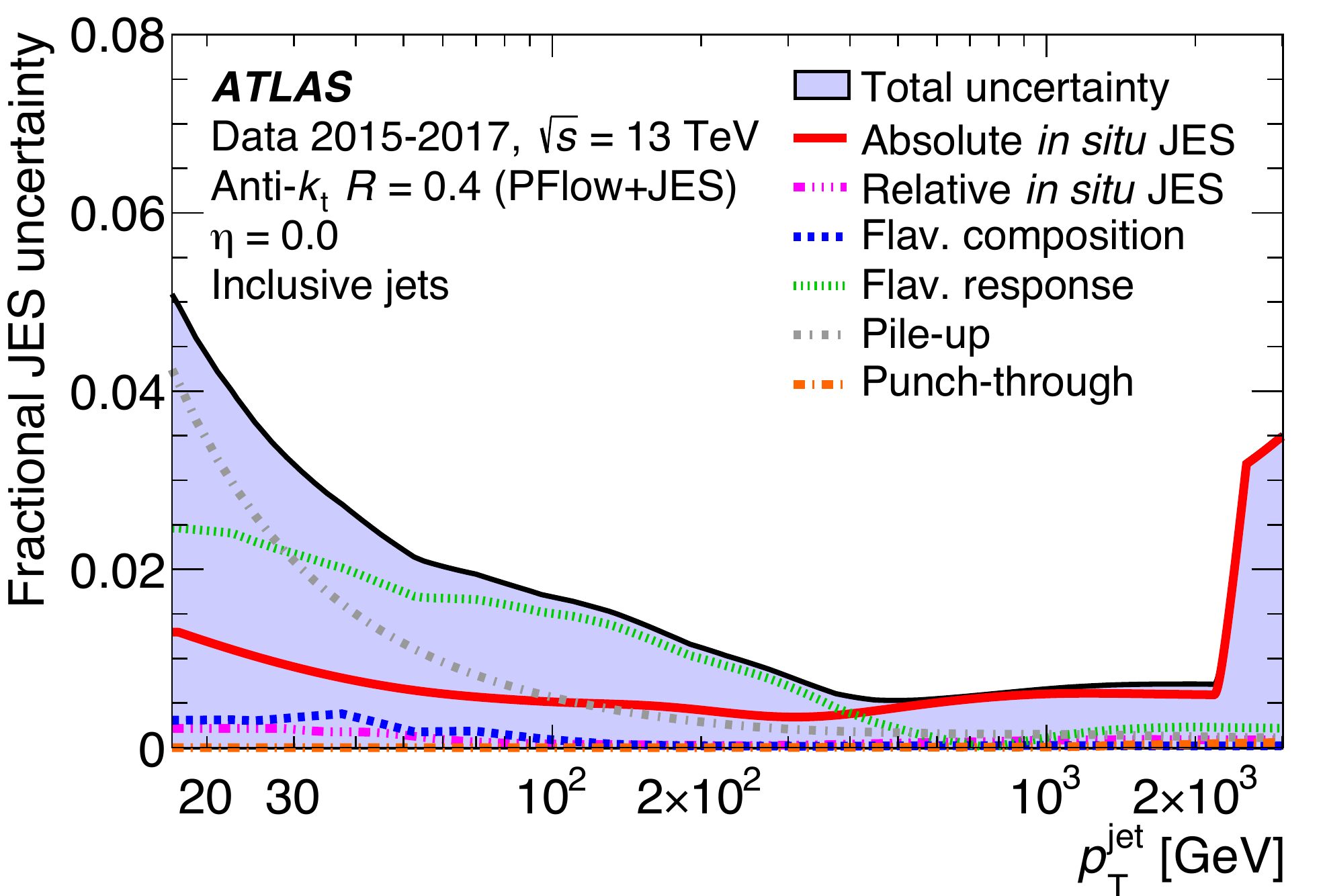}
\caption{}
\label{subfig:vs_pt}
\end{subfigure}
\begin{subfigure}[b]{0.49\textwidth}
\includegraphics[width=\textwidth]{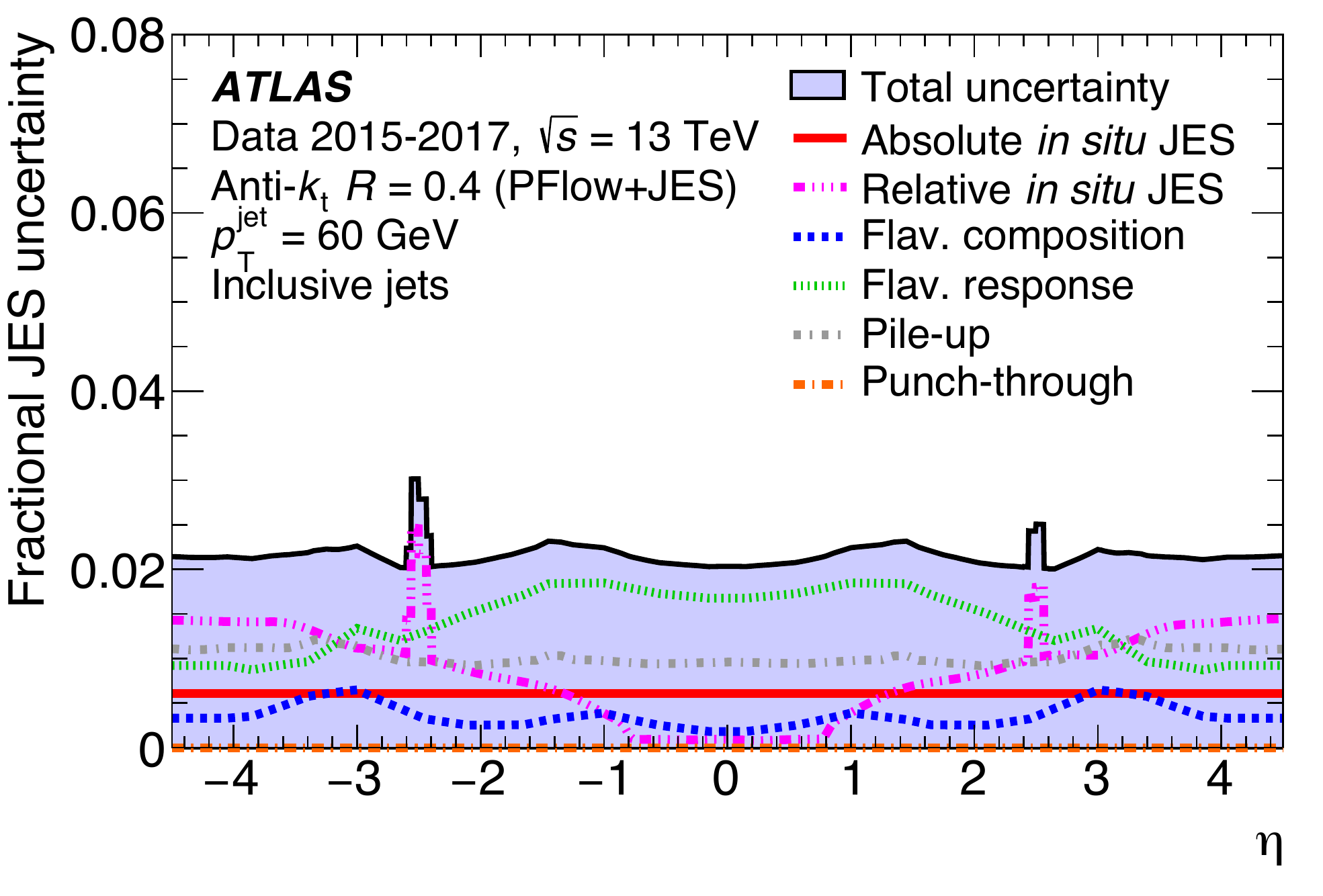}
\caption{}
\label{subfig:vs_eta}
\end{subfigure}
\caption{Fractional jet energy scale systematic uncertainty components for \akt $R=0.4$ jets \subref{subfig:vs_pt} as a function of jet \pt at $\eta = 0$ and \subref{subfig:vs_eta} as a function of $\eta$ at $\pt = 60~\GeV$, reconstructed from particle-flow objects. The total uncertainty, determined as the quadrature sum of all components, is shown as a filled region topped by a solid black line. Flavour-dependent components shown here assume a dijet flavour composition.}
\label{fig:combination_uncertainties}
\end{center}
\end{figure}

\FloatBarrier
\subsubsection{Uncertainty correlations and reductions}
\label{sec:jes-reductions}
The detail contained in 125 independent nuisance parameters is far more than is required by most analyses, so it is necessary to reduce the uncertainty description to a smaller number of terms. One could imagine a single `Jet energy scale' nuisance parameter constructed by adding in quadrature all of the independent components. However, a meaningful set of correlations exist between the jet energy scale uncertainties for two jets at different $\eta$ and \pt as a result of the structures of the nuisance parameters. In the case of reduction to a single component, the entirety of this correlation information would be lost and an unrealistic assumption -- that of full correlation between the jet energy scale uncertainties for any values of $\eta$ and \pt\ -- would be enforced. In practice, a variety of reduced uncertainty schemes are provided to allow simplified descriptions with a minimum loss of correlation information.
 
The 98 uncertainty components stemming from the absolute \insitu analyses are functions only of \pt and thus their behaviour can be easily represented by a smaller number of orthogonal terms. An eigenvector decomposition is performed on a covariance matrix of these uncertainty components and the largest of the resulting orthogonal terms are kept separate as new effective nuisance parameters~\cite{PERF-2012-01}. The remaining terms are combined into a single residual nuisance parameter. To determine how many components to keep independently and how many to combine in the residual term, the covariance matrix for the reduced set is also computed and the difference in correlation in each jet $\eta$ and $\pt$ between the reduced set and the full set is calculated. This difference is taken as a measure of the information loss and the number of combined terms is adjusted so that the difference is below an acceptable bound (usually $0.05$).  Two different reduction schemes are produced: the \textit{global reduction} combines all \pt-dependent \insitu uncertainty components regardless of their sources and results in 8 reduced components for a total of 23 once the two-dimensional terms (not arising from the \insitu analyses and not reduced) are included.
The \textit{category reduction} combines the \pt-dependent \insitu uncertainty components in separate groups based on their origin (detector, statistical, modelling, or mixed) and results in 15 reduced components for a total of 30. The JES correlation matrix for the full set of nuisance parameters is shown in \Fig{\ref{subfig:full_correlation}}. The bin-by-bin correlation loss between the full set of nuisance parameters and the category reduction is shown in \Fig{\ref{subfig:loss_to_category}} and is below 0.05 everywhere as required.
 
\begin{figure}[htp!]
\begin{center}
\begin{subfigure}[b]{0.49\textwidth}
\includegraphics[width=\textwidth]{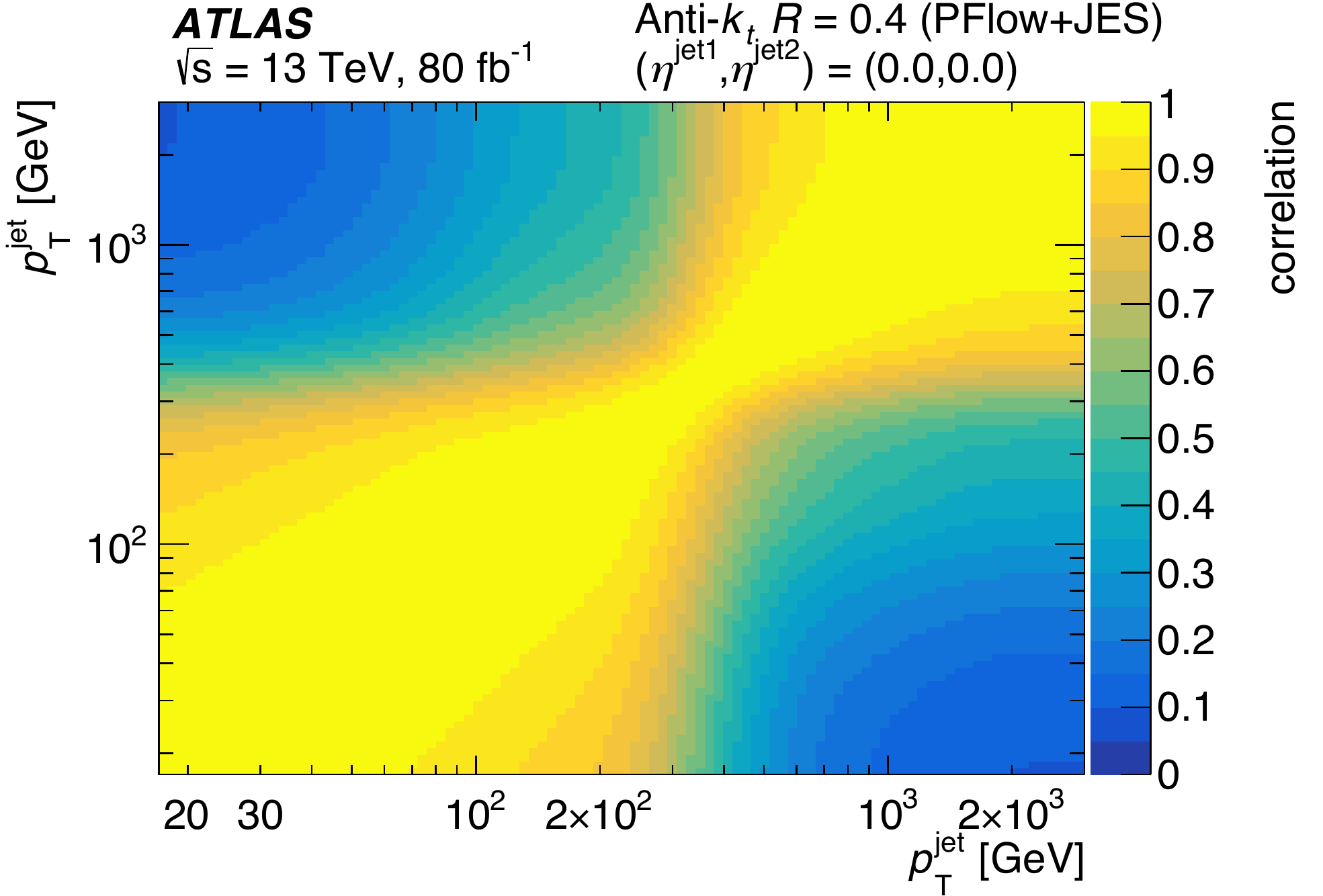}
\caption{}
\label{subfig:full_correlation}
\end{subfigure}
\begin{subfigure}[b]{0.49\textwidth}
\includegraphics[width=\textwidth]{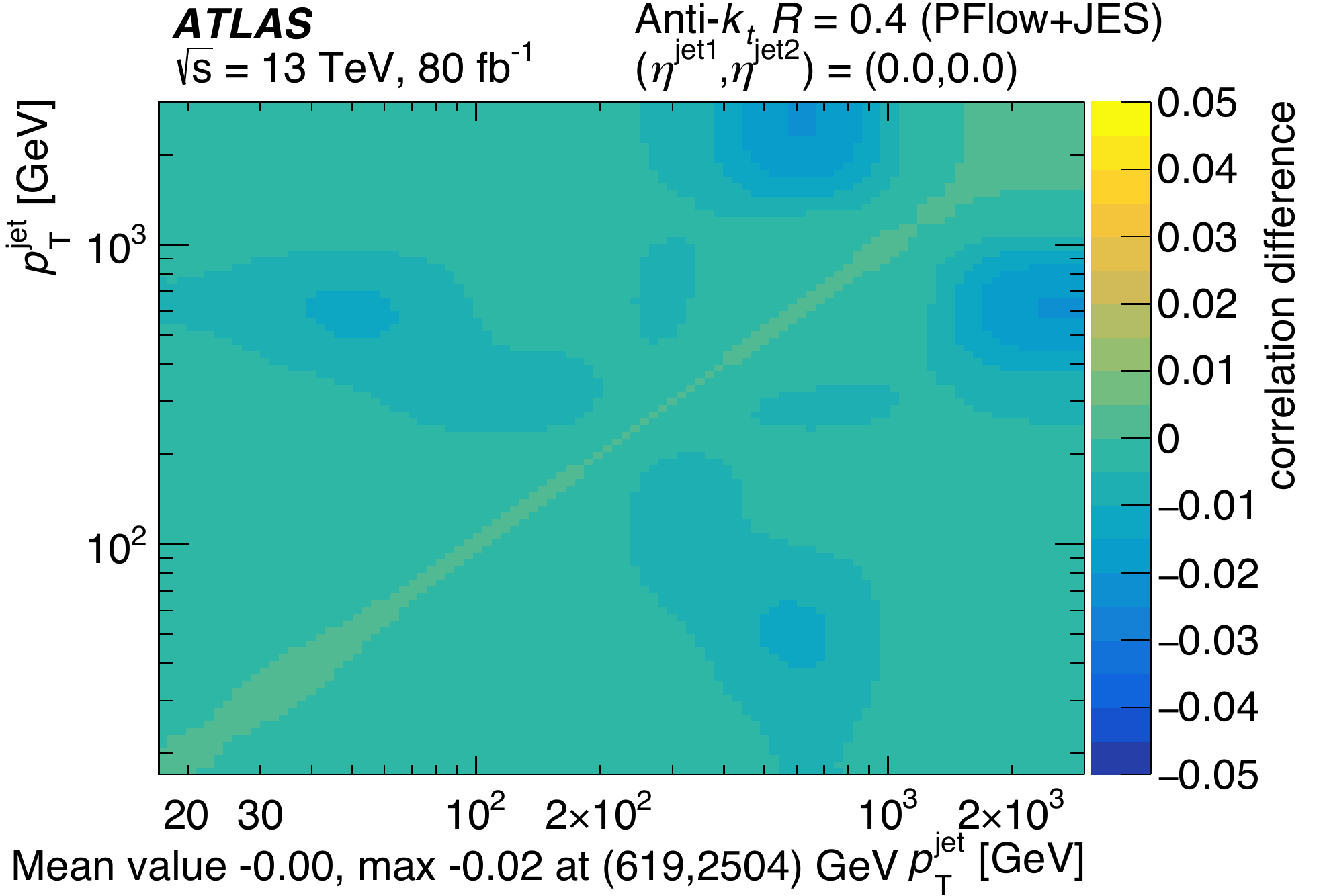}
\caption{}
\label{subfig:loss_to_category}
\end{subfigure}
\caption{\subref{subfig:full_correlation} The jet energy scale correlation matrix for two \EMPFlowJES jets at $\eta = 0$ using the full set of 98 \pt-dependent \insitu nuisance parameters and \subref{subfig:loss_to_category} the difference in correlation information between the full description and the category reduction. The maximum loss of correlation information is 0.02 and occurs at the ($\pt^{j1},\pt^{j2}$) location specified by the text at the bottom of the plot.}
\label{fig:correlations}
\end{center}
\end{figure}
 
While the same procedure could in principle be used for the components which depend on both \pt and $\eta$, the complexity added by the second dimension means that nearly as many eigenvectors would be needed to adequately describe the correlations as there were original terms and so the gain would be minimal. However, many analyses still require fewer than 25 nuisance parameters and are not affected by loss of correlation information. To provide suitable uncertainties for these, a \textit{strong} reduction procedure is used to group the globally reduced versions of the absolute \insitu uncertainties together with the two-dimensional uncertainties into three effective nuisance parameters as detailed in Ref~\cite{PERF-2016-04}. The three terms of the $\eta$ intercalibration non-closure uncertainty are kept separate because their two-dimensional shapes are especially difficult to reduce and would cause an unacceptably large correlation loss.
 
Four different sets (\textit{scenarios}) of the three effective nuisance parameters are created by varying the combinations of terms they contain. The varied sets are chosen such that the correlation loss in each is constrained to an $\eta$--\pt range which is well described by a different set. The metric for assessing performance of the four scenarios is the \textit{uncovered correlation loss}, defined as the maximum difference in correlation between any two reduced scenarios minus the minimum difference in correlation between any reduced scenario and the full set of nuisance parameters. The uncovered correlation loss is calculated for a fine grid of points in $\eta$ and \pt, ensuring no small-scale structures are missed. Contents of the effective nuisance parameters are varied, keeping systematic uncertainties with similar behaviours mostly grouped together, until a set of scenarios is found in which the maximum uncovered correlation loss is kept below 0.25 and confined to sufficiently small regions that the average correlation loss in an $\eta$--$\pt$ plane does not exceed 0.02. A detailed discussion of the application of strongly reduced uncertainties within physics analyses can be found in Ref.~\cite{PERF-2016-04}.
 
\FloatBarrier
\subsubsection{Uncertainties for \EMTopo and \pflow jets}
\label{sec:jes-comparison}
 
Although the scale of individual calibrations may vary between \EMTopo and \pflow jets, the final uncertainties are similar in size. A slightly larger \pileup uncertainty contribution in \pflow jets due to the impact of the underlying event is offset by smaller \insitu uncertainties, leading to a comparable total overall uncertainty. \Fig{\ref{fig:compare_jes_uncert}} shows the total uncertainty in \EMTopo and \pflow jets for a range of \pt values at fixed $\eta=0$ and for a range of $\eta$ values at fixed $\pT = 60~\GeV$. The level of agreement is representative of other $\pt$ and $\eta$ ranges.
 
\begin{figure}[htp!]
\begin{center}
\begin{subfigure}[b]{0.49\textwidth}
\includegraphics[width=\textwidth]{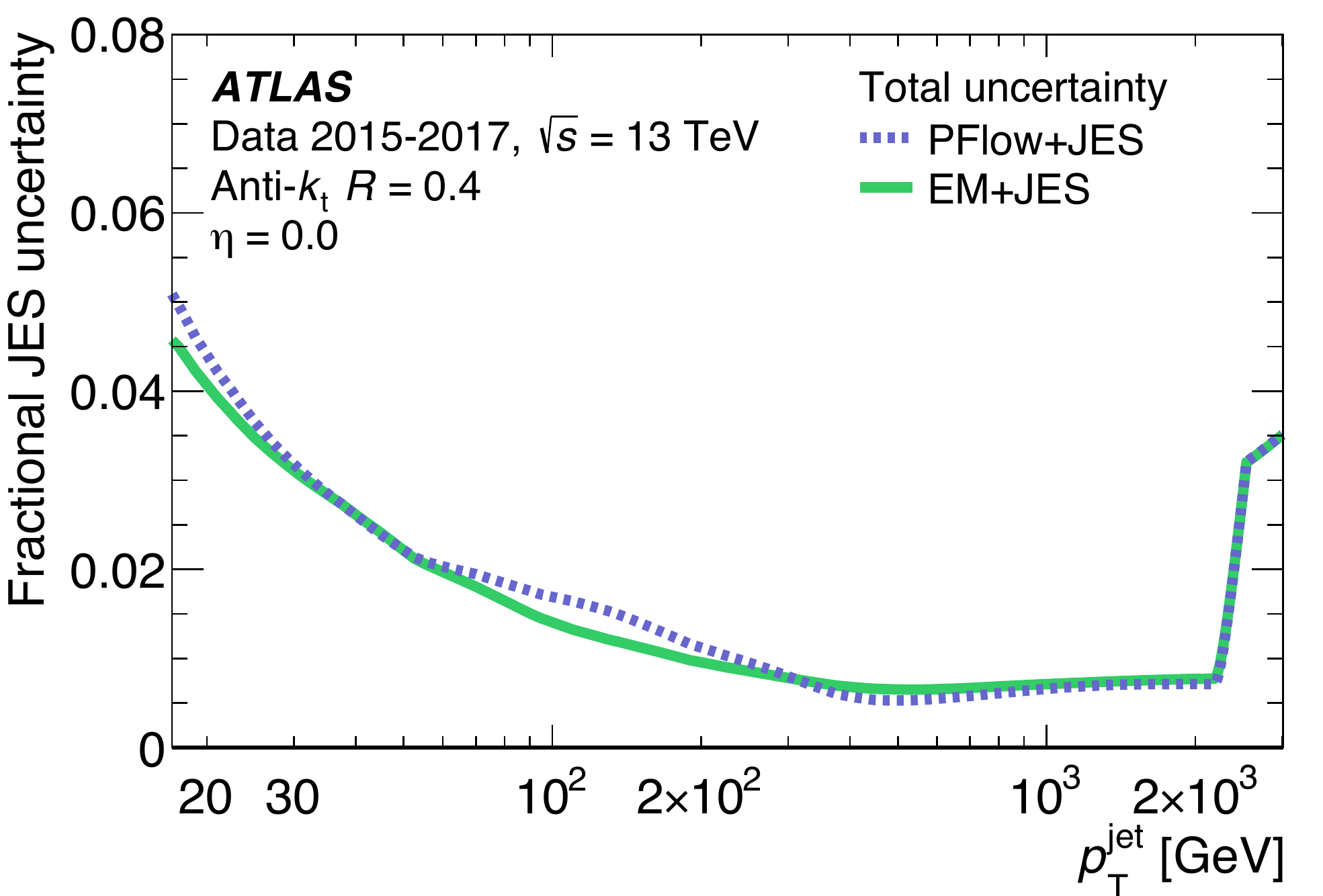}
\caption{}
\label{subfig:comp_vs_pt}
\end{subfigure}
\begin{subfigure}[b]{0.49\textwidth}
\includegraphics[width=\textwidth]{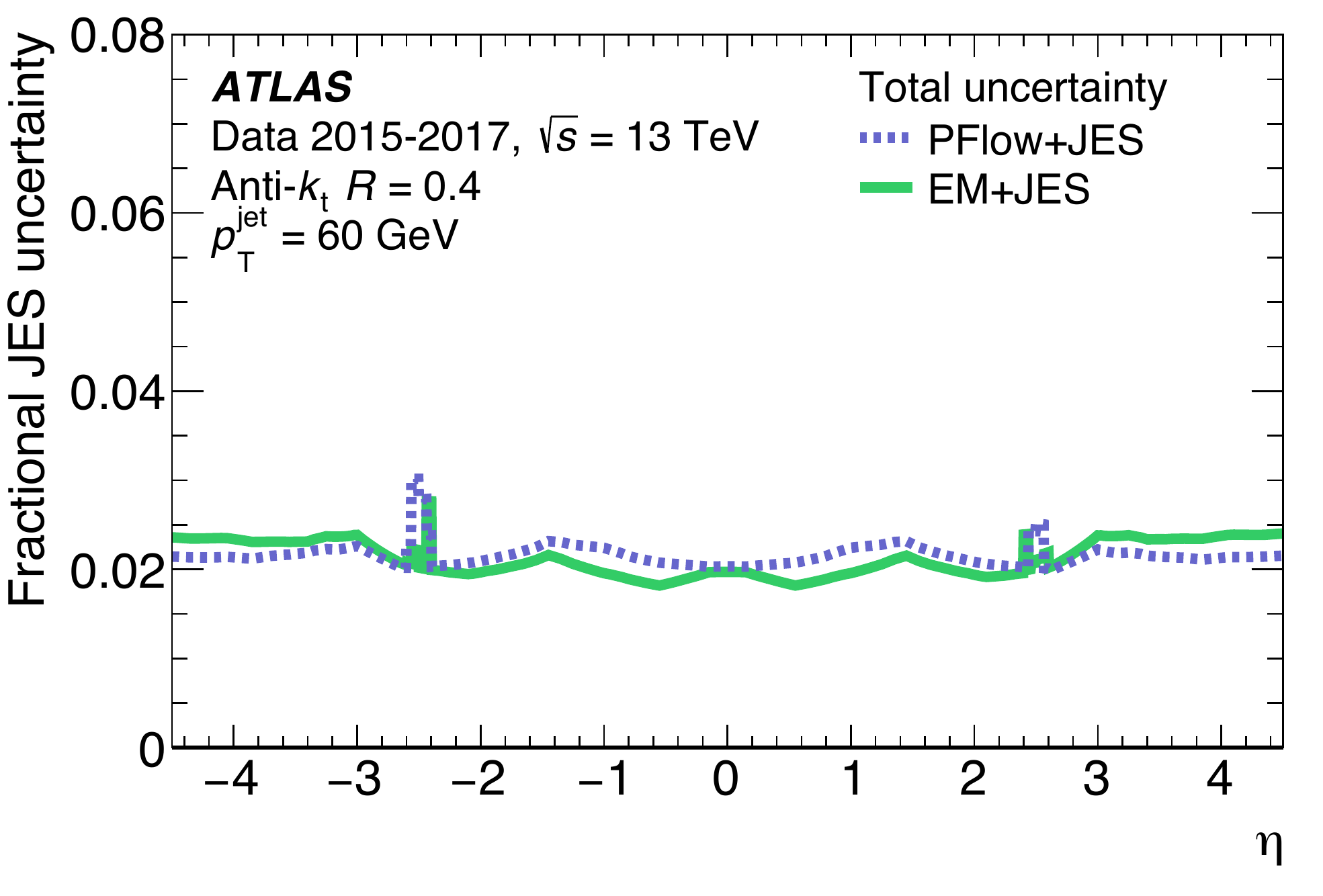}
\caption{}
\label{subfig:comp_vs_eta}
\end{subfigure}
\caption{Fractional jet energy scale systematic uncertainty summed across all components for \akt $R=0.4$ jets \subref{subfig:comp_vs_pt} as a function of jet \pt at $\eta = 0$ and \subref{subfig:comp_vs_eta} as a function of $\eta$ at $\pt = 60~\GeV$. The total uncertainty is shown for both \EMTopo and \pflow jets. Contributions from topology-dependent components are calculated assuming a dijet flavour composition.}
\label{fig:compare_jes_uncert}
\end{center}
\end{figure}
 
 
 
\FloatBarrier
 
\section{Jet energy resolution}
\label{sec:jer}
Precise knowledge of the jet energy resolution (JER) is important for detailed measurements of SM jet production, measurements and studies of the properties of the SM particles that decay to jets (e.g. $\Wboson/\Zboson$ bosons, top quarks), as well as searches for physics beyond the SM involving jets.  The JER also affects the missing transverse momentum, which plays an indispensable role in many searches for new physics and measurements involving particles that decay into neutrinos, and thus rely on well-reconstructed missing momentum.
 
The dependence of the relative JER on the transverse momentum of the jet may be parameterized using a functional form expected for calorimeter-based resolutions, with three independent contributions, namely the noise ($N$), stochastic ($S$) and constant ($C$) terms~\cite{Wigmans:2280786}:
\begin{equation}
\frac{\sigma(\pt)}{\pt} = \frac{N}{\pt}\oplus\frac{S}{\sqrt{\pt}}\oplus\,C\,.
\label{eq:res}
\end{equation}
\noindent The noise ($N$) term is due to the contribution of electronic noise to the signal measured by the detector front-end electronics, as well as that due to \pileup. Since both contribute directly to the energy measured in the calorimeter but are approximately independent of the energy deposited by the showing particles, the contribution to the JER scales like $1/\pt$. The noise term is expected to be significant in the low-$\pt$ region, below $\sim$30~\GeV.
Statistical fluctuations in the amount of energy deposited are captured by the stochastic ($S$) term, which represents the limiting term in the resolution up to several hundred~\GeV\ in jet \pt. The $S$ term contribution to the JER scales like $1/\sqrt{\pt}$.
The constant ($C$) term corresponds to fluctuations that are a constant fraction of the jet \pt, such as energy depositions in passive material (e.g.\ cryostats and solenoid coil), the starting point of the hadron showers, and non-uniformities of response across the calorimeter. The constant term is expected to dominate the high-$\pt$ region, above approximately 400~\GeV.

In order to measure the JER, jet momentum must be measured precisely. This implies that the jets must either recoil against a reference object whose momentum can be measured precisely, or be balanced against one another in a well-defined dijet system~\cite{PERF-2012-01,PERF-2014-02}. Measurements using the latter approach are presented here, as well as a method for measuring the contributions to the resolution from the noise term ($N$) due to both \pileup and electronics. The 2017 data, corresponding to an integrated luminosity of 44~\ifb\ is used for these measurements.

\FloatBarrier
 
\subsection{Resolution measurement using dijet events}
\label{sec:jer-dijets}
 
Dijet events are both plentiful and produced via a set of $2\ra 2$ processes that are theoretically well-understood. JER measurements using these events for the \textit{dijet balance} method rely on the approximate scalar balance between the transverse momenta of the two leading jets. Deviations from exact balance, measured via the asymmetry, given by
\begin{equation}
\asym \equiv \frac{\pTprobe - \pTref}{\pTavg},
\label{eq:asym}
\end{equation}
\noindent are due to a combination of experimental resolution, the presence of additional radiation in the event, and biases due to the event selection used in the measurement. In \Eqn{(\ref{eq:asym})}, \pTref is the \pT of a reference jet which is required to be located in a well-calibrated region of the detector (\etadetRef), taken here to be the central region of the calorimeter $0.2 \leq |\etadetRef| < 0.7$, where the seam at $\etadet=0$ is excluded to ensure the reference jet energy is as cleanly measured as possible. The probe jet, with transverse momentum \pTprobe, may be located either within this central reference region or beyond it, with $|\etadetProbe| < 4.5$. The probe jet is the jet for which the resolution is to be measured and \pTavg is the mean of the probe and reference jet momenta, $\pTavg = (\pTprobe + \pTref)/2$. The standard deviation of \asym for a particular $(\pTavg, \etadetProbe)$ bin is denoted by \sigmaAsym, and in the case of a measurement of the probe jet asymmetry may be expressed as
\begin{equation}\nonumber
\sigmaAsym^\text{probe} = \frac{\sigmaptProbe \oplus\, \sigmaptRef }{\langle \ptavg \rangle} =
\left< \frac{\sigmapt}{\pt} \right>^\text{probe} \oplus\,   \left< \frac{\sigmapt}{\pt} \right>^\text{ref},
\label{eq:dijet-jer}
\end{equation}
\noindent where \sigmaptProbe and \sigmaptRef are the standard deviations of \pTprobe and \pTref, respectively, and are used to denote the JER for each of the relevant objects. For calibrated jets, $\avg{\pTprobe} = \avg{\pTref} = \avg{\pTavg}$ in the reference region.  The reference jet relative resolution, $\langle \sigmapt/\pt \rangle^\text{ref}$, must therefore be subtracted from the measured asymmetry distribution in order to extract the resolution of the probe jet as
\begin{equation}
\left< \frac{\sigmapt}{\pt} \right>^\text{probe} = \sigmaAsym^\text{probe} \ominus\, \left< \frac{\sigmapt}{\pt} \right>^\text{ref}.
\label{eq:dijet-jer-probe}
\end{equation}
Equation~(\ref{eq:dijet-jer-probe}) is valid in the probe region as well, up to a correction factor that accounts for the potential overall imbalance between the reference jet and the probe jet in that region. This correction factor is found to be negligible ($<1\%$) for the measurements performed here. However, the \pt balance of the measured jets, and thus the measured asymmetry distribution, is measurably affected on an event-by-event basis by physics effects such as additional radiation, non-perturbative processes including hadronization and multi-parton interactions, and others that may lead to particle losses and additions in the measured jets. Consequently, the measured dijet balance asymmetry distribution represents a convolution of the intrinsic detector resolution and the particle-level balance affected by the aforementioned effects. The determination of $\sigmaAsym^\text{probe}$ must therefore account for such effects, for example by subtracting the particle-level quantity from the measured quantity in quadrature:
\begin{equation}\nonumber
\left(\sigmaAsym^\text{probe}\right)_{\text{det}} = \left(\sigmaAsym^\text{probe}\right)_{\text{meas}} \ominus\, \sigmaAsym^\text{truth}.
\label{eq:dijet-jer-truthcorrection}
\end{equation}
 
The results presented here use an iterative fitting procedure to extract the impact of these effects and to isolate the intrinsic detector resolution, $\left(\sigmaAsym^\text{probe}\right)_{\text{det}}$, by assuming a Gaussian convolution of detector effects with the particle-level balance. First, the asymmetry distribution measured at particle level in MC simulation is fitted with an \textit{ad~hoc} function $\asym^\text{truth}$ based on exponential curves and found to describe it well. Second, the measured asymmetry distribution, $\asym^{\text{meas}}$, is fitted by the function
\begin{equation}\nonumber
\asym^{\text{meas}} = \asym^\text{truth} \otimes \Response(\mu_{\asym}^{\text{det}}, \sigmaAsym^{\text{det}}),
\label{eq:dijet-jer-deconvolutionfit}
\end{equation}
taking $\asym^\text{truth}$ from the particle-level fit and where $\Response(\mu_{\asym}^{\text{det}})$ is a Gaussian distribution with width $\sigmaAsym^{\text{det}}$ representing the detector resolution for the probe jet and offset $\mu_{\asym}^\text{det}$ accounting for any residual non-closure in the JES calibration.
 
Collision data used for the dijet balance measurement are collected using specific combinations of central and forward jet triggers for each of the 11 \pTavg ranges used in the measurement. Trigger selections are required to be at least 99\% efficient in the range of \pTavg in which a particular combination is used. Jets must also pass JVT selection requirements as described in Section~\ref{sec:jes-vjet}.
 
Topology criteria are applied to select well-defined dijet production processes with minimal contributions due to additional radiation or higher-order processes. The azimuthal angle, $\Deltaphi$, between the two leading jets in the event and the maximum \pt of a potential third jet, \ptjetthree, are constrained by the following two criteria:
\begin{eqnarray}\nonumber
\Deltaphi(\jetone, \jettwo) &\geq& 2.7~\text{rad.}\,, \\
\nonumber
\ptjetthree &<& \max(25~\GeV, 0.25\cdot\pTavg).
\end{eqnarray}
 
Example asymmetry distributions are shown in \Fig{\ref{fig:dijetjer_fit}} in two representative bins of \pTavg and probe jet $\etadet^{\text{probe}}$. An iterative Gaussian fit to the core of the asymmetry distribution is used to extract the JER. The result of the measurement of the relative JER and its systematic uncertainty is shown in \Fig{\ref{fig:resolution_dijet}} for a single narrow range of $\etadet^{\text{probe}}$ and as a function of \pTjet. The JER is observed to be slightly underestimated by MC simulation in this central region of the detector.
 
\begin{figure}
\begin{center}
\begin{subfigure}[b]{0.485\textwidth}
\includegraphics[width=\textwidth]{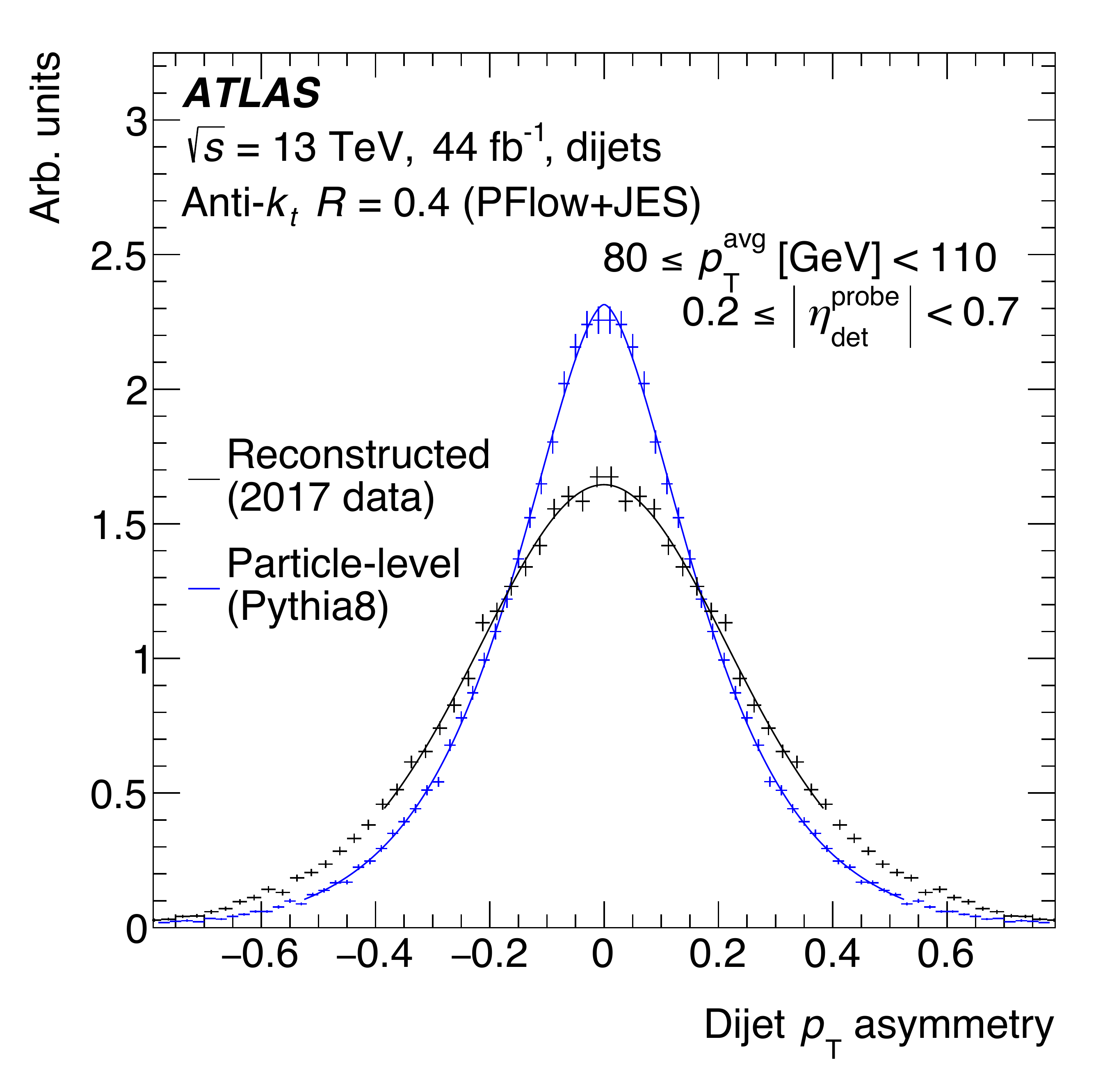}
\caption{}
\label{subfig:examplefit_1}
\end{subfigure}
\begin{subfigure}[b]{0.5\textwidth}
\includegraphics[width=\textwidth]{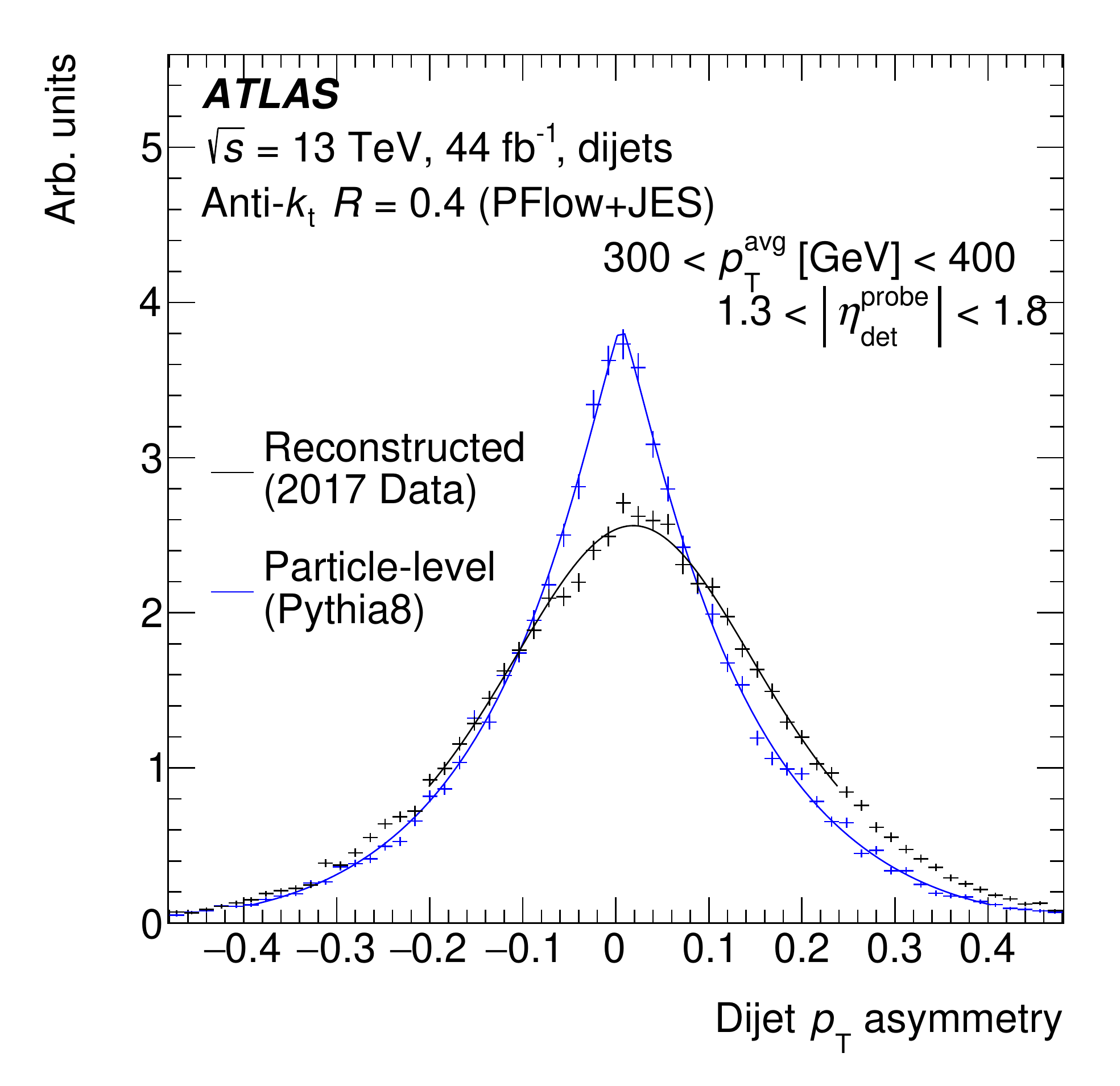}
\caption{}
\label{subfig:examplefit_2}
\end{subfigure}
\caption{Asymmetry distribution measured in data and particle-level \Pythia{}8 for \pflow jets in two example \pt and $\eta$ ranges. Error bars represent the statistical uncertainty. \subref{subfig:examplefit_1} The measured asymmetry is shown for probe jets with $80~\GeV < \pTavg < 110~\GeV$ in the range $0.2 < |\etadet^{\text{probe}}| < 0.7$, where the distributions are symmetric by construction. \subref{subfig:examplefit_2} The measured asymmetry is shown for probe jets with $300~\GeV < \pTavg < 400~\GeV$ in the range $1.3 < |\etadet^{\text{probe}}| < 1.8$. In this $\etadet^{\text{probe}}$ range the distributions can be asymmetric. Two fits are performed iteratively: the particle-level asymmetry is modelled with an \textit{ad~hoc} function which is subsequently convolved with a Gaussian function in order to describe the reconstructed asymmetry. The detector resolution is then extracted from the Gaussian fit parameter.}
\label{fig:dijetjer_fit}
\end{center}
\end{figure}
 
Systematic uncertainties are dominated by imprecise knowledge of the scale of the jets at low \pt, which results in an approximate 1.5\% uncertainty at 40~\GeV, whereas the non-closure of the dijet balance method itself is largely dominant at higher \pT.
The non-closure uncertainty is evaluated as the difference between the resolution measured using the \insitu procedure applied to MC simulation and the particle-level resolution, $\sigma(R)/R$, where  $R=\pTreco/\pTtruth$. Good agreement is found, resulting in an uncertainty in the relative resolution that is approximately 0.4\% and generally increases with \pt due to the non-Gaussian jet response.
At lower \pT the uncertainties propagated from the JES dominate.
The increase in JES uncertainty around 800~\GeV\ is a result of the single-particle uncertainty (see \Sect{\ref{sec:jes-systematics}}): the jet energy scale calibration used for the dijet energy resolution measurement is necessarily based on a smaller dataset than the one presented in this paper, allowing the two measurements to converge simultaneously, and as a result the statistics were lower and the single-particle uncertainty became dominant at a lower \ptjet value than in Figure~\ref{subfig:vs_pt}. Additional systematic uncertainties are estimated by varying the analysis cuts and the JVT selection and by comparing the result with one obtained from an alternative MC generator (\Sherpa 2.1.1).
 
\begin{figure}[htp!]
\begin{center}
\begin{subfigure}[b]{0.49\textwidth}
\includegraphics[width=\textwidth]{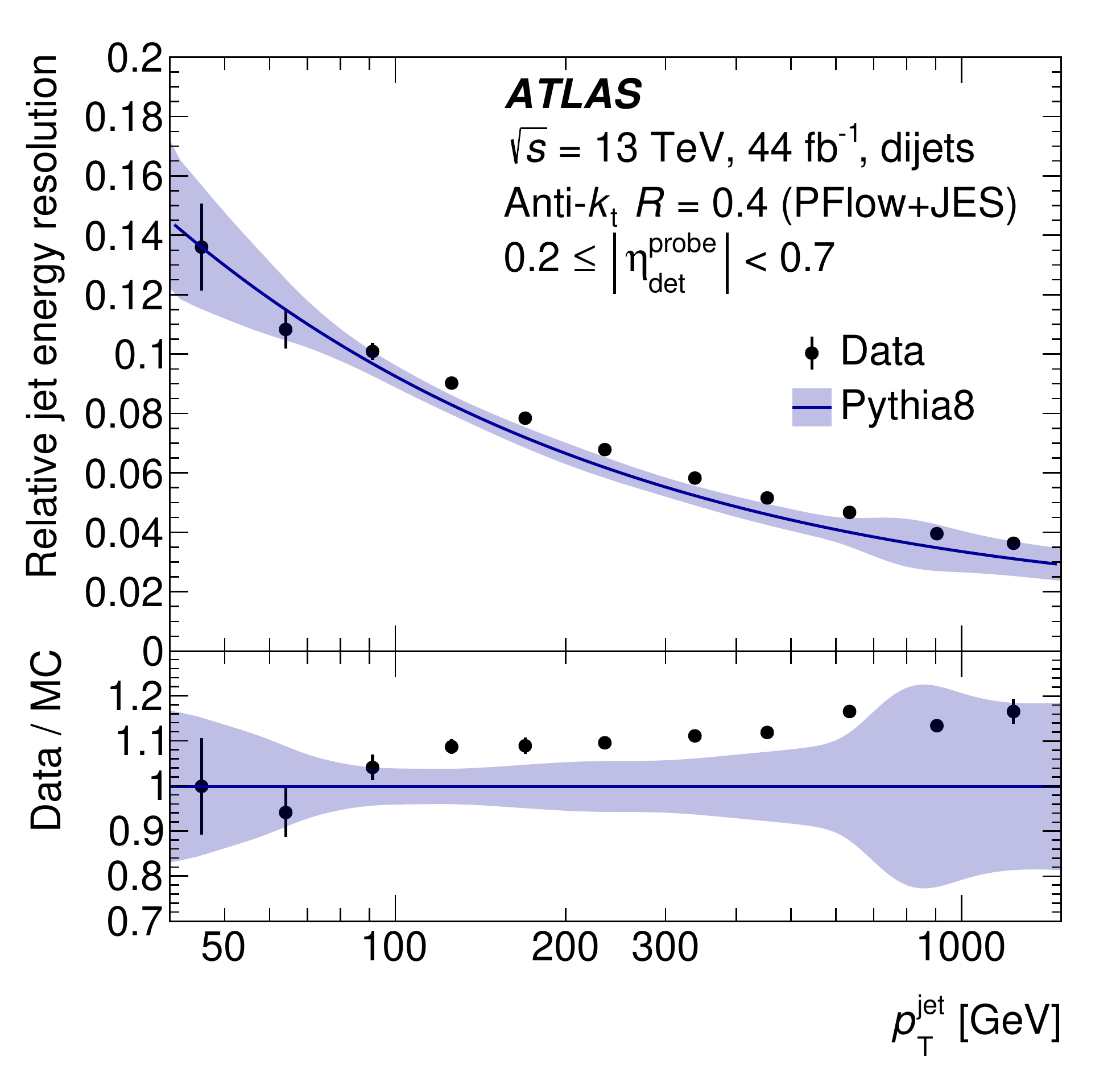}
\caption{}
\label{subfig:res_central}
\end{subfigure}
\begin{subfigure}[b]{0.49\textwidth}
\includegraphics[width=\textwidth]{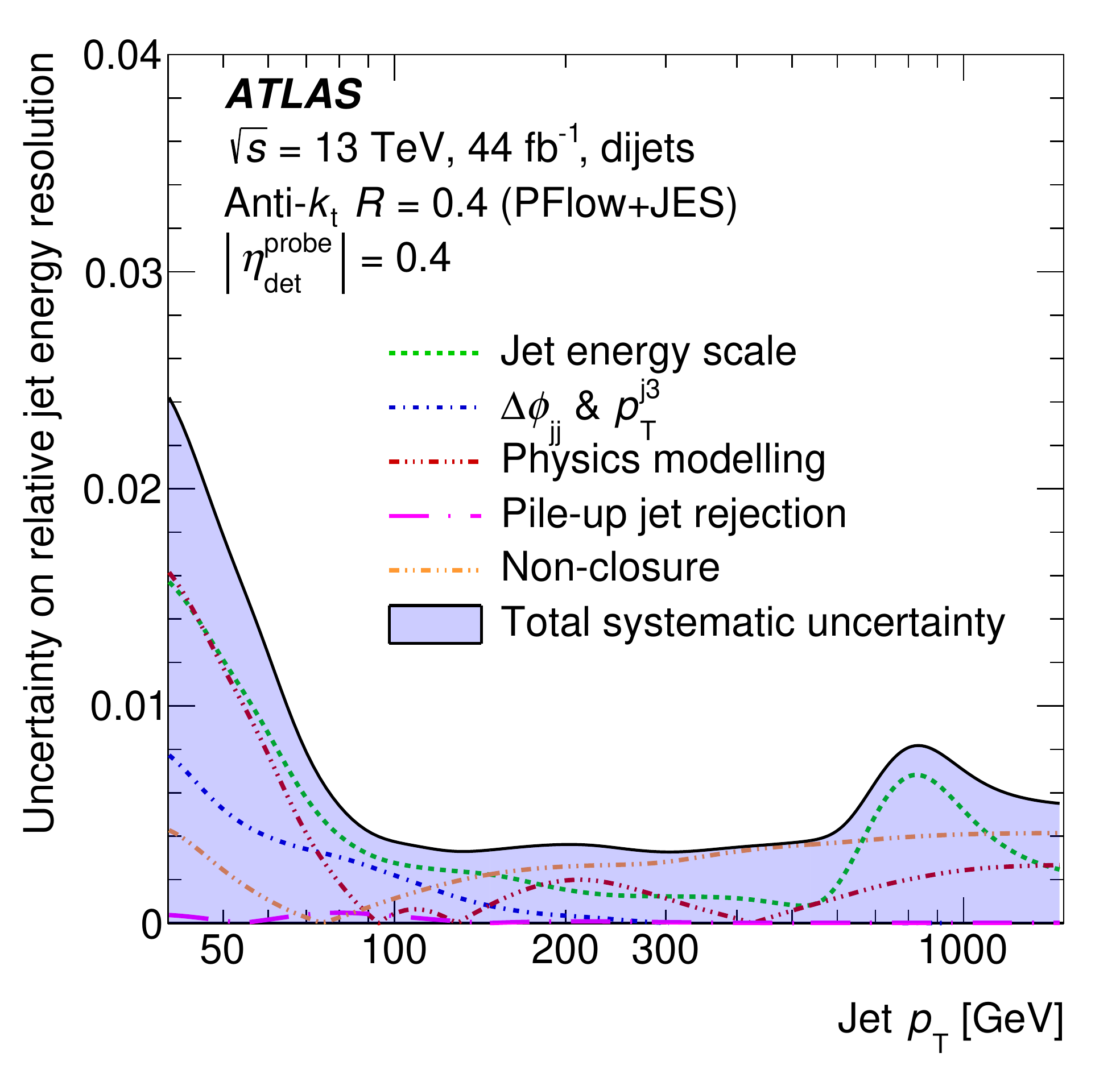}
\caption{}
\label{subfig:res_unc}
\end{subfigure}
\caption{\subref{subfig:res_central} Relative jet energy resolution and \subref{subfig:res_unc} absolute uncertainty in the relative resolution as a function of \pt for \pflow jets in the central region of the detector, measured using the dijet balance method. The resolution in data is shown in black points with error bars indicating statistical uncertainties; the resolution in detector-level simulated events is shown by the blue curve with total systematic uncertainty given by the blue band. The systematic uncertainty is dominated by terms propagated from the JES uncertainty, while additional terms arise from the analysis selection, \pileup rejection (JVT), physics modelling (comparison with alternative generator), and non-closure effects. The bump in uncertainty around 800~\GeV\ comes from the single-particle uncertainty.}
\label{fig:resolution_dijet}
\end{center}
\end{figure}
 
\FloatBarrier
 
\subsection{Noise measurement using random cones}
\label{sec:jer-randomcones}
Direct estimates of the noise term of \Eqn{(\ref{eq:res})} are obtained by measuring the fluctuations in the energy deposits due to \pileup using data samples that are collected by random unbiased triggers. These measurements are performed using the \textit{random cones} method in which energy deposits in the calorimeter are summed at the energy scale of the constituents in circular areas analogous to the jet area for \antikt $R=0.4$ jets. This approach is adopted due to its ability to account for any non-Gaussian behaviour of the noise contributions to the JER.
Two such random cone sums, \ptConeOne and \ptConeTwo, are obtained at random $\phi$ values and within opposite $\pm \Delta\eta$ regions and the difference between them, \deltaPtCone, provides a measure of the random fluctuations of deposited energy. Multiple non-overlapping cones are selected within each event to maximize statistical power; this is demonstrated to cause no bias in the overall result.
This random cone balance is given by
\begin{equation}\nonumber
\deltaPtCone = \ptConeOne - \ptConeTwo,
\label{eq:randomcone-dpt}
\end{equation}
\noindent and the estimated \pileup noise is determined by the central 68\% confidence interval of the distribution of \deltaPtCone, \sigDeltaPtCone, sampled over many events as a function of both $\eta$ and \pileup levels, as indexed by $\mu$. Specifically, the noise term due to \pileup, \NoiseTermPU, is determined as
\begin{equation}
\NoiseTermPU = \frac{\sigDeltaPtCone}{2\sqrt{2}}\,,
\label{eq:randomcone-noiseterm}
\end{equation}
\noindent where the width of the distribution is divided by 2 to obtain the half-width of the distribution, and by $\sqrt{2}$ to obtain the fluctuations corresponding to just a single random cone. The distribution of \deltaPtCone is shown in \Fig{\ref{subfig:randomcones-dpt}}. Updates to the random cone method since its initial description in Ref.~\cite{PERF-2014-02} include removing a restriction to only a pair of back-to-back cones since this was found to have no effect on the result and taking multiple non-overlapping random cone pairs per event to maximise statistics.
 
The energy scale of the noise estimated by \NoiseTermPU in \Eqn{(\ref{eq:randomcone-noiseterm})} is the constituent energy scale and not that of the jets measured in \Sect{\ref{sec:jer-dijets}}. In order to compare the measurement of the noise term \NoiseTermPU using the random cone method with the JER measured at the fully calibrated scale (e.g.\ \EMPFlowJES) a conversion factor is required. The nominal JES calibration factor is used to perform this conversion to the appropriate energy scale. The result is an estimate of the noise due to \pileup that may be directly compared with the measured JER.
 
\begin{figure}
\begin{center}
\begin{subfigure}[b]{0.49\textwidth}
\includegraphics[width=\textwidth]{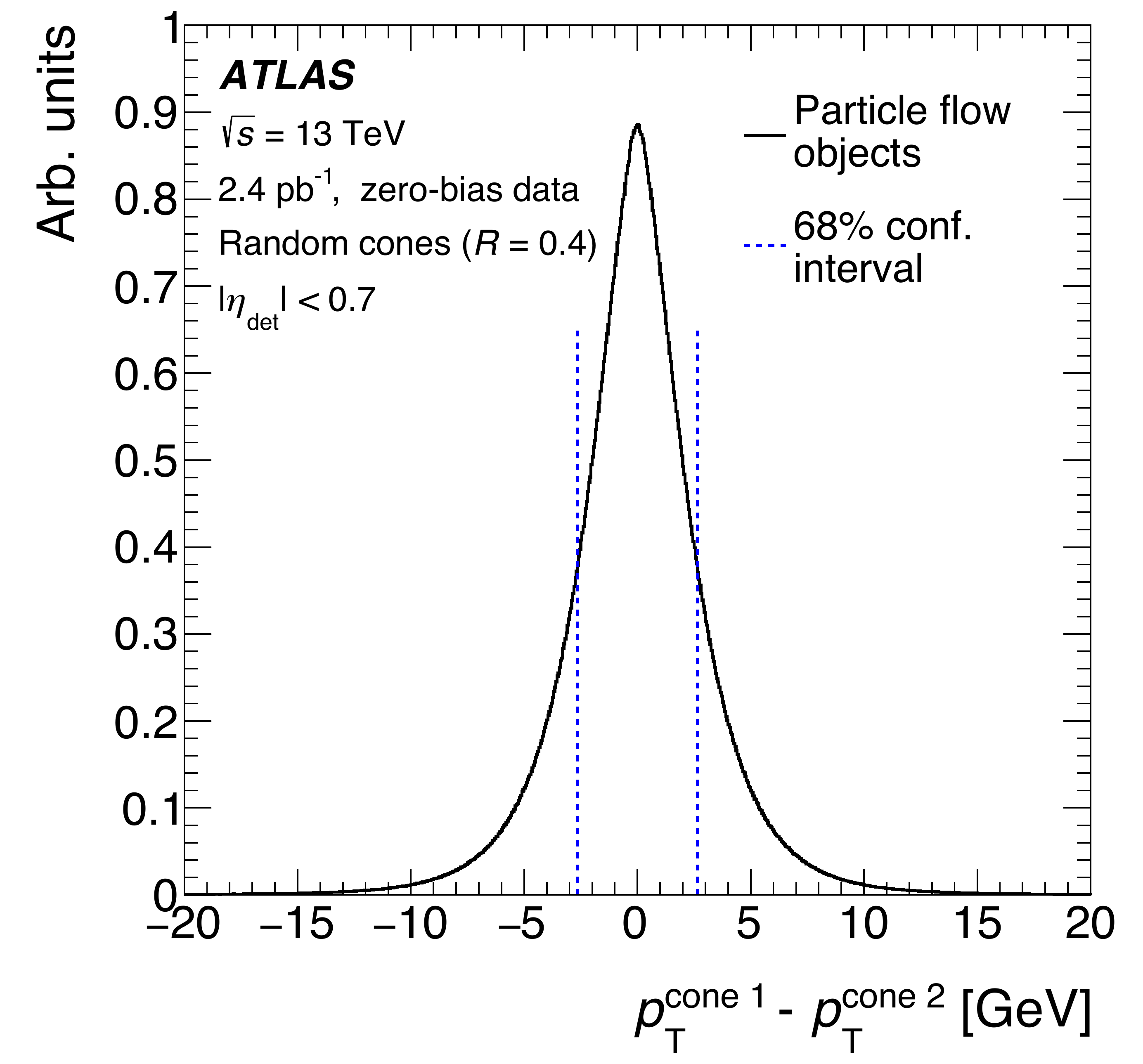}
\caption{}
\label{subfig:randomcones-dpt}
\end{subfigure}
\begin{subfigure}[b]{0.49\textwidth}
\includegraphics[width=\textwidth]{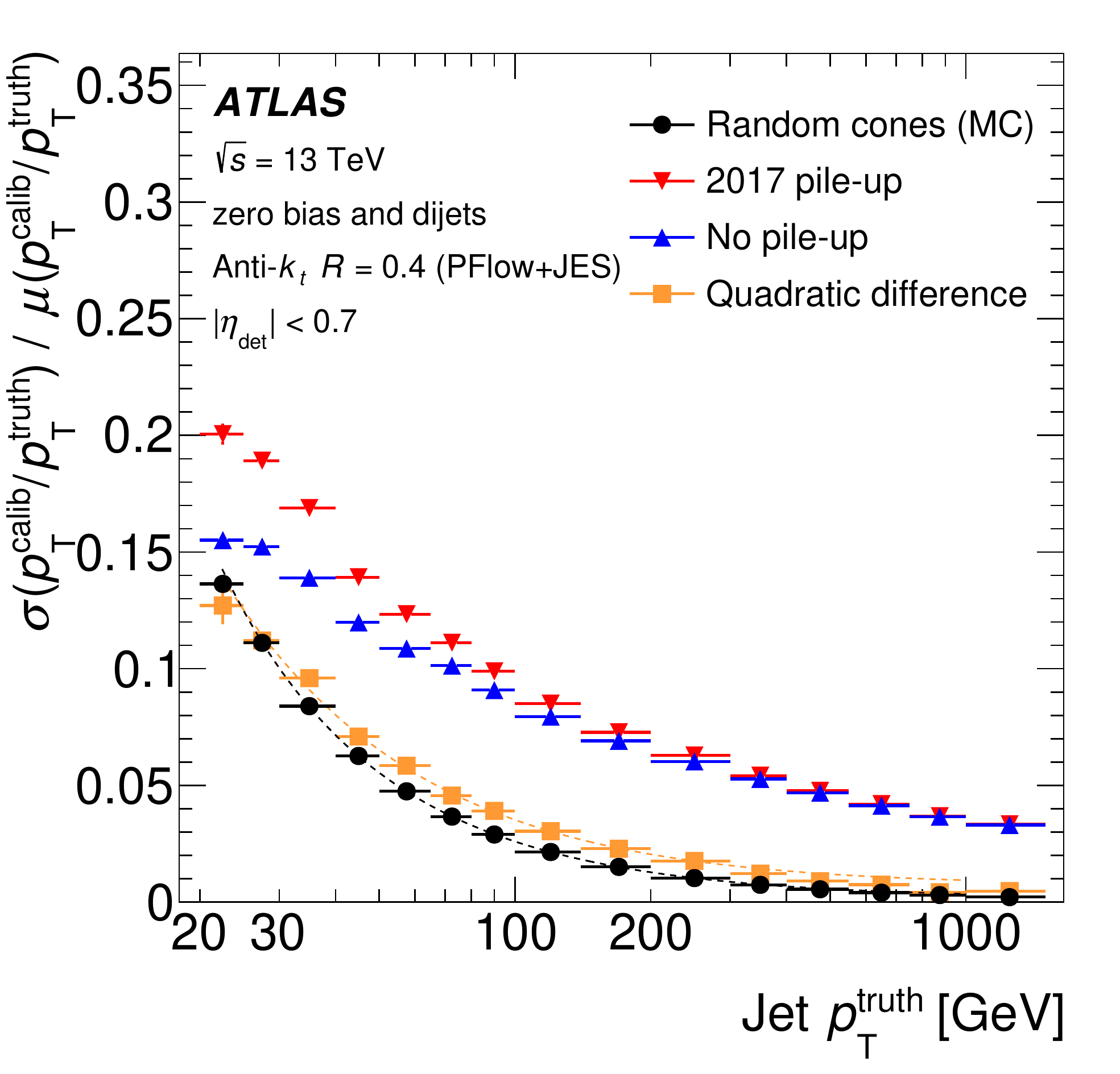}
\caption{}
\label{subfig:randomcones-resolution}
\end{subfigure}
\caption{\subref{subfig:randomcones-dpt} The difference in the random cone sums, \deltaPtCone, measured in the central region ($|\etadet|<0.7$) in randomly triggered data using \pflow objects. \subref{subfig:randomcones-resolution} Comparison between the \pileup noise term \NoiseTermPU determined using the random cone method (black solid circles) and the expectation from MC simulation (orange squares) as extracted from the difference in quadrature of MC simulation with (red downward triangles) and without (blue upward triangles) \pileup. Results are shown at the \EMPFlowJES energy scale for jets in the central region of the detector ($|\etadet|<0.7$).}
\label{fig:randomcones}
\end{center}
\end{figure}

A closure test of the random cone measurements is performed by comparing the \insitu measurement of the calibrated \NoiseTermPU with the expectation from MC simulation. Results are reported here for \pflow jets. To isolate the contribution to the JER from \pileup noise in the MC  simulation, the JER is determined in simulated events both with and without \pileup and a subtraction in quadrature is performed between the extracted resolutions. The two JER determinations in MC simulation events with and without \pileup are shown in \Fig{\ref{subfig:randomcones-resolution}} and their quadratic difference is compared directly to the \insitu measurement from the random cones method. Each is fitted, as shown by the dotted lines in~\Fig{\ref{subfig:randomcones-resolution}}: the random cone measurement is fitted with $N/\pt$ while the quadratic difference is fitted with $N/\pt \oplus S/\sqrt{\pt}$ to account for non-negligible stochastic contributions. The non-closure of the method is largely due to the differences in \topo\ formation sensitivity to \pileup and electronic noise in the presence versus absence of hard-scatter particles, and is taken as a systematic uncertainty in the result. This non-closure uncertainty is the dominant uncertainty in the JER noise term, ranging from approximately 17\% in the most central region to 75\% in the endcap transition region ($2.5 < |\eta| < 3.2$).
 
\begin{figure}
\begin{center}
\includegraphics[width=0.7\textwidth]{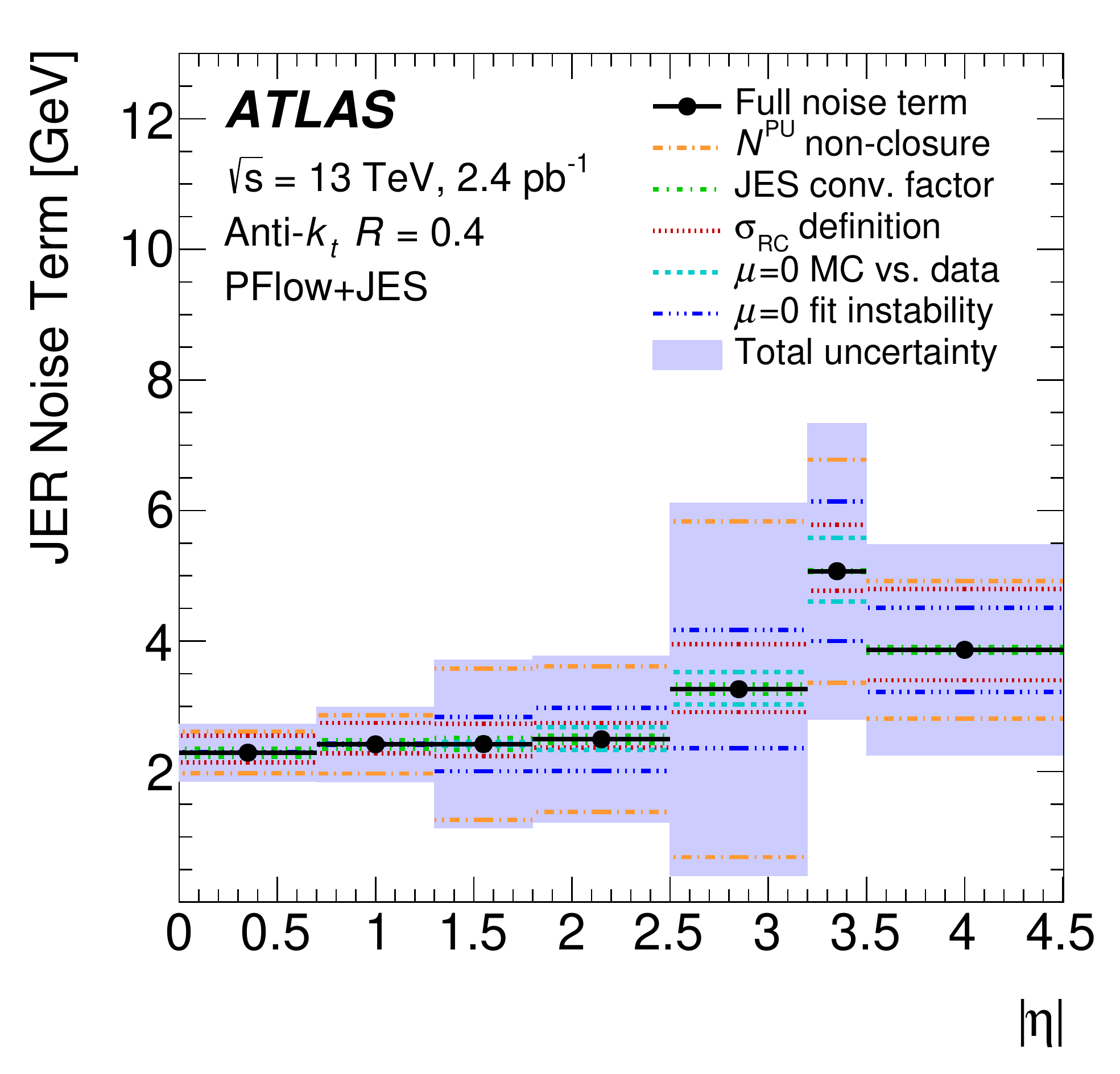}
\caption{Noise term due to \pileup estimated using the random cone method and its uncertainties as a function of $\eta$. The dominant uncertainty is due to non-closure in the method. Additional uncertainties address the \sigDeltaPtCone definition, the JER conversion factor, the differences in JER between data and MC simulation, and the fit stability in extracting the $\mu=0$ noise term. The \sigDeltaPtCone definition uncertainty and $\mu=0$ MC vs data terms are asymmetric in their upwards and downwards components, while all other uncertainties are symmetrized.}
\label{fig:randomcones_v_eta}
\end{center}
\end{figure}
 
The total noise contribution to the JER includes not just \pileup but also electronic noise, to which the random cones are not sensitive due to the \topo ing process. To estimate this electronic contribution, a fit is performed to the JER measured in a dedicated MC simulation sample with $\mu=0$ and the electronic noise term is extracted as \NoiseTermMuZero. The total noise term used in the JER combination is therefore taken to be $\NoiseTerm = \NoiseTermPU \oplus \NoiseTermMuZero$ and is shown as a function of $\eta$ in \Fig{\ref{fig:randomcones_v_eta}} along with its systematic uncertainties. The dominant systematic uncertainty in the random cone measurement of \NoiseTermPU is the previously discussed non-closure uncertainty, but additional terms arise from varying the quantile of the confidence interval used to extract \sigDeltaPtCone and from using a different estimate of the conversion factor to the calibrated JES scale. Two systematic uncertainties apply to \NoiseTermMuZero: a 20\% relative uncertainty conservatively estimating the differences in JER between data and MC simulation and an uncertainty due to the fit parameterization and stability. The systematic uncertainties enter the combined JER fit unsymmetrized in $\eta$ but are symmetrized during the statistical combination, and so the one-sided components are symmetrized in \Fig{\ref{fig:randomcones_v_eta}} to illustrate their final contribution to the total uncertainty.
 
 
\FloatBarrier
 
\subsection{Combination of \insitu jet energy resolution}
\label{sec:jer-combo}
A combined measurement of the JER is obtained by performing a fit to the dijet balance measurements (\Sect{\ref{sec:jer-dijets}}) using a constraint on the noise term (\NoiseTerm) derived from the random cones measurement and $\mu=0$ simulation sample (\Sect{\ref{sec:jer-randomcones}}). The implementation of this statistical combination is performed in a manner nearly identical to that for the JES (\Sect{\ref{sec:jes-combination}}), propagating uncertainties from the dijet measurement in the same way and using a similar eigenvalue decomposition to reduce the final number of nuisance parameters.
 
Instead of using polynomial splines to interpolate across \ptjet, the JER combination uses the functional form from \Eqn{(\ref{eq:res})}. A fit to the dijet measurement data is performed, fixing the noise term to the value measured by the random cone analysis. Dijet measurement uncertainties are taken to be fully correlated between $\eta$ bins.
Uncertainties due to the random cones measurements are determined by propagating the noise term uncertainties and repeating the fit with different values of \NoiseTerm. These uncertainties are taken to be decorrelated between central ($|\eta| < 2.5$) and forward ($|\eta| > 2.5$) regions.
 
\begin{figure}[htp!]
\begin{center}
\begin{subfigure}[b]{0.49\textwidth}
\includegraphics[width=\textwidth]{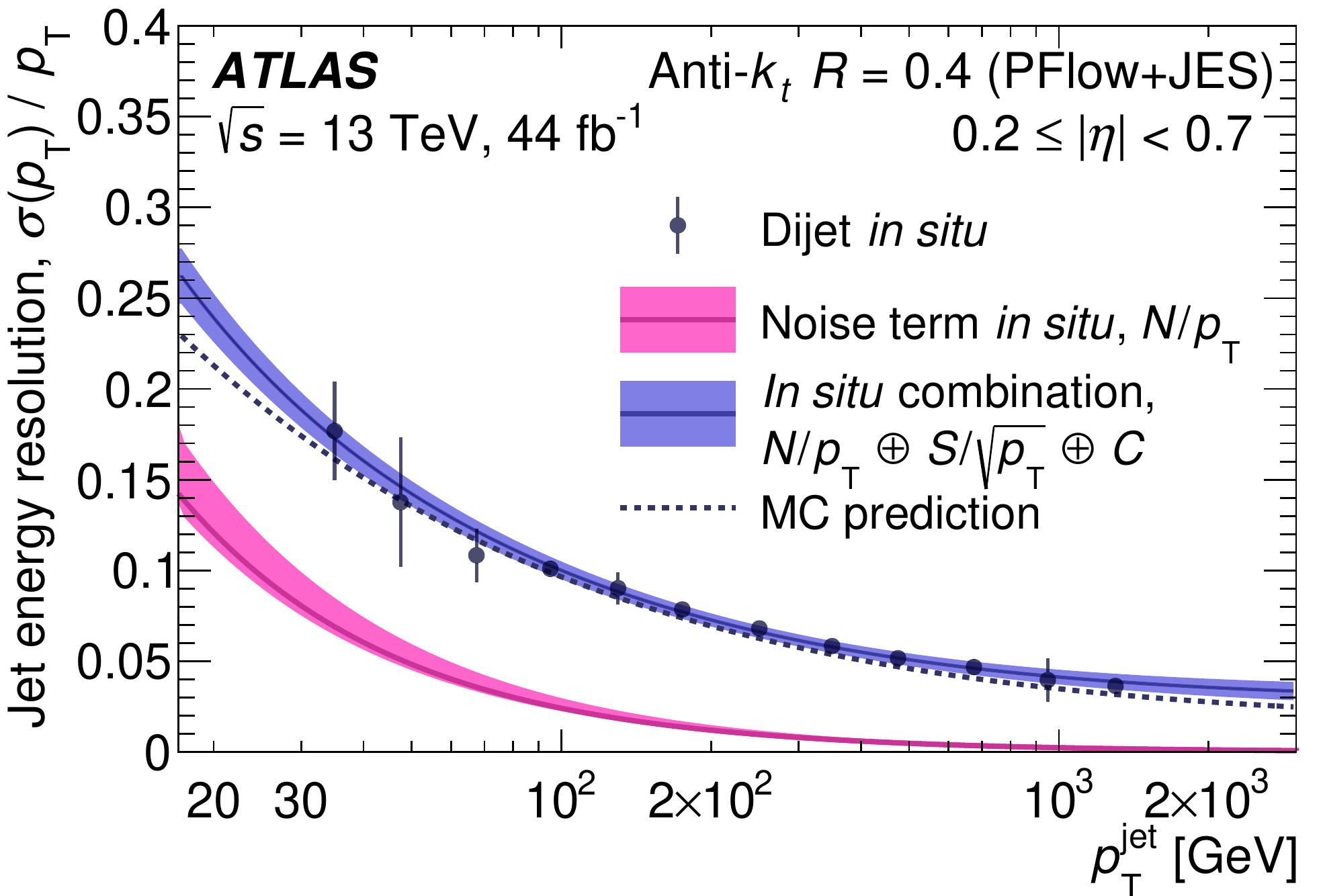}
\caption{}
\label{subfig:jer_combination_plain}
\end{subfigure}
\begin{subfigure}[b]{0.49\textwidth}
\includegraphics[width=\textwidth]{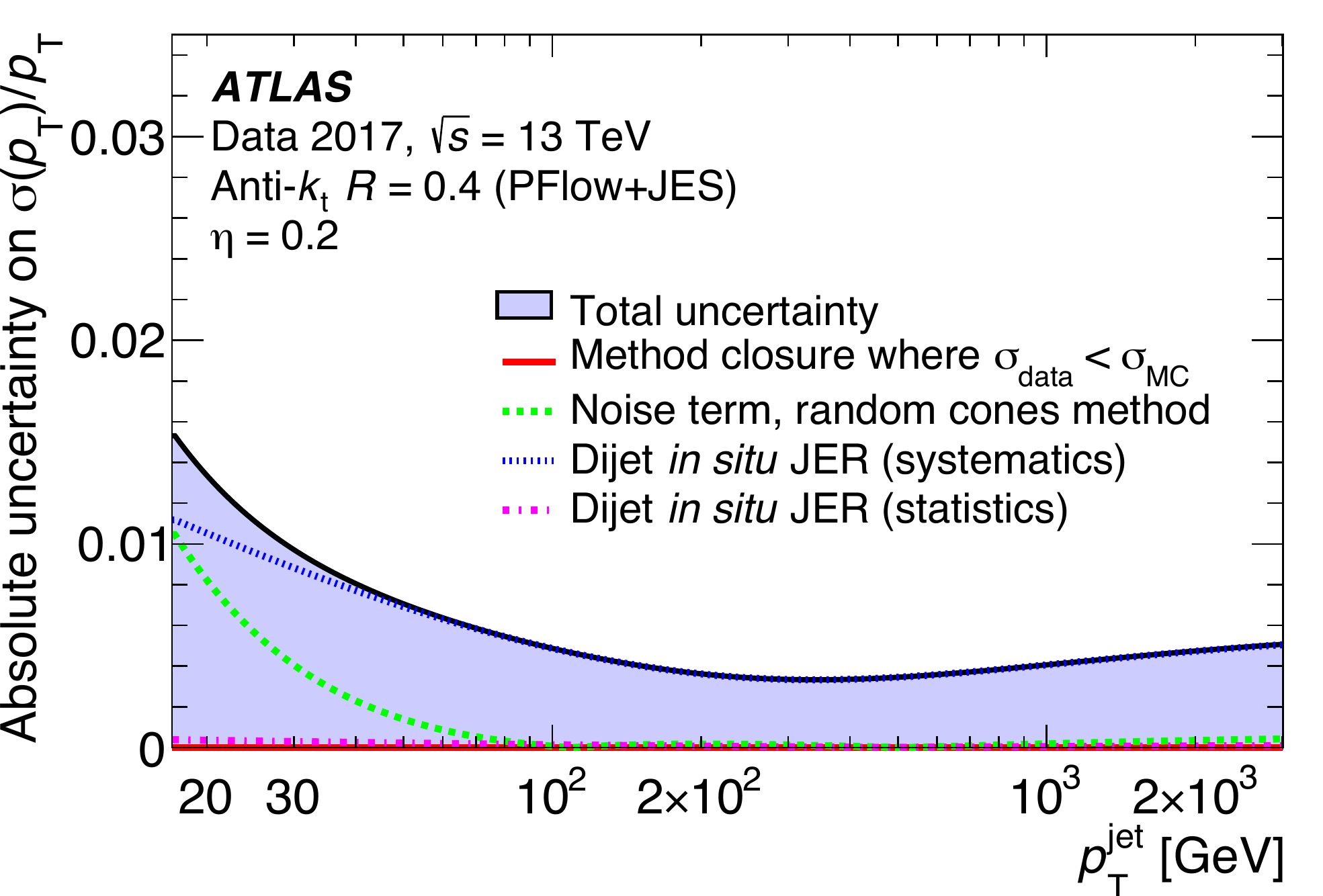}
\caption{}
\label{subfig:jer_combination_comparison}
\end{subfigure}
\caption{\subref{subfig:jer_combination_plain} The relative jet energy resolution as a function of \pt for fully calibrated \EMPFlowJES jets. The error bars on points indicate the total uncertainties on the derivation of the relative resolution in dijet events, adding in quadrature statistical and systematic components. The expectation from Monte Carlo simulation is compared with the relative resolution as evaluated in data through the combination of the dijet balance and random cone techniques. \subref{subfig:jer_combination_comparison} Absolute uncertainty on the relative jet energy resolution as a function of jet \pt. Uncertainties from the two \insitu measurements and from the data/MC simulation difference are shown separately.}
\label{fig:jer_combination}
\end{center}
\end{figure}
 
The resulting combined measurement of the JER for \EMPFlowJES jets is shown in \Fig{\ref{subfig:jer_combination_plain}}. The dijet measurement data points are shown along with the total \insitu combination, while the constraint on the noise term derived from random cones and included in that combination is demonstrated by plotting $N/\pt$ and its uncertainties as a separate curve for illustrative purposes.
\Fig{\ref{subfig:jer_combination_comparison}} shows the absolute uncertainties on the combined JER measurement. For each value of \ptjet and \etadet a toy jet is created and the size of each JER nuisance parameter corresponding to it is retrieved and plotted.
 
Comparisons of the JER measurements for \EMPFlowJES and \EMJES jets, as a function of both $\ptjet$ and $\etajet$, are provided in \Fig{\ref{fig:jer_comparison}}. The fit to the resolution as a function of \pt for the \EMPFlowJES jets shows an improvement in resolution over \EMJES jets at low \pt.
 
\begin{figure}[htp!]
\begin{center}
 
\begin{subfigure}[b]{0.49\textwidth}
\includegraphics[width=\textwidth]{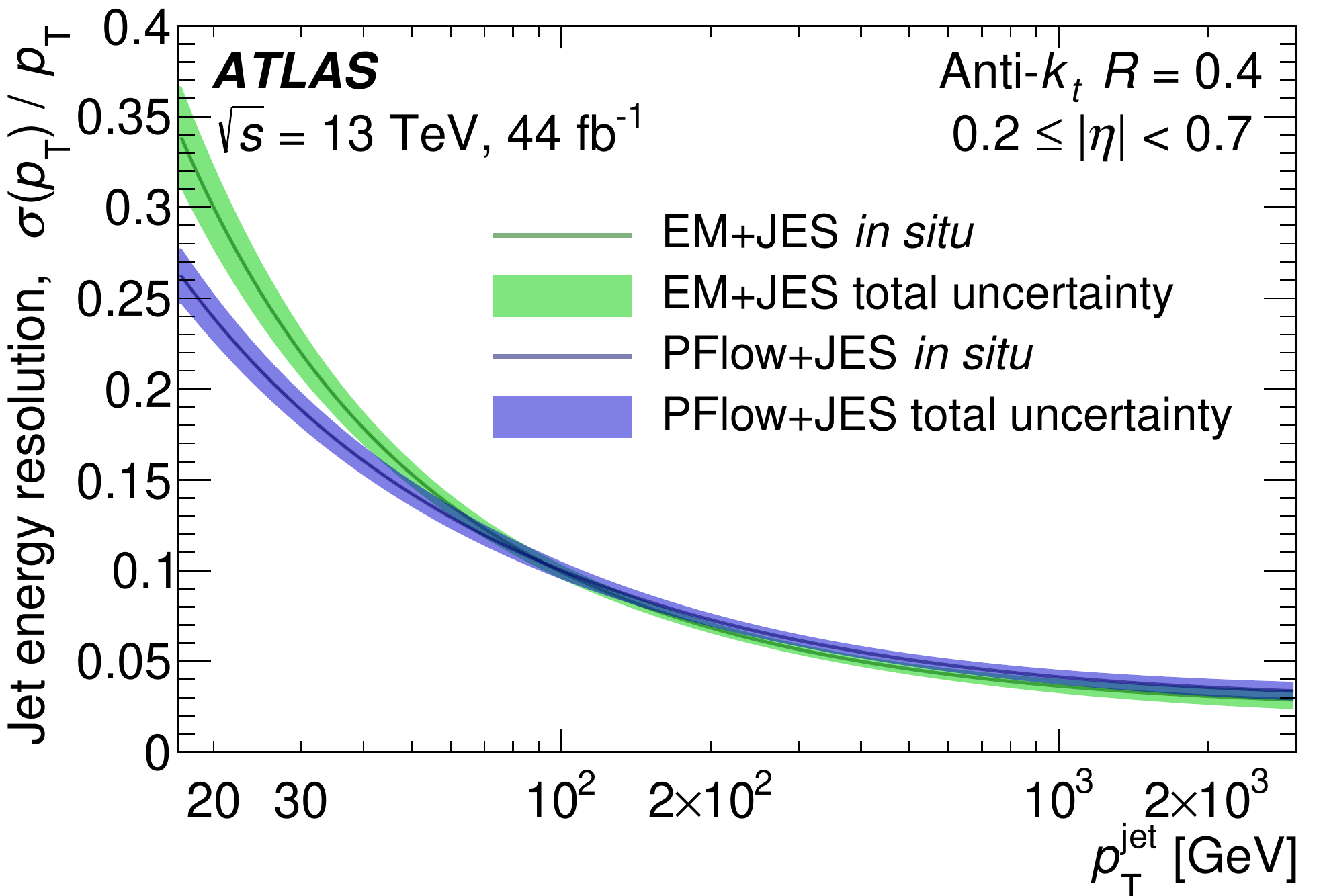}
\caption{}
\label{subfig:jer_comparison_vs_pt}
\end{subfigure}
\begin{subfigure}[b]{0.49\textwidth}
\includegraphics[width=\textwidth]{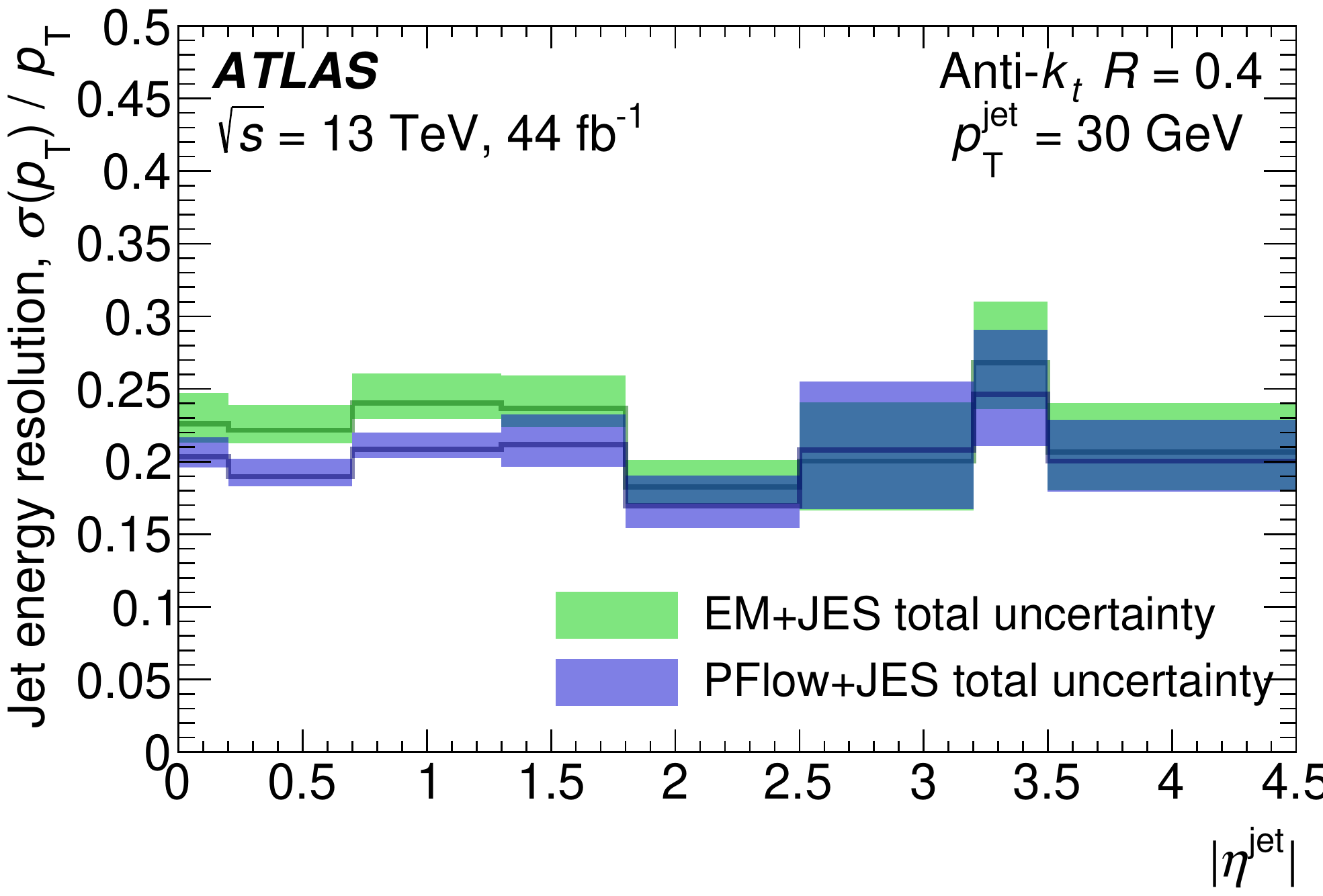}
\caption{}
\label{subfig:jer_comparison_vs_eta}
\end{subfigure}
 
\caption{The relative jet energy resolution for fully calibrated \EMPFlowJES jets (blue curve) and \EMJES jets (green curve) \subref{subfig:jer_comparison_vs_pt} as a function of \ptjet and \subref{subfig:jer_comparison_vs_eta} as a function of $\etajet$. The fit to the resolution as a function of \ptjet for the \EMPFlowJES jets shows an improvement in resolution over \EMJES jets at low-\pt.}
 
\label{fig:jer_comparison}
\end{center}
\end{figure}
 
\Fig{\ref{fig:compare_jer_uncert}} shows the total JER uncertainty in \EMTopo and \pflow jets for a range of \pt values at fixed $\eta=0.2$ and for a range of $\eta$ values at fixed $\pT = 30~\GeV$. The level of agreement is representative of other $\pt$ and $\eta$ ranges.
 
\begin{figure}[htp!]
\begin{center}
 
\begin{subfigure}[b]{0.49\textwidth}
\includegraphics[width=\textwidth]{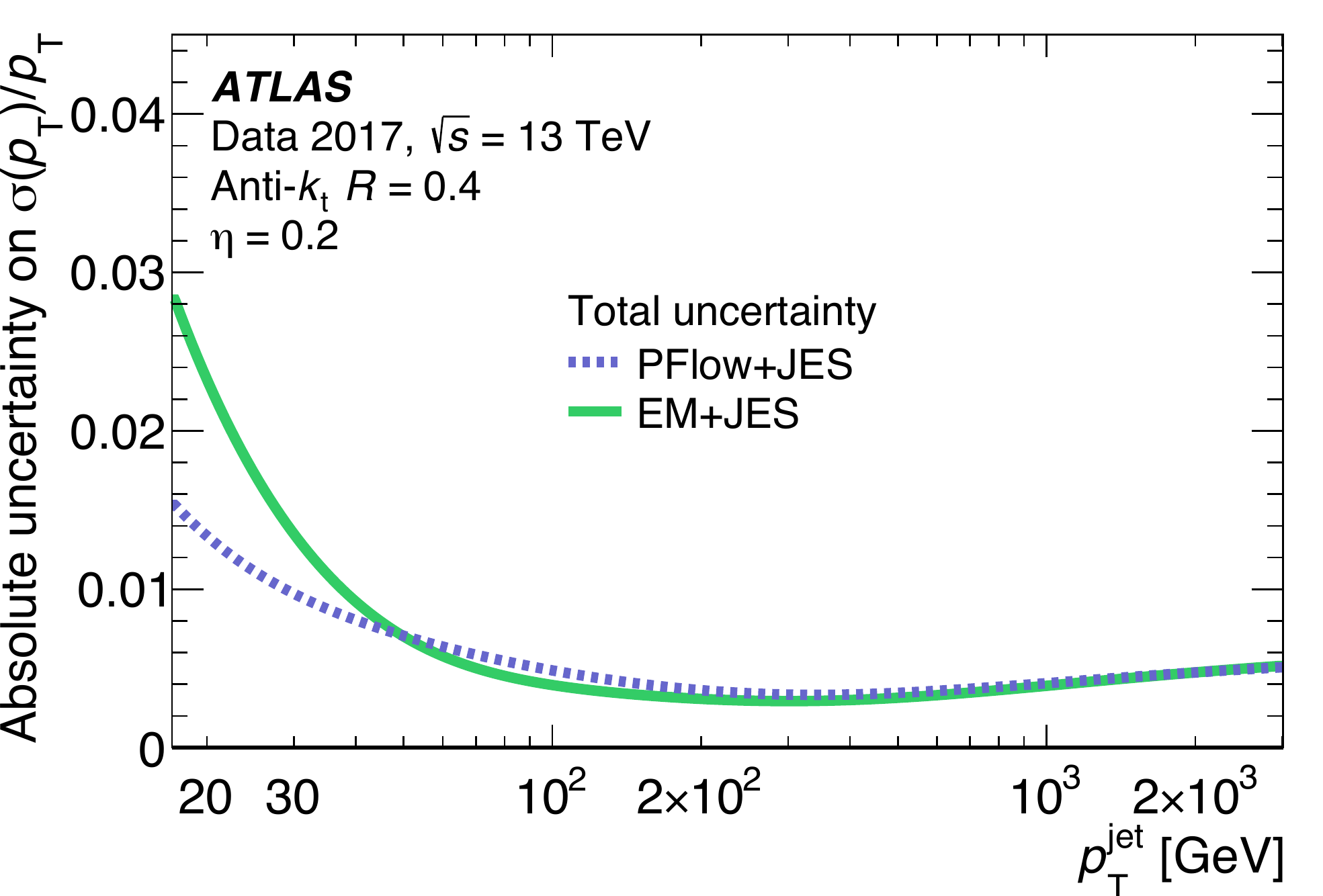}
\caption{}
\label{subfig:jer_unc_comparison_vs_pt}
\end{subfigure}
\begin{subfigure}[b]{0.49\textwidth}
\includegraphics[width=\textwidth]{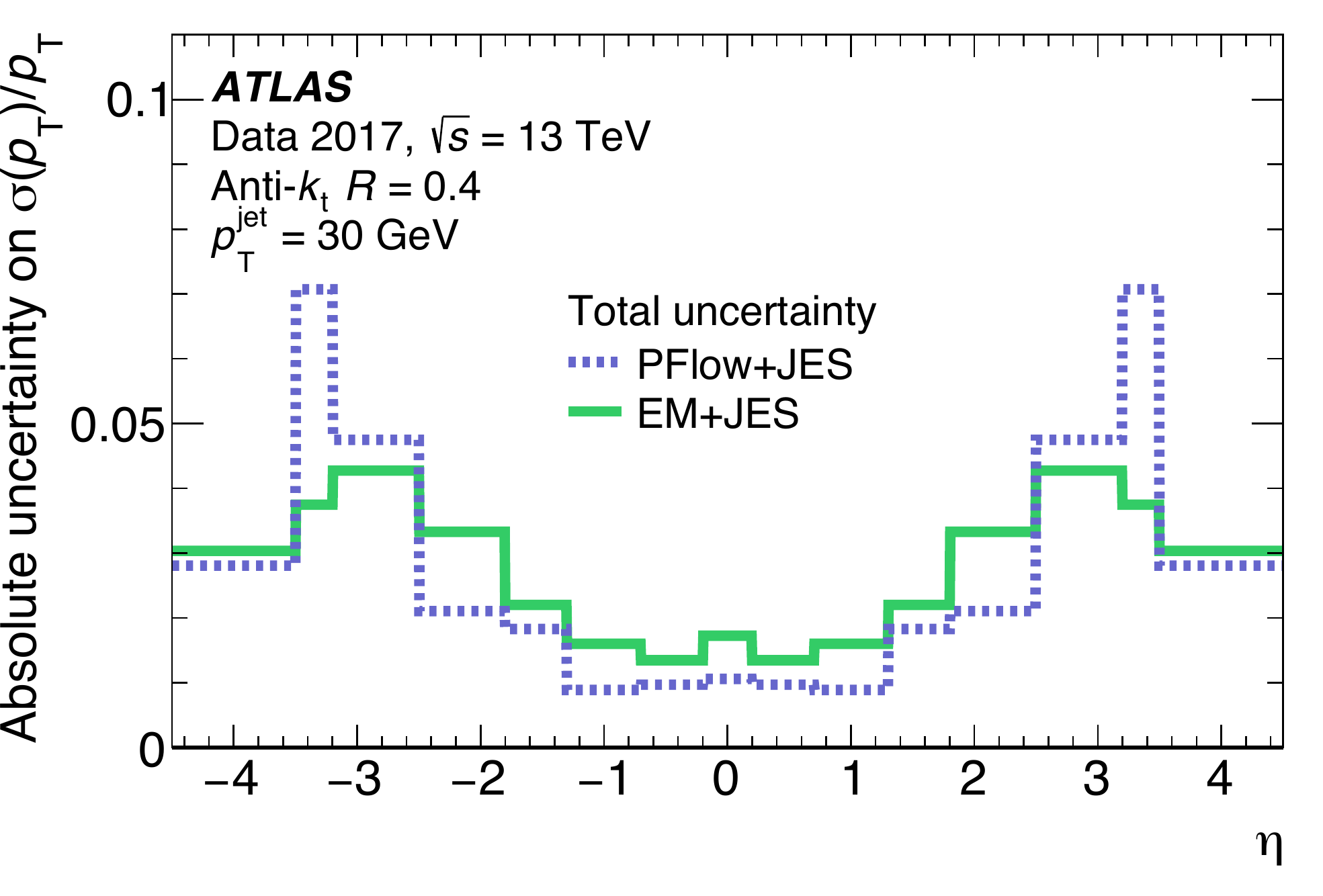}
\caption{}
\label{subfig:jer_unc_comparison_vs_eta}
\end{subfigure}
 
\caption{Fractional jet energy resolution systematic uncertainty summed across all components for \akt $R=0.4$ jets \subref{subfig:jer_unc_comparison_vs_pt} as a function of jet \pt at $\eta = 0.2$ and \subref{subfig:jer_unc_comparison_vs_eta} as a function of $\eta$ at $\pt = 30~\GeV$. The total JER uncertainty is shown for both \EMJES and \EMPFlowJES jets. }
 
\label{fig:compare_jer_uncert}
\end{center}
\end{figure}
 
\FloatBarrier
 
\subsection{Application of JER and its systematic uncertainties}
\label{sec:jer-syst}
In order to ensure that the resolution of the jet energy scale in simulation matches that in data wherever possible, a smearing procedure is recommended. For regions of jet \pt in which the resolution in data is larger than in MC simulation, the simulation sample should be smeared until its average resolution matches that of data. In regions of jet \pt where resolution is smaller in data than in MC simulation, no smearing is performed, since the data should remain unaltered.
 
JER systematic uncertainties are propagated through physics analyses by smearing jets according to a Gaussian function with width $\sigma_{\text{smear}}$. If $\sigma_{\text{nom}}$ is the nominal JER of the sample, after MC simulation smearing if necessary, and $\sigma_{\text{NP}}$ is the one-standard-deviation variation in the uncertainty component to be evaluated, then:
 
\begin{equation*}
\sigma_{\text{smear}}^2 = (\sigma_{\text{nom}}+|\sigma_{\text{NP}}|)^2 - \sigma_{\text{nom}}^2\,.
\end{equation*}
 
Application of JER systematic uncertainties must account for two factors: first, anti-correlations across a single uncertainty component, and second, differences in resolution between data and MC simulation.
 
Anti-correlation becomes an issue when a single JER component is positive in some regions of phase space and negative in others. To propagate such systematic uncertainties to analyses, smearing should be applied to the simulation when $\sigma_{\text{NP}} > 0$ and applied to the data when $\sigma_{\text{NP}}<0$. It should be noted that the nominal data remains unchanged as this applies only to the application of systematic uncertainties. In the case that data statistics are too low to safely smear, pseudo-data may be smeared instead.
 
Differences in resolution between data and MC simulation are already accounted for by the application of additional smearing to the simulation when the resolution in simulation is better than in data. When the JER is smaller in data, this difference is accounted for by applying its full value as an additional systematic uncertainty:
 
\begin{equation*}
\sigma_{\text{NP}} = \sigma_{\text{nom}}^{\text{data}} - \sigma_{\text{nom}}^{\text{MC}}\,.
\end{equation*}
 
This term is defined by the dijet asymmetry measurements of Section \ref{sec:jer-dijets} and is zero for the central $\eta$ slice shown in \Fig{\ref{subfig:jer_combination_comparison}}, but for some \pt ranges in more forward detector regions it can be significant. A large value of this uncertainty for \pflow jets at $\eta\sim3.2$ is the source of the peaks visible in \Fig{\ref{subfig:jer_unc_comparison_vs_eta}}.
 
\FloatBarrier
 
\section{Conclusions}
\label{sec:conclusions}
The calibration of the jet energy scale and resolution for jets reconstructed with the \antikt{} algorithm with radius parameter $R=0.4$ is presented.
Jets are built from either the energy deposits that form topological clusters of calorimeter cells or a combination of charged-particle tracks and topological clusters.
The measurements discussed here use 36--81~\ifb\ of data recorded with the ATLAS detector during 2015--2017 in \pp collisions at a centre-of-mass energy of 13\,\TeV{} at the Large Hadron Collider. It is the first full calibration of \pflow jets performed by the ATLAS collaboration, the first jet energy scale measurement in the high \pileup conditions of late \RunTwo data-taking, and the first jet energy resolution measurement in 13~\TeV\ data.
 
A sequence of simulation-based corrections removes the contribution to the jet energy from additional proton--proton interactions in the same or nearby bunch crossings, corrects the jet so that it agrees in energy and direction with particle-level jets and, improves the jet energy resolution.
Any remaining difference between simulation and data is removed with \insitu techniques using well-measured reference objects, including photons, \Zboson bosons, and other jets, such that the energy scale of fully calibrated jets is unity within uncertainties. The jet energy resolution is measured in a dijet balance system, and the contribution to the resolution from the noise term due to pile-up and electronics is also measured. The relative jet energy resolution ranges from 0.25 (0.35) to 0.04 for \pflow (\EMTopo) jets as a function of jet \pt.
 
Systematic uncertainties in the jet energy scale for central jets ($|\eta|<1.2$) vary from 1\% for a large range of high-\pt jets ($250<\pT<2000~\GeV$), to 5\% at very low \pt ($20~\GeV$) and 3.5\% at very high \pt ($>2.5~\TeV$). The absolute uncertainty on the relative jet energy resolution is found to be 1.5 at 20~\GeV{} decreasing to 0.5 at 300~\GeV.
 
\section*{Acknowledgements}
 

We thank CERN for the very successful operation of the LHC, as well as the
support staff from our institutions without whom ATLAS could not be
operated efficiently.
 
We acknowledge the support of ANPCyT, Argentina; YerPhI, Armenia; ARC, Australia; BMWFW and FWF, Austria; ANAS, Azerbaijan; SSTC, Belarus; CNPq and FAPESP, Brazil; NSERC, NRC and CFI, Canada; CERN; ANID, Chile; CAS, MOST and NSFC, China; COLCIENCIAS, Colombia; MSMT CR, MPO CR and VSC CR, Czech Republic; DNRF and DNSRC, Denmark; IN2P3-CNRS and CEA-DRF/IRFU, France; SRNSFG, Georgia; BMBF, HGF and MPG, Germany; GSRT, Greece; RGC and Hong Kong SAR, China; ISF and Benoziyo Center, Israel; INFN, Italy; MEXT and JSPS, Japan; CNRST, Morocco; NWO, Netherlands; RCN, Norway; MNiSW and NCN, Poland; FCT, Portugal; MNE/IFA, Romania; JINR; MES of Russia and NRC KI, Russian Federation; MESTD, Serbia; MSSR, Slovakia; ARRS and MIZ\v{S}, Slovenia; DST/NRF, South Africa; MICINN, Spain; SRC and Wallenberg Foundation, Sweden; SERI, SNSF and Cantons of Bern and Geneva, Switzerland; MOST, Taiwan; TAEK, Turkey; STFC, United Kingdom; DOE and NSF, United States of America. In addition, individual groups and members have received support from BCKDF, CANARIE, Compute Canada, CRC and IVADO, Canada; Beijing Municipal Science \& Technology Commission, China; COST, ERC, ERDF, Horizon 2020 and Marie Sk{\l}odowska-Curie Actions, European Union; Investissements d'Avenir Labex, Investissements d'Avenir Idex and ANR, France; DFG and AvH Foundation, Germany; Herakleitos, Thales and Aristeia programmes co-financed by EU-ESF and the Greek NSRF, Greece; BSF-NSF and GIF, Israel; La Caixa Banking Foundation, CERCA Programme Generalitat de Catalunya and PROMETEO and GenT Programmes Generalitat Valenciana, Spain; G\"{o}ran Gustafssons Stiftelse, Sweden; The Royal Society and Leverhulme Trust, United Kingdom.
 
The crucial computing support from all WLCG partners is acknowledged gratefully, in particular from CERN, the ATLAS Tier-1 facilities at TRIUMF (Canada), NDGF (Denmark, Norway, Sweden), CC-IN2P3 (France), KIT/GridKA (Germany), INFN-CNAF (Italy), NL-T1 (Netherlands), PIC (Spain), ASGC (Taiwan), RAL (UK) and BNL (USA), the Tier-2 facilities worldwide and large non-WLCG resource providers. Major contributors of computing resources are listed in Ref.~\cite{ATL-SOFT-PUB-2020-001}.
 
 
\printbibliography

\clearpage 
 
\begin{flushleft}
\hypersetup{urlcolor=black}
{\Large The ATLAS Collaboration}

\bigskip

\AtlasOrcid[0000-0002-6665-4934]{G.~Aad}$^\textrm{\scriptsize 102}$,    
\AtlasOrcid[0000-0002-5888-2734]{B.~Abbott}$^\textrm{\scriptsize 128}$,    
\AtlasOrcid[0000-0002-7248-3203]{D.C.~Abbott}$^\textrm{\scriptsize 103}$,    
\AtlasOrcid[0000-0002-2788-3822]{A.~Abed~Abud}$^\textrm{\scriptsize 36}$,    
\AtlasOrcid[0000-0002-1002-1652]{K.~Abeling}$^\textrm{\scriptsize 53}$,    
\AtlasOrcid[0000-0002-2987-4006]{D.K.~Abhayasinghe}$^\textrm{\scriptsize 94}$,    
\AtlasOrcid[0000-0002-8496-9294]{S.H.~Abidi}$^\textrm{\scriptsize 167}$,    
\AtlasOrcid[0000-0002-8279-9324]{O.S.~AbouZeid}$^\textrm{\scriptsize 40}$,    
\AtlasOrcid{N.L.~Abraham}$^\textrm{\scriptsize 156}$,    
\AtlasOrcid[0000-0001-5329-6640]{H.~Abramowicz}$^\textrm{\scriptsize 161}$
% ...
\bigskip
\\

% ...
$^{36}$CERN, Geneva; Switzerland.\\
% ...
$^{40}$Niels Bohr Institute, University of Copenhagen, Copenhagen; Denmark.\\
% ...
$^{53}$II. Physikalisches Institut, Georg-August-Universit\"{a}t G\"ottingen, G\"ottingen; Germany.\\
% ...
$^{94}$Department of Physics, Royal Holloway University of London, Egham; United Kingdom.\\
% ...
$^{102}$CPPM, Aix-Marseille Universit\'e, CNRS/IN2P3, Marseille; France.\\
$^{103}$Department of Physics, University of Massachusetts, Amherst MA; United States of America.\\
% ...
$^{128}$Homer L. Dodge Department of Physics and Astronomy, University of Oklahoma, Norman OK; United States of America.\\
% ...
$^{156}$Department of Physics and Astronomy, University of Sussex, Brighton; United Kingdom.\\
% ...
$^{161}$Raymond and Beverly Sackler School of Physics and Astronomy, Tel Aviv University, Tel Aviv; Israel.\\
% ...
$^{167}$Department of Physics, University of Toronto, Toronto ON; Canada.\\
% ...

\end{flushleft}


\end{document}